%% file: main.tex
\documentclass{article}
\usepackage{natbib}
\usepackage{etex}
\usepackage{ifpdf}
\ifpdf
\usepackage{etex}
\usepackage[pdftex]{graphicx}
\usepackage{eps_and_ps_topdf}
\else
\usepackage[dvips]{graphicx}
\fi
\usepackage{a4wide,amsmath,amsfonts,amssymb,pictex,latexsym}
\include{header_macros}

\newcommand{\CCB}{}
\newcommand{\CCBII}{}

\begin{document}

\title{%
Force and torque acting on particles in a transitionally rough open
channel flow}  

\author{%
  Clemens Chan-Braun, 
  Manuel Garc\'{\i}a-Villalba\footnote{%
    Present address:
    Bioingenier\'{\i}a e Ingenier\'{\i}a Aeroespacial, Universidad
    Carlos III de Madrid, Legan\'es 28911, Spain} 
  and Markus Uhlmann
  \\[1ex]
  Institute for Hydromechanics,\\
  Karlsruhe Institute of Technology,\\
  76131 Karlsruhe, Germany
}


\date{\today}

\maketitle

\input{abstract_mod_v1}

\input{introduction_mod_v1}
\input{setup_mod_v1}

\input{results_mod_v1}
\input{results_for_mod_v1}
\input{results_torq_mod_v1}

\input{discussion_mod_v1}
\input{conclusions_mod_v1}

\input{appendix_paper_mod_v1}


\input{main.bbl}
\end{document}

%% file: header_macros.tex
\newcommand{\T}{\mathcal{T}}

\newcommand{\U}{\left\langle u \right\rangle}

\newcommand{\uv}{\left\langle u^\prime  v^\prime  \right\rangle}

\usepackage[usenames,dvipsnames]{color}
\usepackage{tikz}
\usetikzlibrary{shapes,arrows,positioning,calc,through}
\tikzstyle{every picture}+=[remember picture]
\usetikzlibrary{patterns}

%% file: abstract_mod_v1.tex
\begin{abstract}
  Direct numerical simulation of open channel flow over a 
  geometrically rough wall has been performed at a bulk Reynolds number
  of $Re_b \approx 2900$. 
  The  wall consisted of a layer of spheres in a square
  arrangement. Two cases have been considered. In the first case
  the spheres are small (with diameter equivalent to $10.7$ wall
  units) and the limit of the hydraulically smooth 
  flow regime is  approached.  
  In the second case the spheres are 
  more than three times larger
  ($49.3$ wall
  units) and the flow is 
  in the transitionally rough flow regime.
  Special emphasis is given on the characterisation of the
  force and torque acting on a particle due to the turbulent flow.
  It is found that in both cases the mean drag, lift and spanwise
  torque are to a large extent produced at the top region of the
  particle surface. 
  The intensity of the particle force fluctuations
  is significantly larger in the large-sphere case, while the trend 
  differs for the fluctuations of the individual components of the
  torque.  
  A simplified model is used to show that the torque fluctuations
  might be explained by the spheres acting as a filter with respect to
  the size of the flow scales which can effectively generate torque
  fluctuations. 
  Fluctuations of both force and torque are found to exhibit strongly
  non-Gaussian probability density functions with particularly long
  tails, an effect which is 
  more pronounced in the small-sphere case.
  Some implications of the present results for sediment erosion are
  briefly discussed. 
\end{abstract}

%% file: introduction_mod_v1.tex
\section{Introduction}
Sediment erosion by turbulent open-channel flow is an important aspect
for fluvial engineering 
applications as it can, for example, cause the collapse of
bridges. 
The mechanisms that lead to sediment erosion however are far from
being completely 
understood due to the complex interactions between the turbulent flow and the
sediment particles. 
Turbulent flow in an open-channel is statistically
inhomogeneous in at least one spatial direction, 
and the Reynolds numbers of interest are typically high, which leads
to a wide range of velocity and length scales.
In addition, the 
presence of a range of sediment sizes, shapes and compositions
complicates the description. 
Under certain conditions, the turbulent motions
might erode the bed and entrain sediments as a result of the
hydrodynamic force and torque acting on the particles. 
In order to improve the understanding of the mechanisms that lead to
sediment erosion it 
appears necessary to simplify the problem under
consideration.  
Therefore, as a first step we study 
the statistical properties of hydrodynamic force and torque acting on
fixed spherical particles adjacent to the wall-plane in
fully-developed, open-channel flow.  
For this configuration, we have performed direct numerical simulations
(DNS).

A wide body of literature exists that focuses on the hydrodynamic force
acting on spherical 
objects placed in a fluid flow. 
In the low Reynolds number range analytical solutions have been proposed
for various flow configurations, e.g.~the case of a particle in a
linear shear flow
\citep{Saffman_JFM_1965, Auton_JFM_1987}, in a non-uniform rotational
flow \citep*{Auton_Hunt_Prudhomme_JFM_1988}, and of a particle in the
vicinity of a smooth wall \citep{Krishnan_Leighton_POF_1995}.
In order to gain information on the mechanism that leads to lift and
drag on a particle in the range from small to moderate Reynolds
numbers, similar flow 
configurations have also been explored
by means of experimental studies \citep{King_Leighton_POF_1997}
and by means of direct numerical simulations
\citep{Kim_Elghobashi_Sirignano_JFM_1993, Bagchi_Balachandar_JFM_2002,
  Zeng_Najjar_Balachandar_Fischer_POF_2009, Lee_Balachandar_JFM_2010}.
In the high Reynolds number limit numerous studies can be found that
describe  the flow around spheres in unbounded flow 
\cite*[see][for an overview]{Yun_Kim_Choi_POF_2006}. 

The studies mentioned above have focused on situations in which the
flow field approaching the sphere is laminar
in nature. It is well known, however, that turbulence can have a
significant effect on the 
statistics of the forces acting on a sphere.
A review on the effect of turbulence on an isolated sphere can be found in
\cite{Bagchi_Balachandar_POF_2003}. 
The authors studied the forces on an isolated sphere subject to
free-stream isotropic turbulence for small and moderate Reynolds
numbers by means of direct numerical simulation. 
They found that turbulence
had only little effect on the mean drag and that the fluctuations of
lift and drag scaled linearly with both the mean drag and the
turbulence intensity. 

In contrast, turbulence 
appears to have a significant effect in the case of a sphere positioned 
close to a wall, 
the lift being particularly affected 
\citep{Willetts_Murray_JFM_1981, Hall_JFM_1988,
  Zeng_Balachandar_Fischer_Najjar_JFM_2008}. 
The experimental evidence shows, that similar to the low Reynolds
number regime, significant positive values for mean lift (directed
away from the wall) are obtained
for a sphere touching the wall plane \citep*{Willetts_Murray_JFM_1981,
  Hall_JFM_1988, Mollinger_Nieuwstadt_JFM_1996,
  Muthanna_Nieuwstadt_Hunt_EIF_2005}. 
When the sphere is not touching the wall, the picture is less clear and 
still a matter of discussion: 
both positive and negative values of the lift 
are reported. \cite{Willetts_Murray_JFM_1981} found changes in sign
for the value of the mean lift when increasing the wall distance;
\cite{Hall_JFM_1988} measured consistently positive values for various wall
distances; \cite{Zeng_Balachandar_Fischer_Najjar_JFM_2008} obtained
negative values (directed towards the wall) in case the sphere is
placed in the buffer layer.  
\cite{Zeng_Balachandar_Fischer_Najjar_JFM_2008} note that the
classical formulae based on unbounded shear flow fail to predict their
DNS results correctly, stating that further investigations are required to
understand the discrepancy. 

\cite{Hall_JFM_1988} showed that the effect of a nearby wall on the lift
experienced by a spherical body
differs significantly depending on the wall being rough or smooth. 
In particular, it was found that 
the lift significantly decreased when the sphere was positioned
in between of spanwise oriented, rod-shaped roughness elements.
When the sphere was positioned on top of
the array of wall-mounted rods, however, the measured lift was
comparable to the corresponding smooth-wall values. 

The difficulties related to the direct measurement of particle forces
as in the studies above 
have been discussed by
\cite{Muthanna_Nieuwstadt_Hunt_EIF_2005}. Another, more indirect approach 
was taken by \cite{Einstein_Elsamni_RMP_1949}.
They approximated the force exerted on
hemispheres in an open channel flow 
by pressure measurements on top and near the bottom of the
hemispheres. They reported positive lift on the hemispheres,
and were among the first who stated the  
relevance of the forces on particles in a rough wall to the
understanding of sediment erosion.
More recent studies following this approach 
present approximations of lift and drag on cubes, spheres and naturally
shaped stones by local pressure measurements
\citep*{Hofland_etal_JHE_2005, Hofland_Battjes_JHE_2006, 
  Detert_Weitbrecht_Jirka_JHE_2010}. 
These studies have focused on the higher
Reynolds number regime with particle Reynolds numbers
of the order of thousands. 

The investigations discussed so far have for the most part
concentrated on the flow around single spherical objects. 
Beyond these studies, a large body of literature exists which deals
with the characteristics of flow over rough surfaces. Although the
precise nature of the fluctuating forces acting on individual
roughness elements is often not of interest in the context of
studies of roughness effects, findings from that research area
are of relevance here. 
A reference for the earlier work on roughness is
\cite{Schlichting_1965}; a more recent review on the subject,
including numerical studies is given by
\cite{Jimenez_ARFM_2004}. 
Some aspects of rough-wall flows at high Reynolds number have
recently been reviewed by \cite{Marusic_etal_POF2010}, in particular
the question whether 
roughness does indeed modify the 
turbulence structure in the outer flow or simply provides a modified 
friction velocity.

The research on rough wall turbulence focuses 
almost exclusively on the effect of the rough wall on the fluid. Some of the
key questions of interest are how roughness influences the turbulence
structure, what are the consequences for scaling, 
and how can the effect on the fluid be estimated from 
the roughness geometry.

Numerical studies of rough wall flows are
very demanding in terms of computational cost,
much more so than comparable
simulations of flow over smooth walls. 
Therefore, far less direct numerical studies of flow over rough walls
have been carried out so far in contrast to flows over smooth wall. 
Recently a direct numerical study of a boundary layer over surfaces
roughened by rectangular spanwise bars and cubes has been carried out by
\cite{Lee_sung_JFM_2007} and \cite*{Lee_sung_krogstad_JFM_2011}. 
 Direct numerical simulation of channel flow over a wall
  similarly roughened by spanwise-oriented square bars has been
  carried out by  
\cite{leonardi_orlandi_smalley_djenidi_antonia_JFM_03, 
leonardi_orlandi_antonia_POF_2007} among others,
while \cite{Orlandi_Leonardi_JFM_2008} have simulated plane
channel flow including different layouts of wall-mounted cubes.
Direct numerical simulations of channel flow with wall velocity
disturbances (acting as artificial roughness) have been carried out by
\cite{orlandi_lonardi_tuzi_antonia_POF_2003} and \cite{Flores_Jimenez_JFM_06}. 
More in line with the present setup,
\cite*{Singh_sandman_williams_JHR_07} have performed simulations of open
channel flow over spheres in hexagonal arrangement, albeit at
considerably coarser resolution than the one employed in the present
study. 

As a first step to understand the mechanism leading to sediment
erosion 
here we present high-fidelity data on the flow over a rough wall with
a regular array of fixed spheres.  
In contrast to most previous studies on roughness, 
the emphasis of the present work is 
on the effect of the turbulence on the spherical elements which form
the rough wall, including the characteristics of hydrodynamic force
and torque.  

The article is structured as follows. 
In \S\ref{sec:num_setup} the setup of the simulation is described and
basic definitions are given. The section also includes a brief
discussion of the numerical scheme used. 
In \S\ref{sec:results_discussion} the results are discussed and
compared with previous findings in the literature when possible.
The results of the time and spatially averaged flow field statistics
are discussed in \S\ref{ssec:stat_flow}, followed by a discussion of
the time-averaged three-dimensional flow field statistics in
\S\ref{ssec:3d_flow}. 
The statistics of the particle force and particle torque respectively
are presented in \S\ref{ssec:stat_forces} and 
\S\ref{ssec:stat_torq}, jointly with
their probability density function (pdf) and the local surface
distribution of the contribution to the mean values.  
In the discussion the results are related to some degree to forces and
torque on an area element in a smooth wall channel flow. 
Conclusions and an outlook are given in \S\ref{sec:conclusion}. 
%

%% file: setup_mod_v1.tex
\section{Flow configuration} \label{sec:num_setup}
\begin{figure}
  \begin{center}
    \begin{minipage}{.4\linewidth}
      \includegraphics[clip,bb=  100 25 450 250,
      width=1.\linewidth]{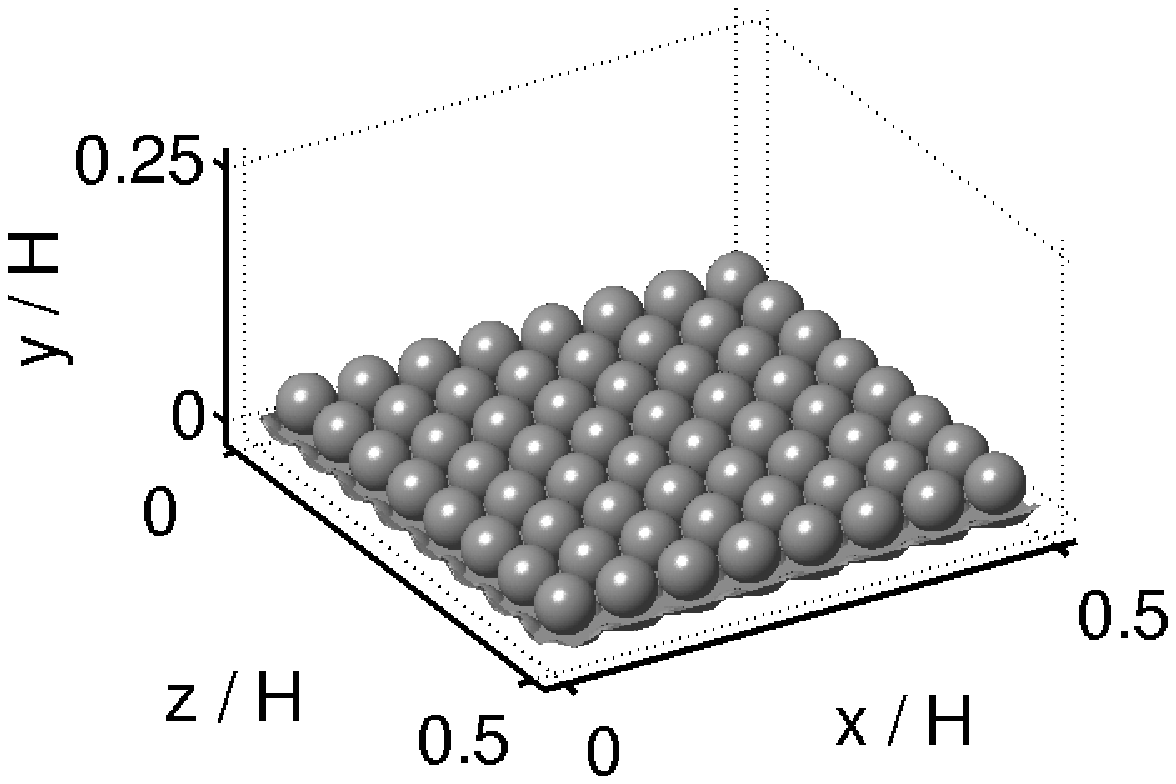}
      \hspace{-0.85\linewidth}\raisebox{0.6\linewidth}{$(a)$}%
    \end{minipage}
    \begin{minipage}{.4\linewidth}
      \includegraphics[clip,bb=  100 25 450 250,
      width=1.\linewidth]{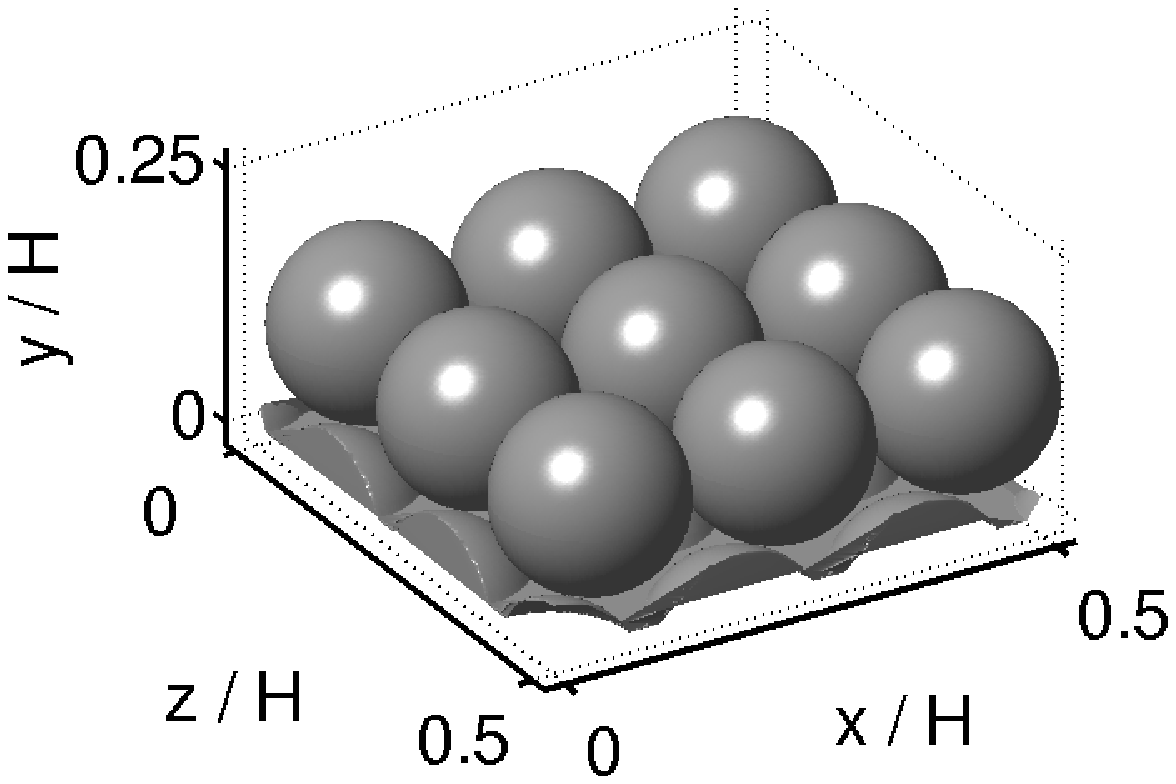} 
      \hspace{-0.85\linewidth}\raisebox{0.6\linewidth}{$(b)$}%
    \end{minipage}
  \end{center}
  \caption{Close-up of a section of
    the computational domain with
    the geometry of the bottom
    wall consisting of a layer of fixed spheres arranged on
    a square lattice; 
    $(a)$ case F10; $(b)$ case F50.} 
  \label{fig:domain_s}
\end{figure}

The flow configuration consists of turbulent open channel flow over a
geometrically rough wall.  
The  wall is formed by one layer of fixed spheres which are
packed in a square arrangement (see figure \ref{fig:domain_s}). The
distance between the particle centres is $D + 2\Delta x$, where $D$ is the
particle diameter and $\Delta x$ is the grid spacing. 
At $y=0$ a rigid wall is located below the layer of spheres. As
can be seen in figure \ref{fig:domain_s} this 
rigid wall is roughened by spherical caps
that can be defined as the part above $y=0$ of spheres located at 
$y= D/2 -\sqrt{2} (D/2  +\Delta x )$, 
staggered in the streamwise and spanwise
direction with respect to the layer of spheres above.  

The physical and numerical
parameters of the simulations are summarised in table \ref{tab:setup_param}. 
The computational domain dimensions are $L_x / H \times
L_y / H \times L_z / H = 12 \times 1 \times 3$, in streamwise,
wall-normal and spanwise direction, respectively.  
An equidistant Cartesian grid with $3072 \times 256 \times 768$ grid points
is employed.

One important parameter is the ratio between the domain height, $H$, 
and the spheres diameter, $D$. Ideally, a large $H/D$ is desirable
to ensure that the spheres can be considered as roughness and not as
obstacles in a channel \citep{Jimenez_ARFM_2004}. However,
from a practical point of view it is difficult to reach large values
of $H/D$
without increasing excessively the computational cost. In this work, two 
cases are considered: case F10 with $H/D=18.3$ and a total of 9216 
particles, and case F50 with
$H/D=5.6$ and a total of 1024 particles above the bottom wall. 

Periodic boundary conditions are applied in streamwise and spanwise
directions. At the upper boundary a free-slip condition is employed. At the 
bottom boundary a no-slip
condition is applied. The spheres are resolved using the immersed
boundary method 
which is described in \S\ref{sec:method}.

%
\begin{table}
\begin{center}
\begin{tabular}{lcccccccc}
Case &   $U_{bh} / u_\tau $  & $Re_{b}$ &
$Re_\tau$ & $D^+$ & $D/\Delta x$ & 
$\Delta x^+$ & $N_p$ &  $\T_c U_{bH} / H$\\ 
F10 &  15.2 & 2870 & 188 & 10.7 & 14 & 0.77 & 9216 &
$120$ \\ 
F50 &  12.2 & 2880 & 235 & 49.3 & 46 & 1.07 & 1024 &
$120$ 
\end{tabular}
\caption{Setup parameters of simulations;
  $U_{bH}$ is the bulk velocity based on
  the domain height $H$, $U_{bh}$ is the bulk velocity based on
  the effective flow depth $h$ defined as $h=H-0.8D$,
  $u_\tau$ is the friction velocity, 
  $Re_{b}=U_{bH}H/\nu$ is the bulk Reynolds number, $Re_\tau=u_\tau h
  / \nu$ is the friction Reynolds number, $D^+= D u_\tau/\nu$
  is the particle diameter in viscous units, $D / \Delta x$ 
  is the resolution of a particle, $\Delta x^+$ is the grid
  spacing in viscous units,
  $N_p$ is the total number of
  particles in a layer, 
  $\T_c$ is the time over which statistics were collected.} 
\label{tab:setup_param}
\end{center}
\end{table}
%
%
In order to scale the results,
two quantities need to be specified: the friction
velocity $u_\tau$ and the location of the virtual wall, $y_0$, since
for a  geometrically
rough wall the 
position of the wall cannot be unambiguously
defined \citep[cf.][]{Townsend_1971,raupach_antonia_rajagopalan_AMR_1991}. 
As discussed in detail in appendix \S\ref{ssec:appendix_utau_y0}, we
choose to define the position of the virtual wall as $y_0=0.8D$ 
throughout this study.
The value of $u_\tau$ is defined by extrapolating the total shear
stress  $\tau_{tot}=
\rho\nu\partial\langle u\rangle/\partial y
-\rho\langle u^\prime v^\prime\rangle$
from above the roughness layer (where it varies linearly)
down to the location of the virtual wall $y_0$.
%
The effective
flow depth, $h$, can be defined as the distance from the virtual
wall to the top boundary, $h=H-y_0$. 
The bulk velocity based on the domain height, $H$, is defined
as $U_{bH}=1/ H \int_0^H \langle u \rangle \mathrm{d}y$; the bulk
velocity based on the effective flow depth is defined as $U_{bh}= 1 /
h \int_{y_0}^H \langle u \rangle \mathrm{d}y\approx U_{bH}H/h$.
Angular brackets are used for the notation of the averaging
operator jointly 
with sub-indexes $t,x_i,p$ 
in order to specify averaging
in time, along the direction $x_i$ or over the periodically
repeating cells of the geometry, respectively. 
Angular brackets without additional indices refer to
quantities which are averaged over time 
as well as spatially over wall-parallel planes, i.e.\ along the $x$ and
$z$ directions. 
%
The bulk Reynolds number, $Re_{b} = U_{bH} H / \nu$,
was kept constant at a value of $2870$ and $2880$ in cases F10 and F50,
respectively. 
%
This corresponds to a friction Reynolds number, $Re_\tau=u_\tau h /
\nu\simeq180$ in case of a smooth wall. In the present simulation the
value for 
$Re_\tau$ increases to 188 in case F10 and to 235 in case F50.
The grid resolution of the simulation 
was in both cases approximately equal to the viscous length
$\delta_\nu=\nu/u_\tau$ in all spatial directions.  
The resolution can therefore be qualified as exceptionally fine away
from the wall and as reasonably fine in the vicinity of the wall. 
In the following, normalisation with wall units will be denoted by a
superscript $+$.

%
%
%
The initial turbulent flow field of each simulation was taken from a
similar simulation on a coarser grid. Subsequently, the simulation 
was run until the flow reached a statistically-stationary state. 
The simulation was 
then continued for 120 $H/U_{bH}$
during which flow field statistics as well
as particle data such as forces and torques were collected. 
Entire flow fields jointly 
with the particle data were stored at intervals of about
$H/U_{bH}$. 
Based on these data a statistical analysis has been carried out. 
If not explicitly stated otherwise, the statistics shown in the following
stem from the data collected during the runtime of the simulation and
are averaged over the entire domain including the region within the
particles. 
Details on the different averaging procedures used and how they compare
are provided in \S\ref{ssec:appendix_ave1} and \S\ref{ssec:appendix_ave2}.  

\subsection{Numerical scheme} \label{sec:method}
%
In order to discretise the complex shape of the wall 
(cf.~figure \ref{fig:domain_s}),
the present simulations were carried out with the aid of a 
variant of the immersed boundary technique
\citep{peskin_phd_72,peskin_JFM_02} proposed by \cite{Uhlmann_JCP_2005}. 
This method employs a direct forcing approach, where a localised
volume force term is added to the momentum equations. The additional
forcing term is explicitly computed at each time 
step as a function of the no-slip condition at the fixed particle 
surface, without resorting to a
feed-back procedure.
The necessary interpolation of variable values from 
Eulerian grid positions to particle-related Lagrangian positions (and
the inverse operation of spreading the computed force terms back to
the Eulerian grid) are performed by means of 
the regularised delta function given by~\cite*{Roma_Peskin_Berger_JCP_1999}.   

A Cartesian grid 
with uniform isotropic mesh width $\Delta x=\Delta y=\Delta z$ is
employed which 
ensures that the regularised delta function
verifies important identities (such as the conservation of the total
force and torque during interpolation and spreading). 
For reasons of efficiency, forcing is only applied to the surface of
the spheres, leaving the flow field inside the particles to develop
freely. 

The immersed boundary technique is implemented in a
standard fractional-step method for the incompressible Navier--Stokes
equations. 
The temporal discretisation is semi-implicit, based on the
Crank--Nicholson scheme for the viscous terms and a low-storage 
three-step Runge--Kutta procedure for the non-linear part
\citep*{verzicco:96}. The spatial operators are evaluated by central
finite-differences on a staggered grid. The temporal and spatial
accuracy of this scheme are of second order. 

An important benefit for the present simulation is that the
hydrodynamic forces acting upon a particle are readily obtained 
by summing the additional volume forcing term over all discrete
forcing points.
The analogue procedure is applied for the computation of the
hydrodynamic torque.

The present numerical method has been submitted to exhaustive
validation tests
\citep{uhlmann_rep04,Uhlmann_JCP_2005,uhlmann:05a,uhlmann_rep06}, as
well as grid convergence studies \citep{uhlmann:06c}. In addition, the
computational code has been applied to the case of vertical plane
channel flow 
with many moving
particles \citep{Uhlmann_POF_2008}. 
In particular, this reference also includes a validation against
the benchmark case of \cite{Kim_Moin_Moser_JFM_1987}.
Recently, the
present immersed boundary method has been successfully implemented
and employed in different numerical codes by other researchers
\citep*[e.g.][]{Lucci_Ferrante_Elghobashi_JFM_2010, 
  Lee_Balachandar_JFM_2010}.
%

%% file: results_mod_v1.tex
\section{Results and discussion}\label{sec:results_discussion}
\subsection{Flow field statistics}
\label{ssec:stat_flow}
\begin{figure}
  \begin{center}
    \begin{minipage}{2ex}
      \rotatebox{90}{\hspace{3ex}$\left( y-y_0 \right) / h$}
    \end{minipage}
    \begin{minipage}{.45\linewidth}
      \includegraphics[height=.8\linewidth]
      {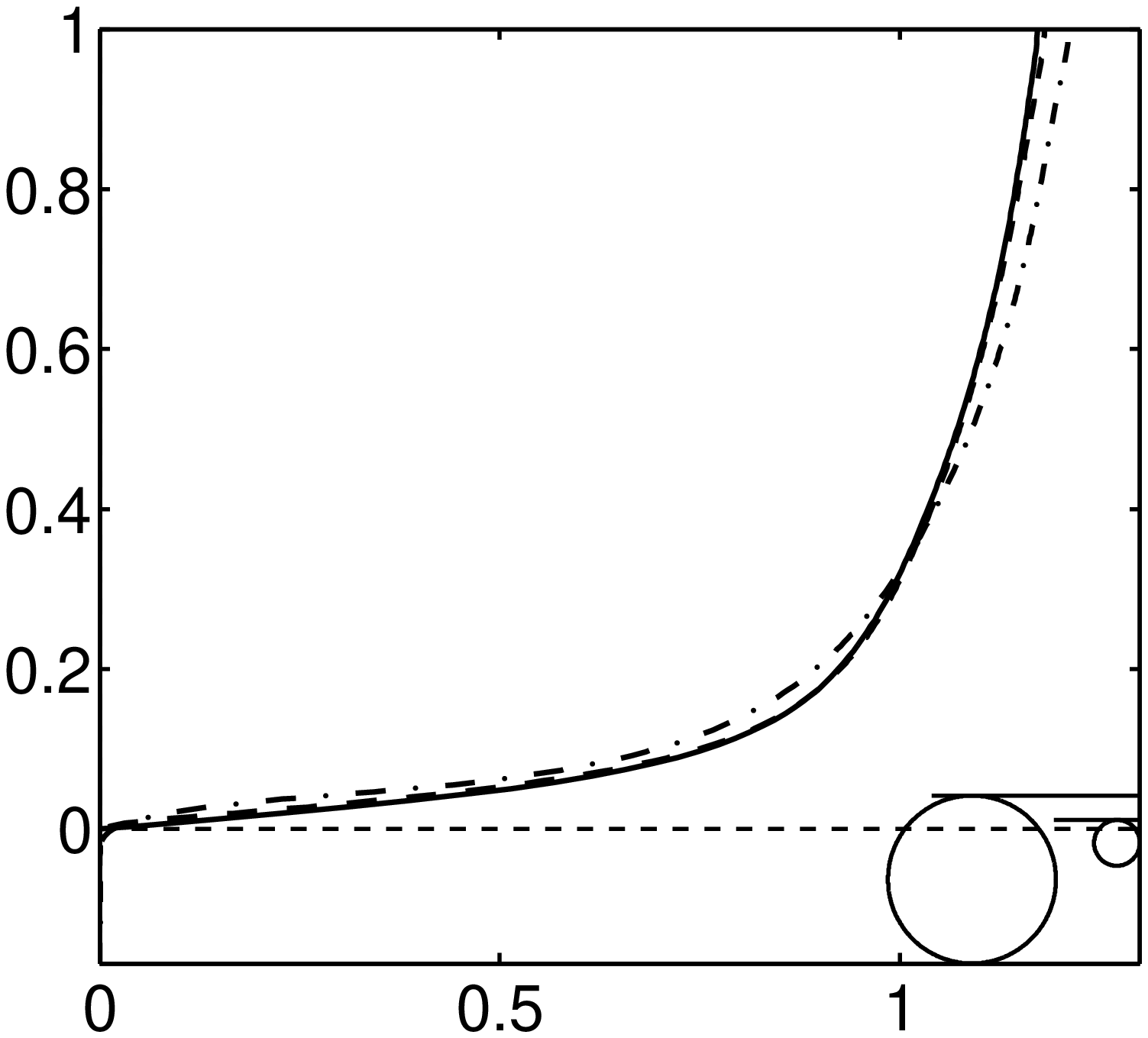}
      \hspace{-0.7\linewidth}\raisebox{0.65\linewidth}{$(a)$}\\
      \centerline{$\langle u \rangle / U_{bh}$}
    \end{minipage}
    \begin{minipage}{2ex}
      \rotatebox{90}{\hspace{3ex}$\langle u \rangle^+$}
    \end{minipage}
    \begin{minipage}{.45\linewidth}
      \includegraphics[height=.8\linewidth]
      {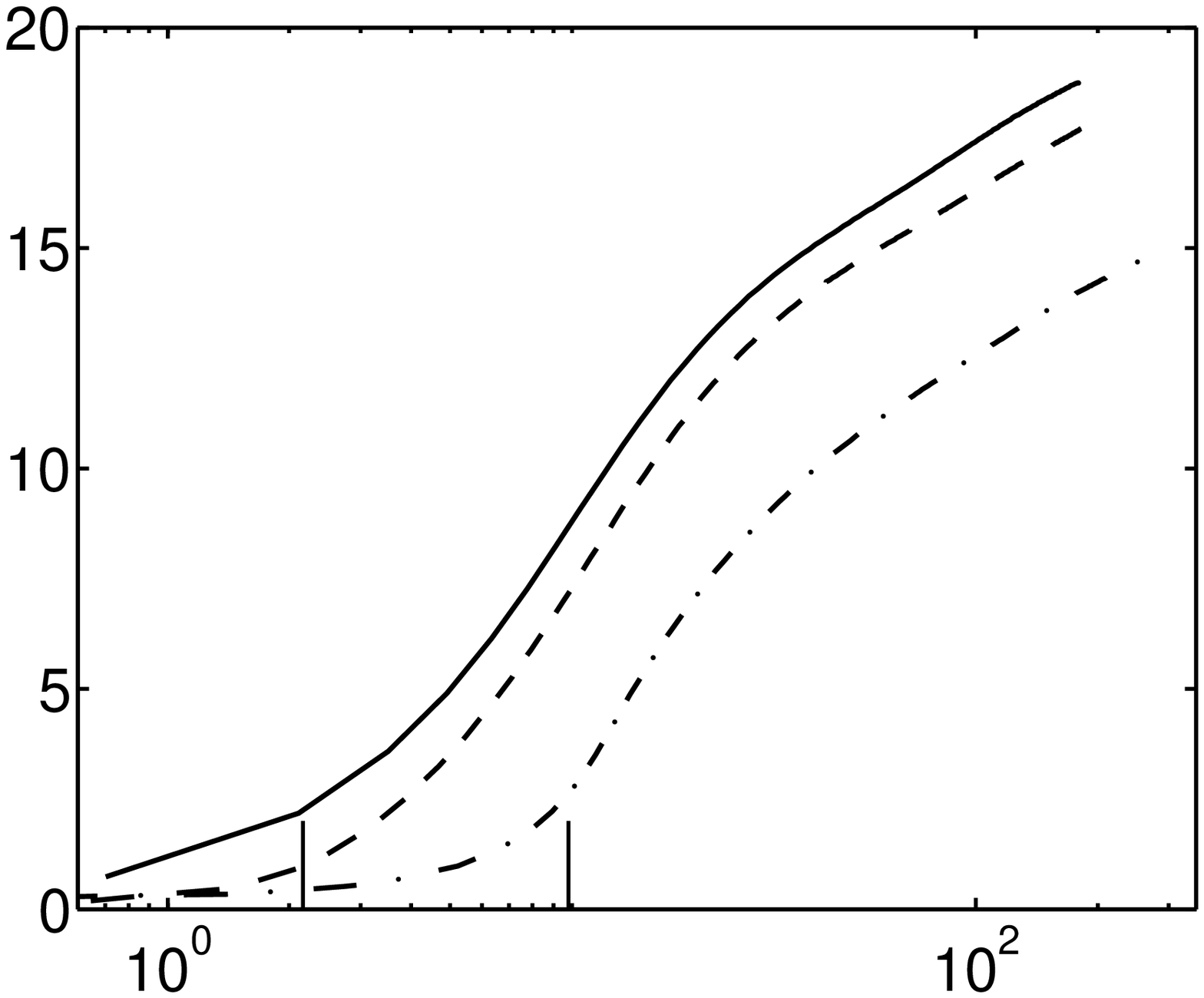} 
      \hspace{-0.8\linewidth}\raisebox{0.65\linewidth}{$(b)$}\\
      \centerline{$\left( y-y_0 \right) / \delta_\nu$}
    \end{minipage}
  \end{center}

  \caption{Time and spatially averaged streamwise velocity component
    $\U$
    of case F10 (dashed line) and case F50 (dashed dotted line) in
    comparison with smooth wall open channel flow
    (continuous line);   
    $(a)$: normalised with $U_{bh}$ as a function of
    $(y-y_0) /  h $; $(b)$: 
    in semi-logarithmic scale normalised by $\delta_\nu$ and $u_\tau$; 
    the position of the particles top are
    marked with 
    horizontal $(a)$ and vertical $(b)$ solid lines.
  } 
  \label{fig:umean_h}
\end{figure}

Figure \ref{fig:umean_h} shows the profiles of the time and plane
averaged streamwise velocity 
component, $\langle u \rangle$, as a function of the vertical coordinate.
The results of case F10 and case F50 
are compared with the reference case S180 of a smooth-wall 
open-channel flow at $Re_b=2880$ 
and $Re_\tau=183$, which has been recomputed for the present study.
%
%
The profiles show the expected effect of roughness that
is described 
in various textbooks \citep{Schlichting_1965,Pope_2000}. 
As the particle diameter $D$ increases, while keeping the value of the
bulk Reynolds number $Re_b$ constant, the friction velocity increases;
the  
streamwise velocity profile normalised by outer scales flattens
(figure \ref{fig:umean_h}$a$), and the streamwise velocity profile
normalised by viscous
scales increasingly shifts towards lower values
of $\langle u \rangle^+$ (figure \ref{fig:umean_h}$b$). 
Figure \ref{fig:umean_h} shows that a logarithmic
layer exists, if at all, only over a small range due to the low Reynolds
number considered. The logarithmic law for
{\CCB the flow over}
a rough wall can be written 
as in the case of a smooth wall with an additional offset $\Delta U^+$
that accounts for the roughness effect
\begin{eqnarray}
  \label{eqn:loglaw_r}
  \U^+ = \frac{1} {\kappa} \ln \left(
    \frac{y-y_0}{\delta_\nu} \right) +A - \Delta U^+ \,,  
\end{eqnarray}
where $\kappa$ and $A$ are constants obtained empirically to be
$\kappa\approx 0.4$ and $A\approx 5.1$ 
{\CCBII 
\citep[according to
experimental findings summarized e.g.\ in][]{Jimenez_ARFM_2004}.}
From the profiles in figure \ref{fig:umean_h} it appears that in case
F10 the roughness effect is weak, while 
in case F50 a stronger roughness effect can be seen. 

\begin{figure}
\begin{center}
  \begin{minipage}{2ex}
    \rotatebox{90}{\hspace{4ex}$\Delta U^+$}
  \end{minipage}
  \begin{minipage}{.6\linewidth}
    \includegraphics[width=1.\linewidth,clip]
    {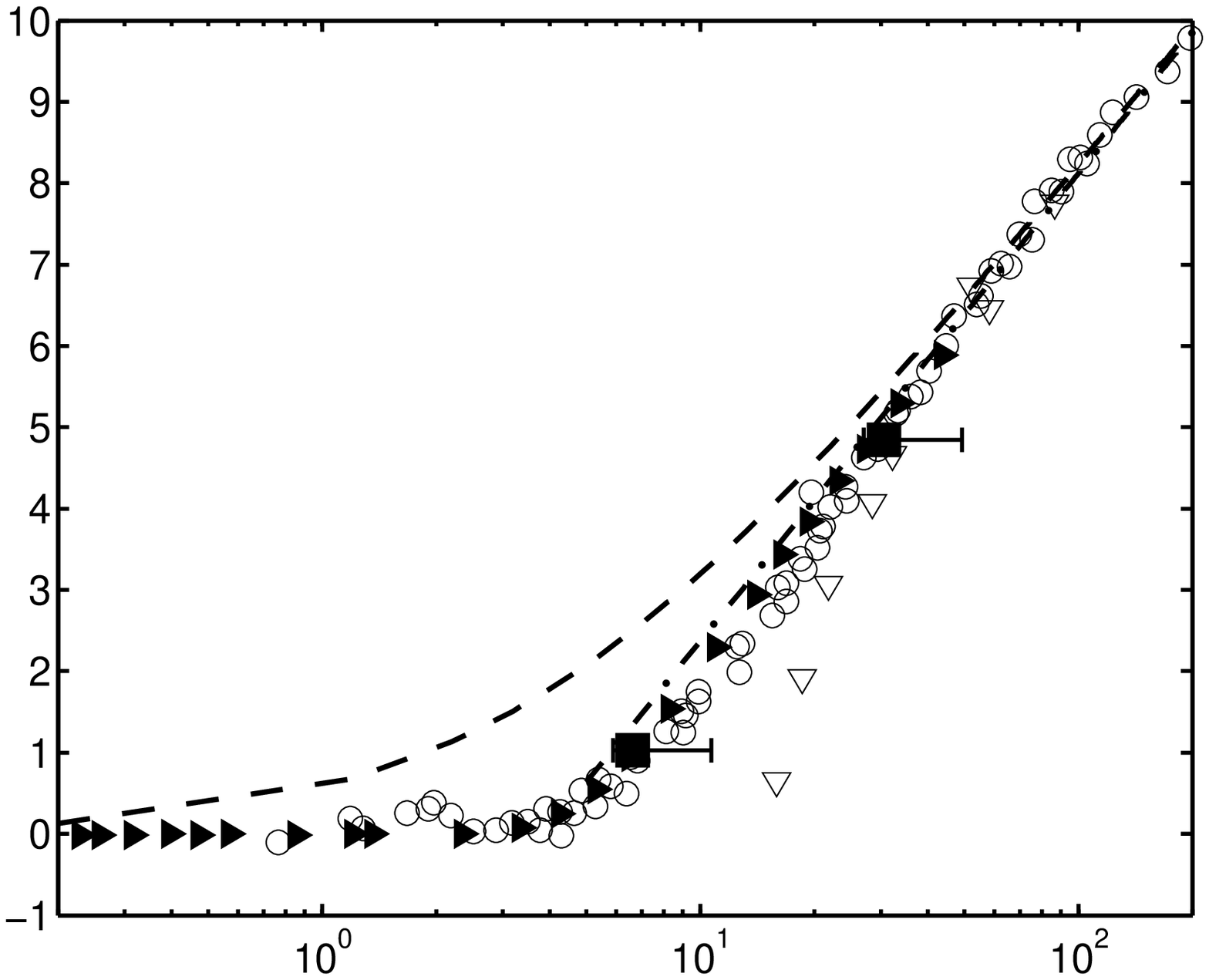}
    \hspace{-0.5\linewidth}\raisebox{0.7\linewidth}\\
    \centerline{$k_{s\infty}^+$}
  \end{minipage}

\caption{Roughness function for several transitionally rough surfaces
  as a function of the Reynolds number based on $k_{s\infty}^+$ adapted
  from \cite{Jimenez_ARFM_2004}. 
  $\circ$, \cite{Nikuradse_VDI_1933}, uniform sand, pipe flow;
  $\triangledown$,
  \cite{Ligrani_Moffat_JFM_1986},  uniform densely-packed spheres,
    boundary layer; $\blacktriangleright$,  
  \cite{Shockling_Allen_Smits_JFM_2006}, honed aluminium, pipe flow;
  $\blacksquare$, 
  present simulation with $k_{s\infty}/ D = 0.63$, 
  error-bars show the
  range of $k_{s\infty}/ D = 0.55$ to $1$; 
  dashed line, 
  $\Delta U^+ = 5.75 \log \left(1+0.26 k_{s\infty}^+\right)$ 
  proposed by \cite{Colebrook_1939}; 
  dashed dotted line, relation (\ref{eqn:deltau}) with $A=5.1, B=8.48$
  and $C=5.75$. 
} 
\label{fig:duplus}
\end{center}
\end{figure}

It is customary to quantify  
the roughness effect by using the equivalent sand grain roughness
$k_s$ \citep{Schlichting_IngArch_1936}. 
It can be obtained by a fit
to the mean velocity profile in the logarithmic layer using the
following equation 
\begin{eqnarray}
  \label{eqn:loglaw_ks}
  \U ^+ =  C ~\log_{10}{\left(
      \frac{y-y_0}{k_s}\right)} + B \,, 
\end{eqnarray}
where $B\approx 8.48$ and $C= 1 / \kappa ~\ln{(10)}\approx 5.75$ are
empirically obtained values \citep*[cf.][]{Shockling_Allen_Smits_JFM_2006}.
At high enough Reynolds numbers $k_s$ becomes constant, i.e.~$\lim_{Re
  \rightarrow \infty}k_s=k_{s\infty}$, defining the so-called 
fully rough flow regime.  
The specific value of $k_{s\infty}$ is a property of the surface that
depends on the characteristics of the roughness, such as shape, arrangement or
roughness area ratio.
Flow over roughness can be classified as hydraulically smooth, transitionally 
rough or fully rough according to
a small, moderate or high value of $k^+_{s\infty}$.  
\cite{Nikuradse_VDI_1933} gave values of $5<k^+_{s\infty}<70$ to
define the transitionally rough flow regime.
However, these values should be taken with care as the transition
might be influenced by the specific characteristics of the roughness
\cite[cf. discussion in][among
others]{bradshaw_POF_2000,Jimenez_ARFM_2004,
  Shockling_Allen_Smits_JFM_2006}.  
In particular, it has been speculated that a uniformly sized,
structured arrangement of roughness elements as in the present case
might lead to 
a sharp transition from hydraulically smooth to the fully rough regime
\citep{Colebrook_1939, Jimenez_ARFM_2004}.
Figure \ref{fig:duplus} shows the transition from the hydraulically smooth
flow
regime to the fully rough flow regime obtained in different experiments. 
It shows the offset $\Delta U^+$ as a function of $k_{s\infty}^+$. 
At low values of $k_{s\infty}^+$ the effect of roughness should be
negligible and correspondingly $\Delta U^+$ approaches zero. 
In the fully rough flow
regime the roughness effect should purely depend on $k_{s\infty}^+$. 
Comparing equations (\ref{eqn:loglaw_r}) and (\ref{eqn:loglaw_ks}) 
a formula for $\Delta U^+$ can be derived for the fully rough regime,
\begin{eqnarray}
  \label{eqn:deltau}
  \Delta U ^+ =  C ~\log_{10}{\left(k_{s\infty}^+\right)} - B + A \,,
\end{eqnarray}
with the constants $A,B$ and $C$ as above. 
The relation (\ref{eqn:deltau}) above is shown in figure
\ref{fig:duplus} jointly with results from experiments.

For the present simulations the values of $\Delta U^+$ can be obtained
by the vertical shift of the mean velocity profiles in the log-region
of figure \ref{fig:umean_h}$(b)$. They are $1.03$ and $4.85$ for
cases F10 and F50, respectively.
However, the value of $k_{s\infty} / D$ for the present arrangement of
spheres is unknown. 
\cite{Schlichting_IngArch_1936},
\cite{Ligrani_Moffat_JFM_1986} and 
\cite*{Pimenat_Moffat_Kays_rep_1975} found a value of 
 $k_{s\infty} / D \sim 0.63$ for flow over spheres in
hexagonal packing, 
while somewhat larger values were obtained by 
\cite{Singh_sandman_williams_JHR_07} ($k_{s\infty}/D=0.77$) and 
\cite{Detert_Nikora_Jirka_JFM_2010} ($k_s/D=0.81$). For flow over
spheres in random packing the values generally obtained vary 
in the range of 
$k_{s\infty}/D= 0.55$ to $0.85$ \citep[cf.][]{Goma_Gelhar_MITrep_1968,
  Grass_Stuart_Mansour-Tehrani_PRSA_1991}. 
Few studies exist that
use values of $k_{s\infty}/D= 1$ for structured \citep{Einstein_Elsamni_RMP_1949} 
or random arrangements \citep{Nakagawa_Nezu_JFM_1977}. 
Figure \ref{fig:duplus} shows the pair of values $(k^+_{s\infty},
\Delta U^+)$ for cases F10 and F50 
 when approximating $k_{s\infty}/D$
by the value found for a hexagonal packing, i.e.\
$ k_{s\infty}/D=0.63.$
The error-bars indicate the range of  $k_{s\infty}/D = 0.55$ to $1$ as
found in the literature. 
It can be seen that case F10
approaches the hydraulically smooth  flow
regime while case F50 is 
in the transitionally rough flow regime.
%
%
%
%

%
%

%

\begin{figure}
\begin{center}
  \begin{minipage}{2ex}
    \rotatebox{90}{\hspace{3ex}$(y-y_0)/h$}
  \end{minipage}
  \begin{minipage}{.5\linewidth}
    \includegraphics[height=5.25cm]
    {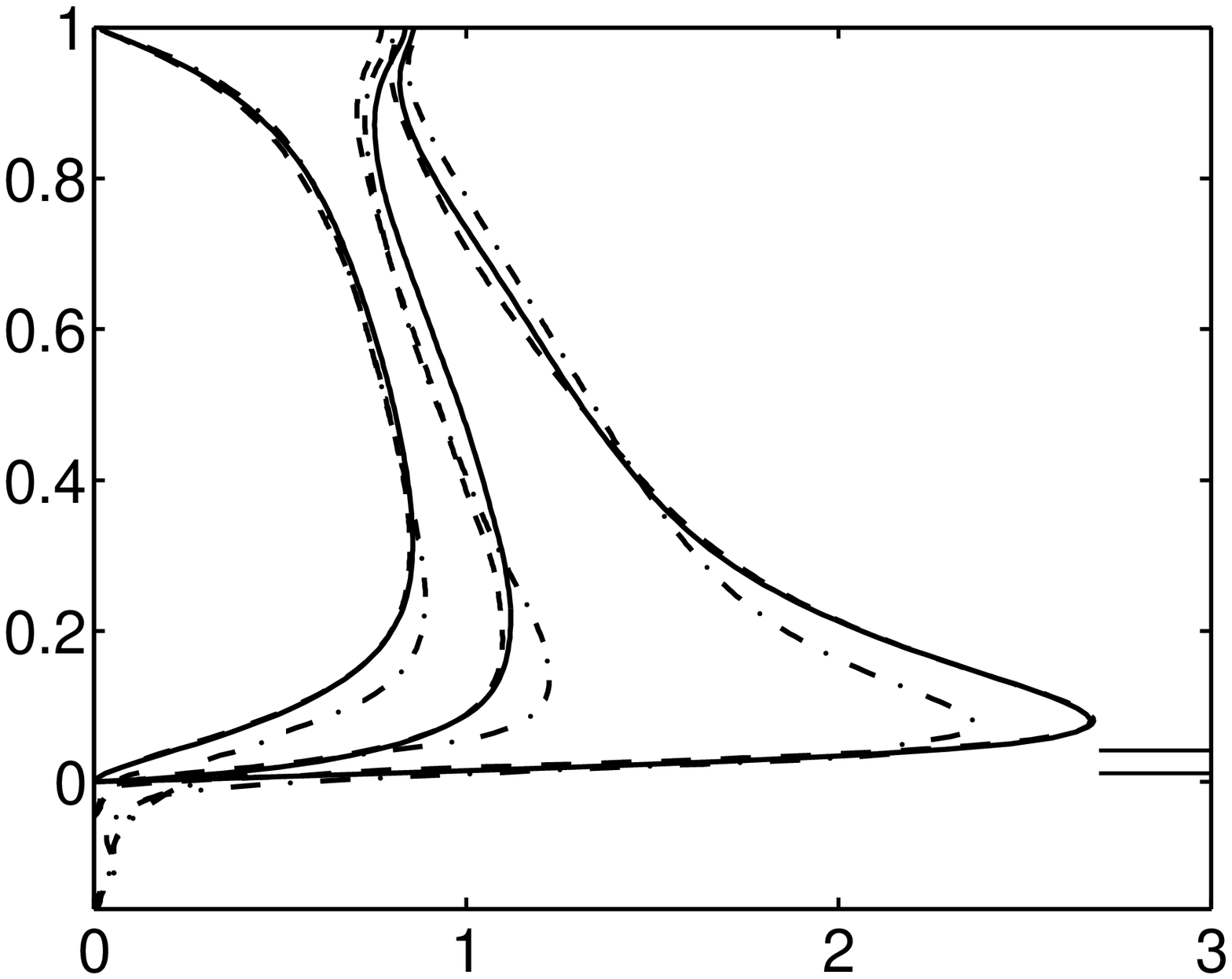}
    \hspace{-0.2\linewidth}\raisebox{4.25cm}{$(a)$}\\
    \centerline{$u^i_{rms} / u_\tau$}
  \end{minipage}
  \begin{minipage}{2ex}
    \rotatebox{90}{\hspace{3ex}$\left( y-y_0 \right) / h $}
  \end{minipage}
  \begin{minipage}{.4\linewidth}
    \includegraphics[height=5.25cm]
    {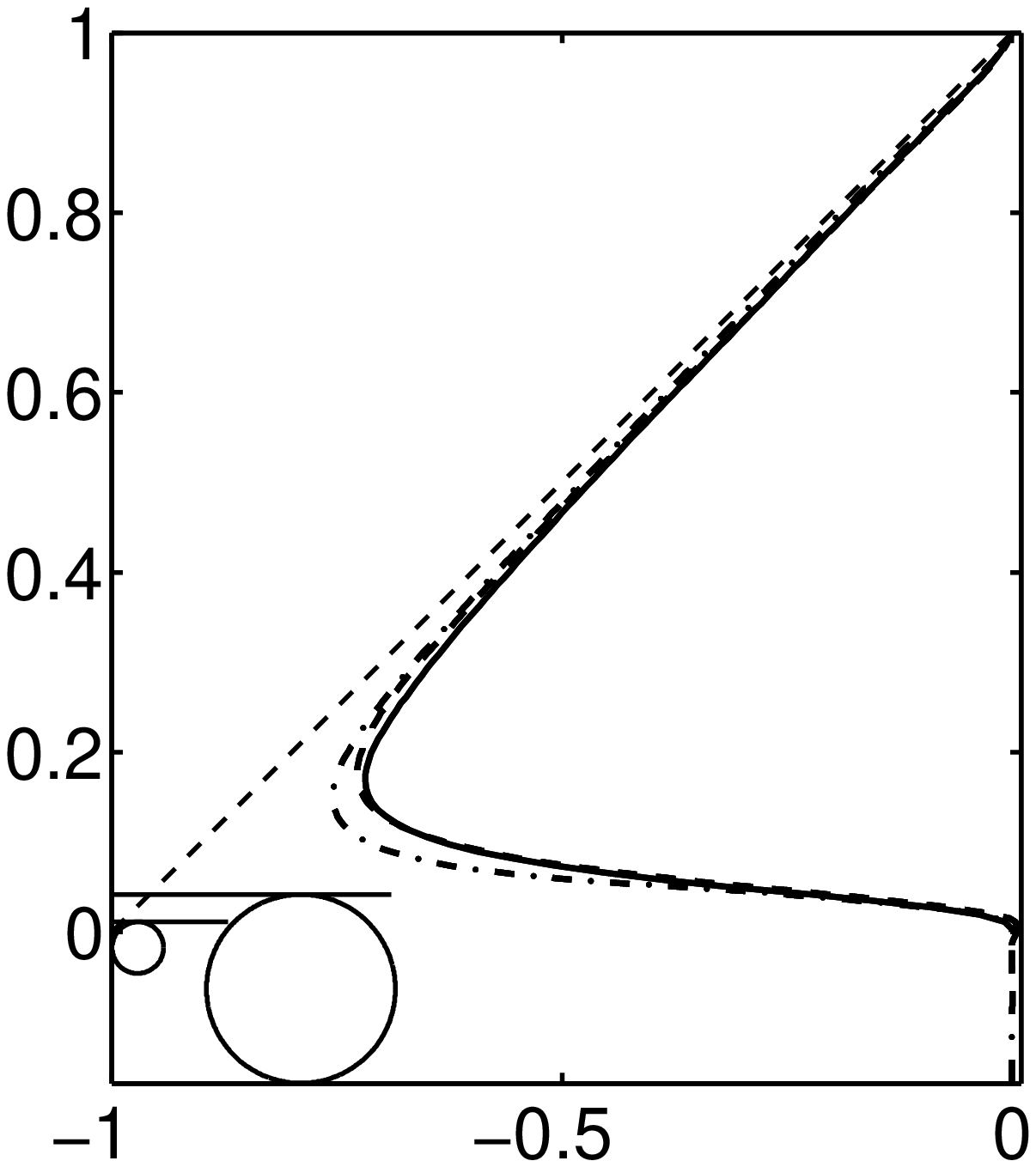}
    \hspace{-0.7\linewidth}\raisebox{4.25cm}{$(b)$}\\
    \centerline{$\langle u^\prime v^\prime \rangle / u_\tau^2$}
  \end{minipage}
\end{center}

  \caption{$(a)$ Root-mean-square of velocity fluctuations of case F10
    and case F50 normalised by $u_\tau$ in comparison with results of
    smooth wall as a function of  wall distance; curves from left to right 
    wall-normal ($v_{rms}/u_\tau$), spanwise
    ($w_{rms}/u_\tau$) and streamwise ($u_{rms}/u_\tau$); $(b)$:
    distribution of Reynolds shear
    stress $\langle u^\prime v^\prime \rangle$ as a function of
    wall distance. Legend as in figure \ref{fig:umean_h}.  
    The position of the particle tops are marked with  horizontal solid
    lines. In $(b)$ the 
    straight dashed line is included to guide the eye.
  }    
  \label{fig:velrms_norm}
\end{figure}

A similar conclusion can be reached by analysing the profiles of
the root-mean-square of the velocity fluctuations normalised with
$u_\tau$ 
that are shown in figure \ref{fig:velrms_norm}$(a)$. In case F10, the
profiles 
of the three velocity components 
almost collapse with the smooth-wall results, indicating that, indeed, 
the flow over the relatively small roughness elements 
in this case can be considered as nearly hydraulically smooth. 
In case F50, some differences with respect to the smooth wall case are
evident.  
The near-wall peak in the streamwise fluctuation profile decreases but
it is still visible. 
This indicates that the flow is 
in the 
 transitionally rough flow
regime since experiments in the fully rough 
flow regime present no clear peak 
\citep[see for example figure 5 of][]{Jimenez_ARFM_2004}. 
The wall-normal and spanwise fluctuations present slightly higher
values near the wall than the corresponding ones in the smooth-wall
case. 
Therefore, the anisotropy of the
fluctuations near the wall is smaller than in the smooth-wall
case. This tendency of roughness to make the fluctuations more
isotropic is a phenomenon which
has been often reported in the literature 
\citep*[e.g.][]{Poggy_Porporato_Ridolfi_POF_2003, 
  Orlandi_Leonardi_JFM_2008}. 
Also in case F50, above $(y-y_0)/h\sim0.4$ all three components
agree well with the 
values of the smooth-wall case.

\cite{Orlandi_Leonardi_JFM_2008} discuss the velocity shift $\Delta
U^+$ as a function of $v_{rms}$ at the roughness
crest.
The present simulations result in pairs
$\left( v_{rms}\,,\,\Delta U^+\right)$ of
$(0.10\,,\,1.03)$  $\left( 0.46\,,\,4.85\right)$ for case F10 and F50
respectively which agree well within the scatter of the reported data
(graph not shown).

%
Figure \ref{fig:velrms_norm}$(b)$ shows the profiles of the Reynolds stress,
$\uv$, normalised by $u_\tau^2$. 
The Reynolds stress profile $\langle u^\prime v^\prime \rangle$ of
case F10 nearly
collapses with the profile of the smooth-wall case. In case F50 a
slight increase and a 
small shift towards the wall of the near-wall peak can be seen which
could be an effect of the higher value of $Re_\tau$ in this case.

%
%

\begin{figure}
  \begin{center}
    \begin{minipage}{2ex}
      \rotatebox{90}{\hspace{3ex}$(y-y_0)/D$}
    \end{minipage}
    \begin{minipage}{.6\linewidth}
      \includegraphics[width=1.0\linewidth]
      {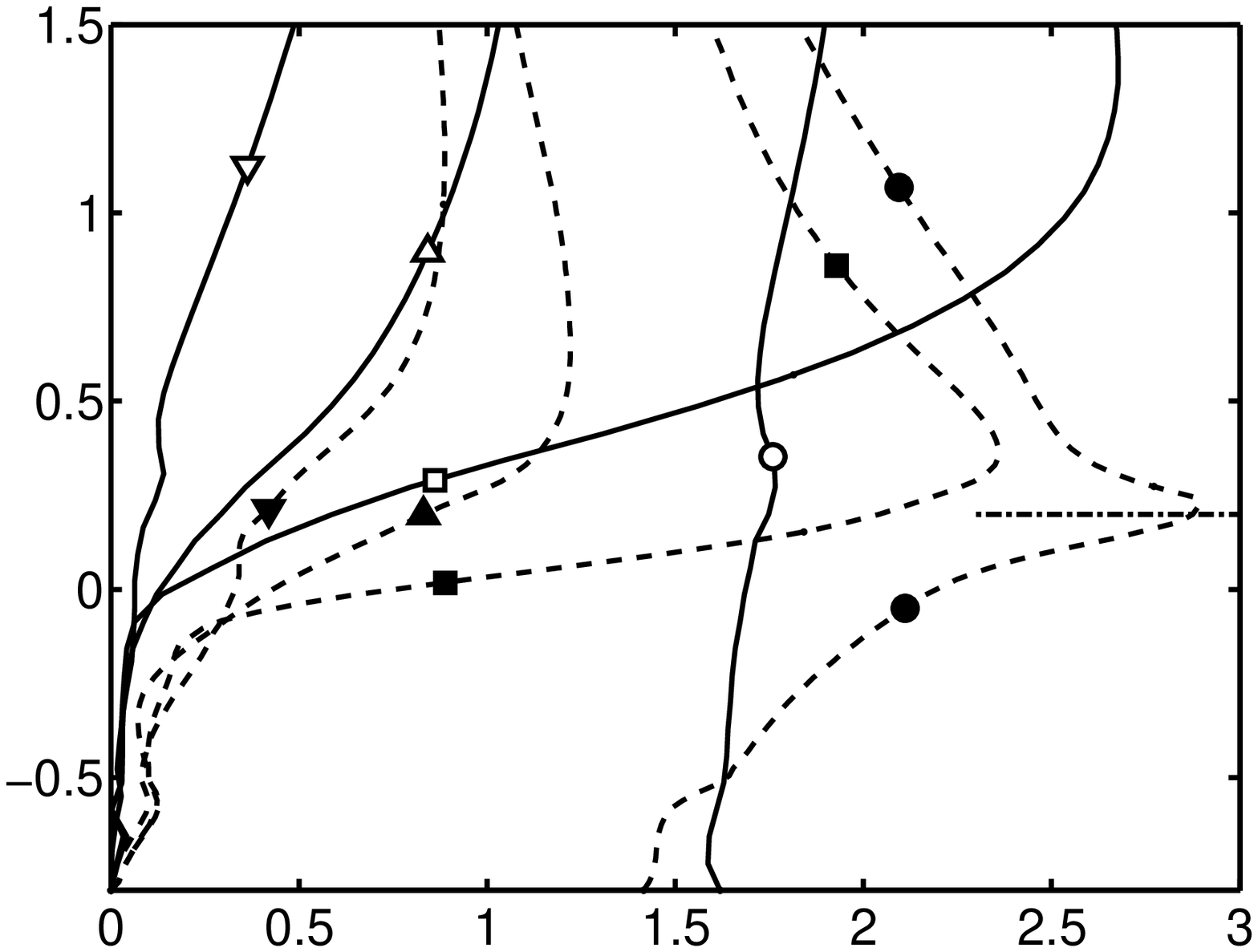} 
      \\
      \centerline{$u^i_{rms} / u_\tau ~ ;\quad p_{rms} / (\rho u_\tau^2)$}
    \end{minipage}
  \end{center}
  
\caption{Root-mean-square of velocity and pressure fluctuations of case F10
  and case F50, normalised by $u_\tau$ and  $\rho u_\tau^2$
  respectively, as a function of $(y - y_0) /
  D$;
  averaging has been carried out over fluid cells only,
  for details see
  \S\ref{ssec:appendix_ave1};
  $\square: u_{rms}/u_\tau$,
  $\bigtriangledown: v_{rms}/u_\tau$, 
  $\bigtriangleup: w_{rms}/u_\tau$, 
  and 
  $\bigcirc: p_{rms}/\rho u^2_\tau$; 
  solid lines and empty symbols: case F10, 
  dashed  lines and full symbols: case F50;
  horizontal line: position of particle tops in both cases.
}   
\label{fig:velrms_norm_D}
\end{figure}

In order to study the near wall behaviour of the velocity fluctuations,
a close-up of the profiles shown in
figure \ref{fig:velrms_norm}$(a)$,
is plotted
as a function of $(y-y_0)/ D$ in
figure \ref{fig:velrms_norm_D}.
Additionally, the profiles of the root-mean-square of the pressure
fluctuations, $p_{rms}/(\rho u^2_\tau)$, are included.  
Note that in contrast to figure \ref{fig:velrms_norm} the profiles
shown in figure \ref{fig:velrms_norm_D} are
obtained from snapshots of the flow field and
obtained by averaging over cells outside of the particles
as described in
detail in \S\ref{ssec:appendix_ave1} and \S\ref{ssec:appendix_ave2}. 
The amplitudes of the fluctuations of the three velocity components 
present similar values below the virtual wall, $(y-y_0)/D<0$. These
are much smaller than the values above the virtual wall, and they are 
somewhat larger in case F50 ($u^i_{rms}\sim0.1 u_\tau$)
compared to F10 ($u^i_{rms}<0.05 u_\tau$).
On the contrary, the pressure fluctuations within the roughness layer
for both cases 
present values which are 
similar to the values above the roughness layer.

Recall that also in the case of 
a smooth wall the pressure fluctuations are non-zero at the wall
\citep*{Kim_Moin_Moser_JFM_1987,Kim_JFM_1989}. 
Near the top of the roughness elements, i.e.~around $(y-y_0)/D=0.2$,  
the profiles of $p_{rms}$ 
exhibit a peak which is barely visible in case F10 and 
more pronounced in case F50. 
In case F50 the value of the peak is higher by a factor of two
compared to case F10. 
 Note that the peak is, to some extent, a consequence of the  
three-dimensionality of the time-averaged flow field around the
particle; this point is further elaborated in
\S\ref{ssec:3d_flow}.
 The pressure fluctuation profiles 
of case F10 and F50 (when plotted as a function of $(y-y_0) / h$)
approach each other with increasing wall distance (not shown).
They converge to the profile 
obtained in the case of a smooth wall in the outer part of the flow. 
%

\subsection{Three-dimensional time-averaged flow  field distribution}
\label{ssec:3d_flow}

%

Since the geometry of the roughness is three-dimensional the
time-averaged flow field in the near wall region also varies in all
three directions. 
In the following some characteristics of the time-averaged
flow field obtained from 90 snapshots are discussed. 
In addition to the averaging in time the fields were averaged over
periodically repeating boxes centred on the particles (henceforth
indicated by the symbol 
$\langle\cdot\rangle_{tp}$). For simplicity this will be simply
referred to as time averaging below. 
%

\begin{figure}
  \begin{center}
    \begin{minipage}{2ex}
      \rotatebox{90}{\hspace{3ex}$y/H$}
    \end{minipage}
    \begin{minipage}{.45\linewidth}
      \includegraphics[height=1.\linewidth,
      clip, bb= 35 27 380 412]
      {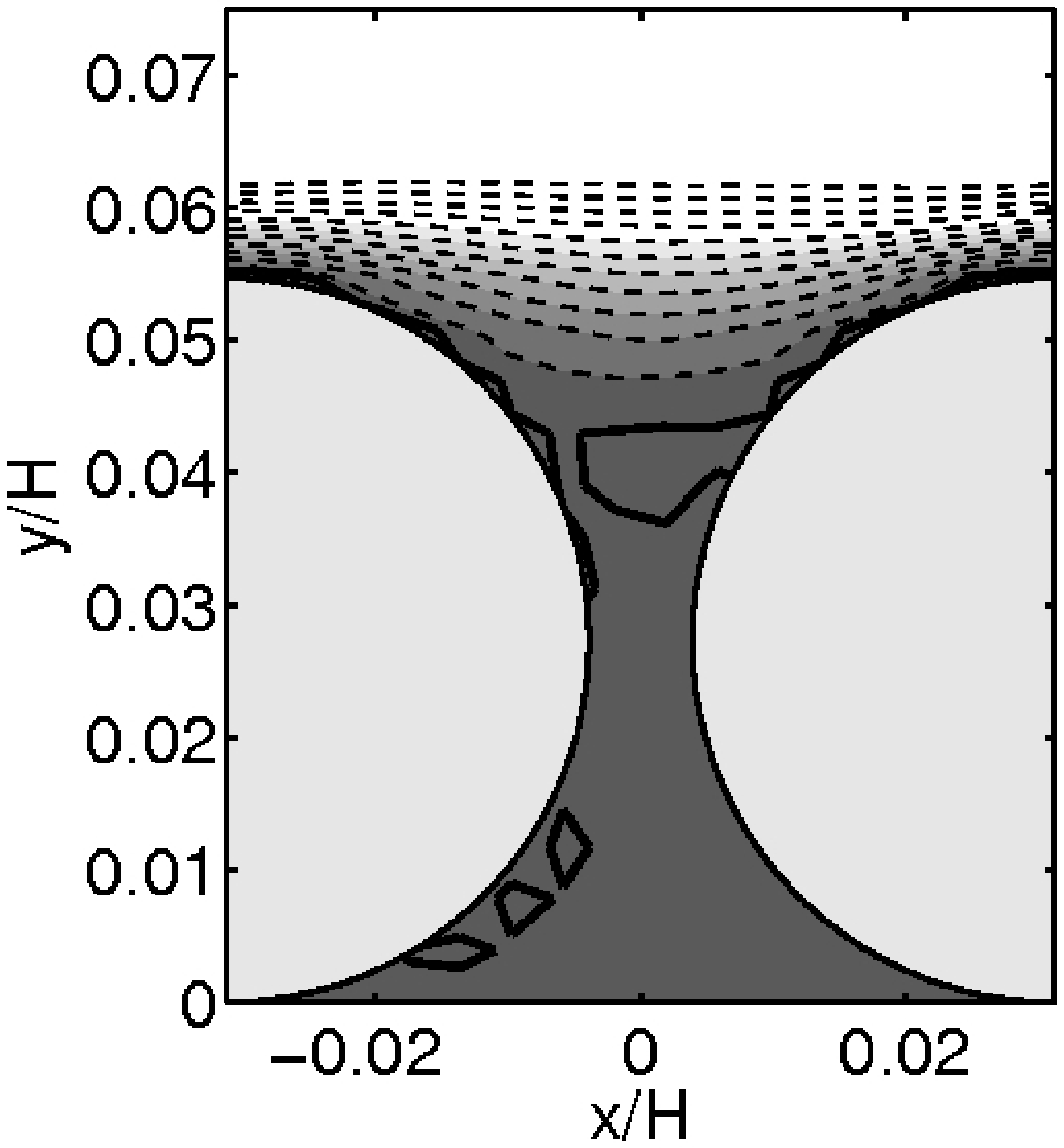}\qquad
      \hspace{-0.75\linewidth}\raisebox{0.88\linewidth}{$(a)$}
      \\
      \centerline{$x/H$}
    \end{minipage}
    \begin{minipage}{.45\linewidth}
      \includegraphics[height=1.\linewidth,
      clip, bb= 0 27 380 412]
      {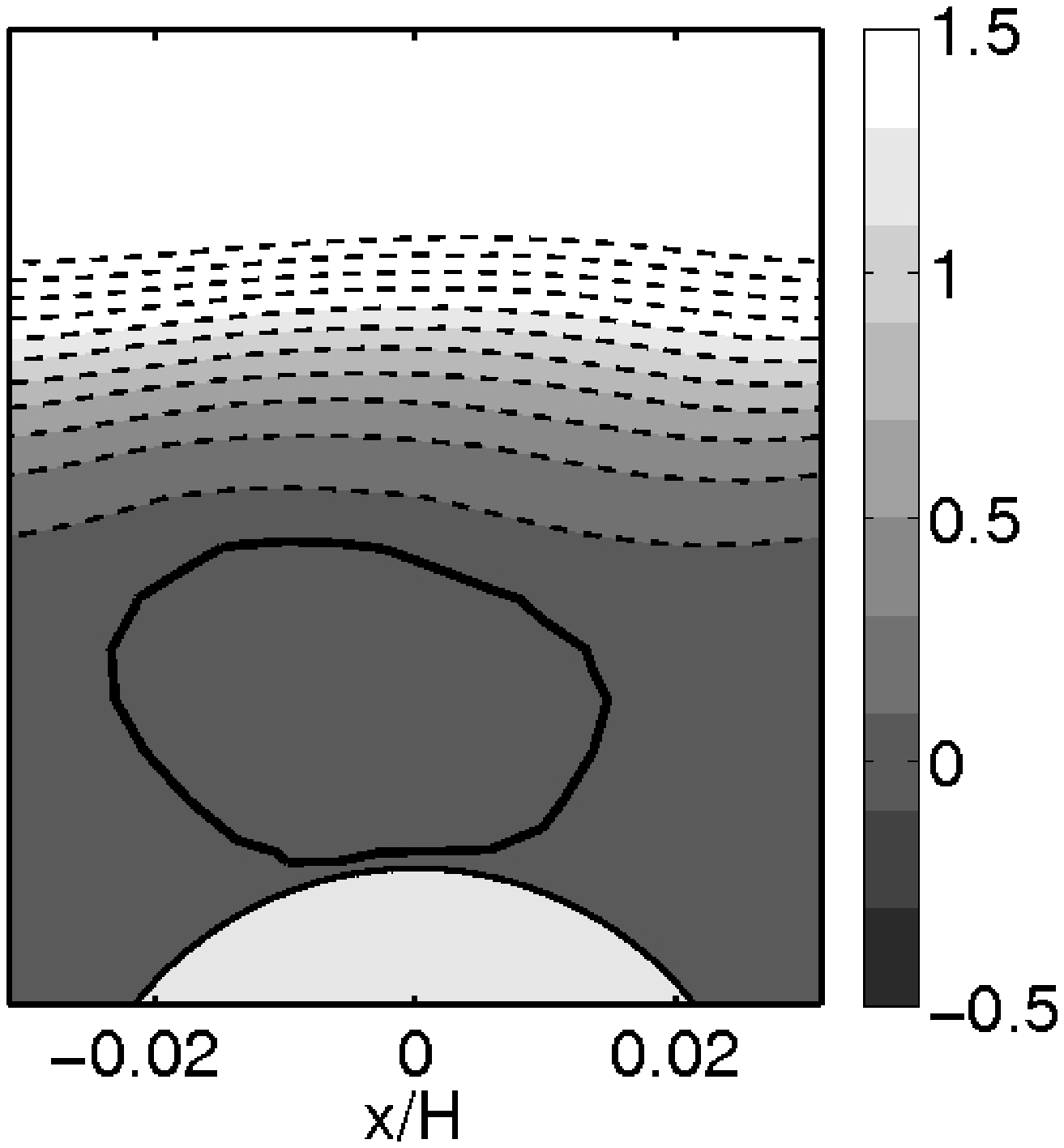}
      \hspace{-0.85\linewidth}\raisebox{0.88\linewidth}{$(b)$}
      \\
      \centerline{$x/H$\hspace{1.5cm}}
    \end{minipage}
    
    \begin{minipage}{2ex}
      \rotatebox{90}{\hspace{3ex}$y/H$}
    \end{minipage}
    \begin{minipage}{.45\linewidth}
      \includegraphics[height=1.\linewidth,
      clip, bb= 35 27 380 412]
      {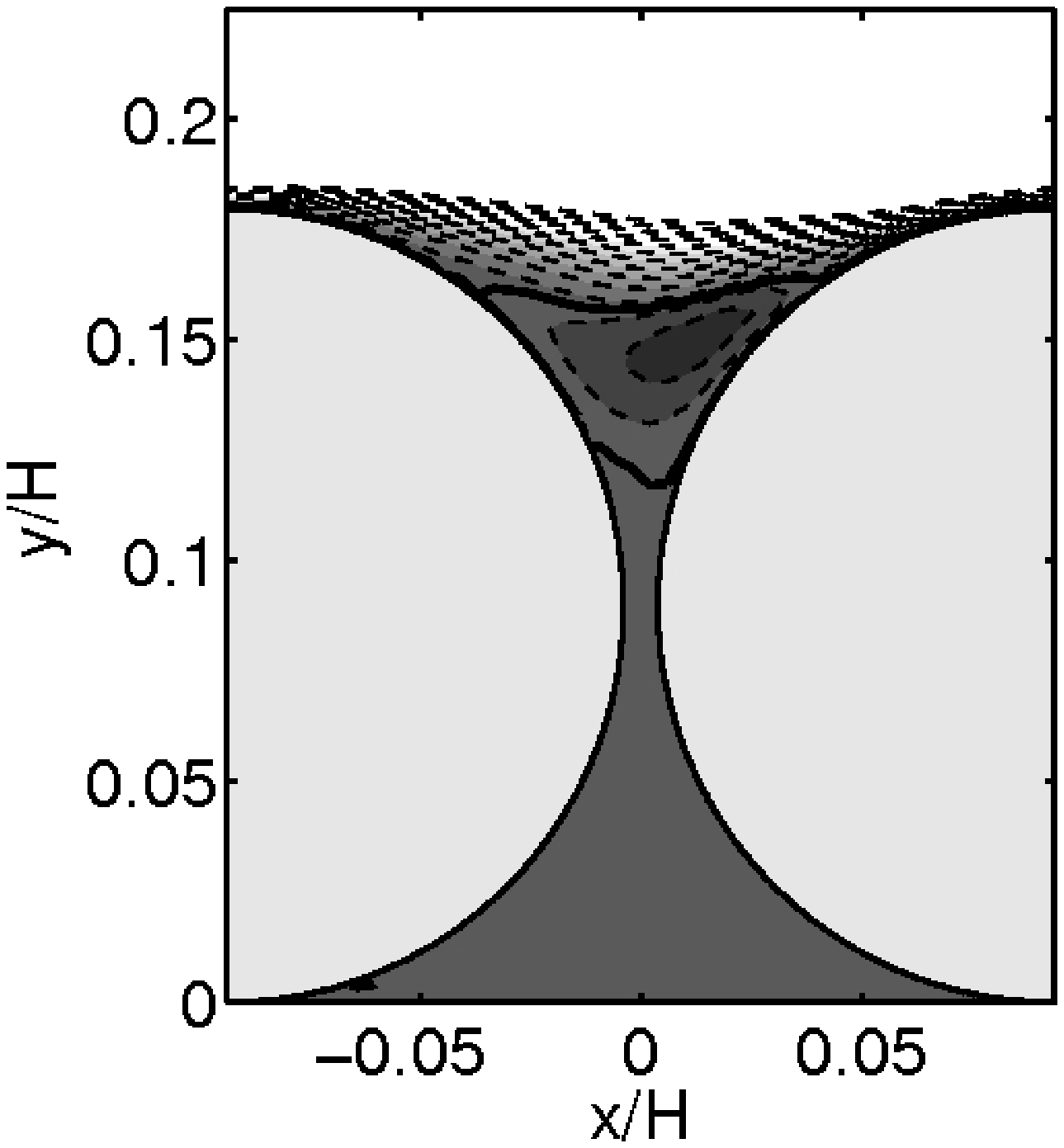}\qquad
      \hspace{-0.75\linewidth}\raisebox{0.88\linewidth}{$(c)$}
      \\
      \centerline{$x/H$}
    \end{minipage}
    \begin{minipage}{.45\linewidth}
      \includegraphics[height=1.\linewidth,
      clip, bb= 0 27 380 412]
      {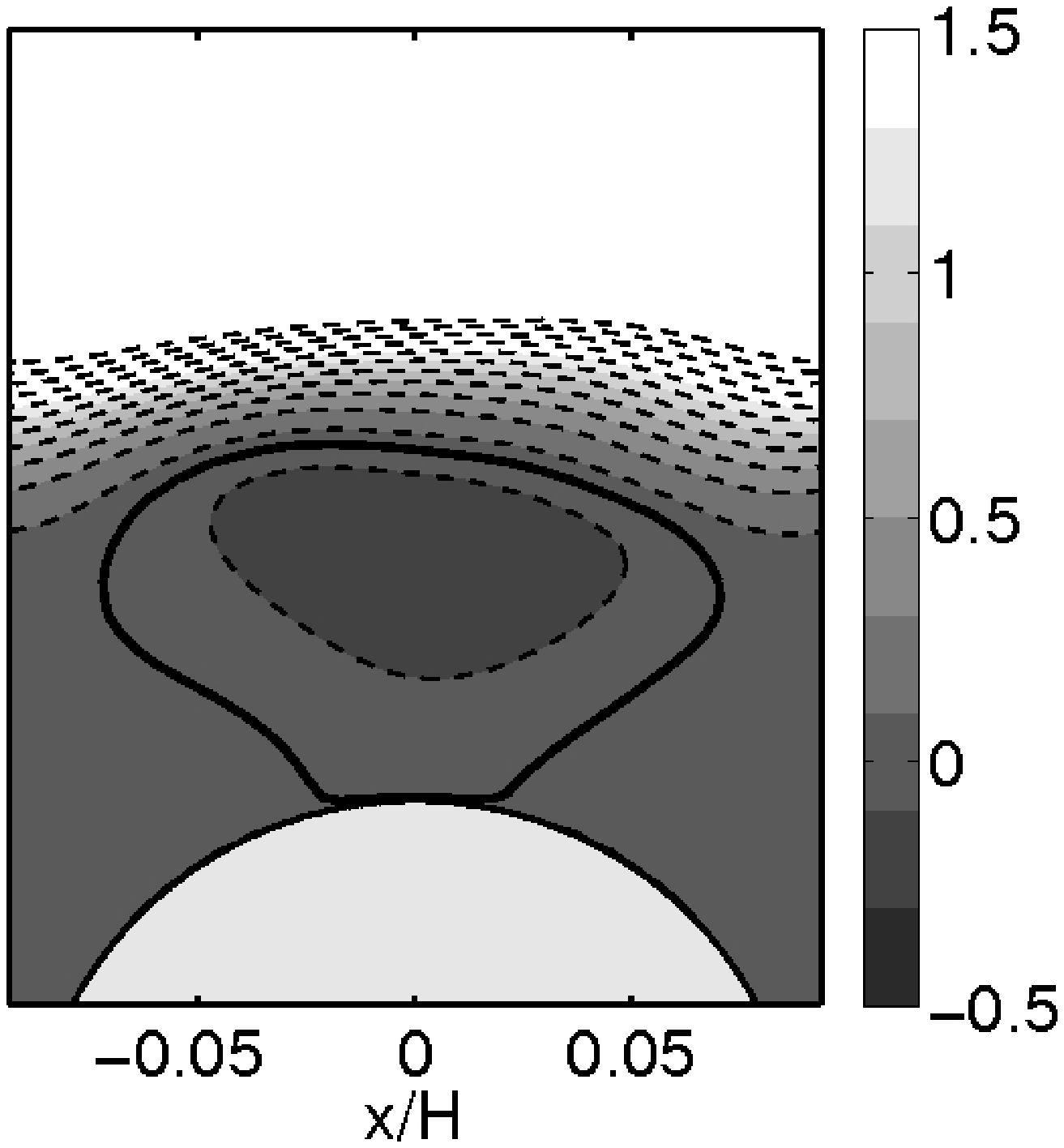}
      \hspace{-0.85\linewidth}\raisebox{0.88\linewidth}{$(d)$}
      \\
      \centerline{$x/H$\hspace{1.5cm}}
    \end{minipage}
  \end{center}
  
  \caption{
    Distribution of the time-averaged streamwise
    velocity in a periodic cell, $\langle u \rangle_{tp}$,  normalised 
    by $u_\tau$ in $x$-$y$ planes;
    $(a,c)$ show a plane through particle centres, 
    $(b,d)$ show a plane centred between spheres.
    Dashed lines: iso-contours of $\langle u \rangle_{tp}$ at
    values of -0.5 to 2.1 in steps of 0.2; 
    black solid line: streamwise velocity contour at $-10^{-3}$. 
    Panels $(a)$ and $(b)$ show case F10, 
    $(c)$ and $(d)$ show case F50.
    The direction of the bulk velocity is from left to right in all panels.
  } 
  \label{fig:3d-mean}
\end{figure}

%
Figure \ref{fig:3d-mean} shows the distribution of the three-dimensional
time-averaged streamwise velocity, $\langle u \rangle_{tp}$,
for both cases.  
Two different $(x,y)$-planes are shown. The first one contains the centre of the particles
of one streamwise row (figure \ref{fig:3d-mean}$a$ and
figure \ref{fig:3d-mean}$c$). The second one is located in between two
streamwise rows of particles (figure \ref{fig:3d-mean}$b$ and
figure \ref{fig:3d-mean}$d$).  
As can be expected the flow field is very different from a single sphere
in an unbounded turbulent flow or in a channel close to a wall
\citep{Bagchi_Balachandar_JFM_2004, Zeng_Balachandar_Fischer_Najjar_JFM_2008}.

\begin{figure}
  \begin{center}
    \begin{minipage}{2ex}
      \rotatebox{90}{\hspace{3ex}$y/H$}
    \end{minipage}
    \begin{minipage}{.45\linewidth}
      \includegraphics[height=0.85\linewidth,
      clip, bb= 35 27 437.76 412]
      {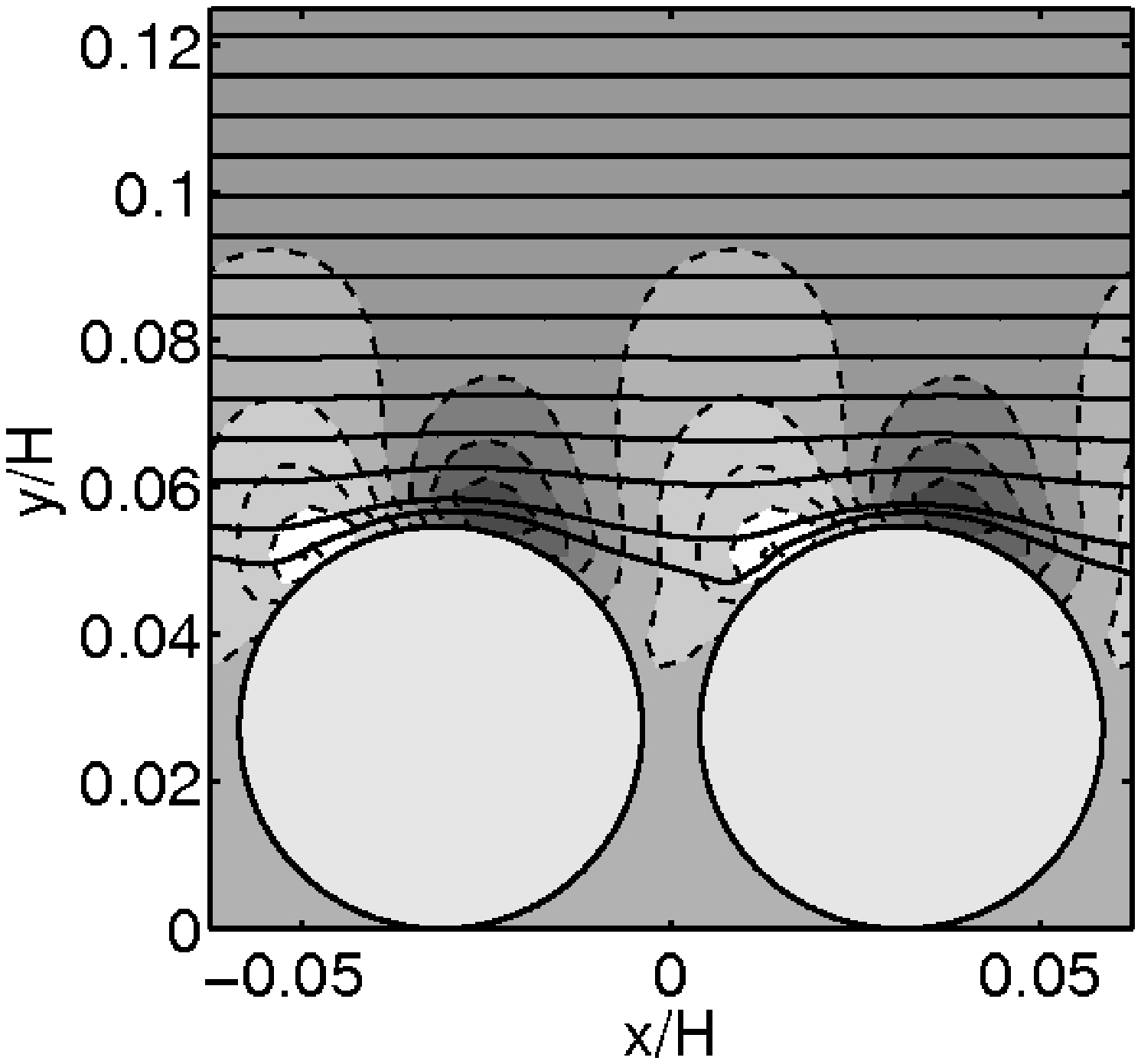}
      \hspace{-0.75\linewidth}\raisebox{0.75\linewidth}
      {\colorbox{white}{$(a)$}}
      \\
      \centerline{$x/H$}
    \end{minipage}
    \begin{minipage}{.45\linewidth}
      \includegraphics[height=0.85\linewidth,
      clip, bb= 0 27 437.76 412]
      {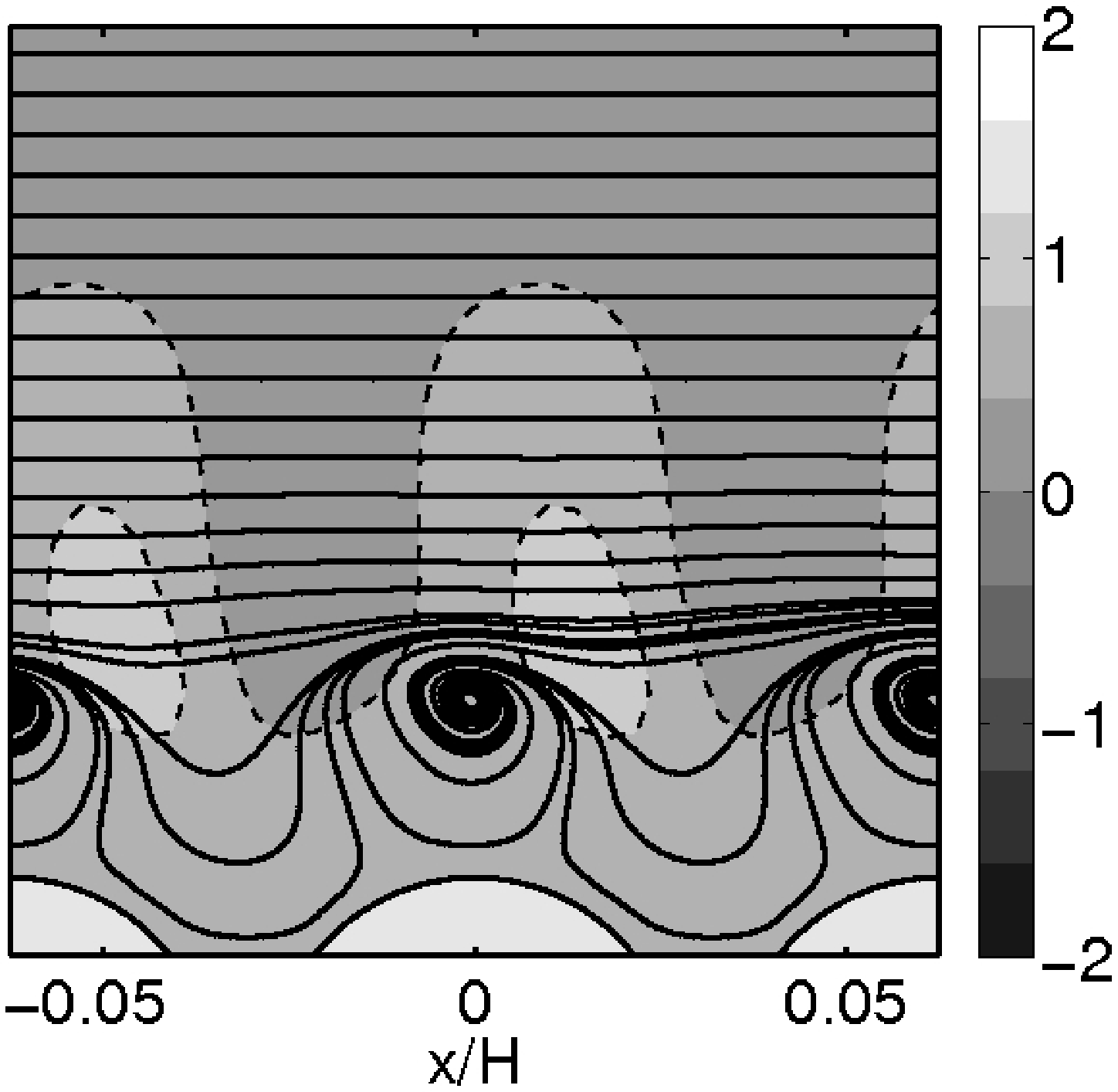}
      \hspace{-0.92\linewidth}\raisebox{0.75\linewidth}
      {\colorbox{white}{$(b)$}}
      \\
      \centerline{$x/H$\hspace{1.5cm}}
    \end{minipage}
    
    \begin{minipage}{2ex}
      \rotatebox{90}{\hspace{3ex}$y/H$}
    \end{minipage}
    \begin{minipage}{.45\linewidth}
      \includegraphics[height=0.85\linewidth,
      clip, bb= 35 27 437.76 412]
      {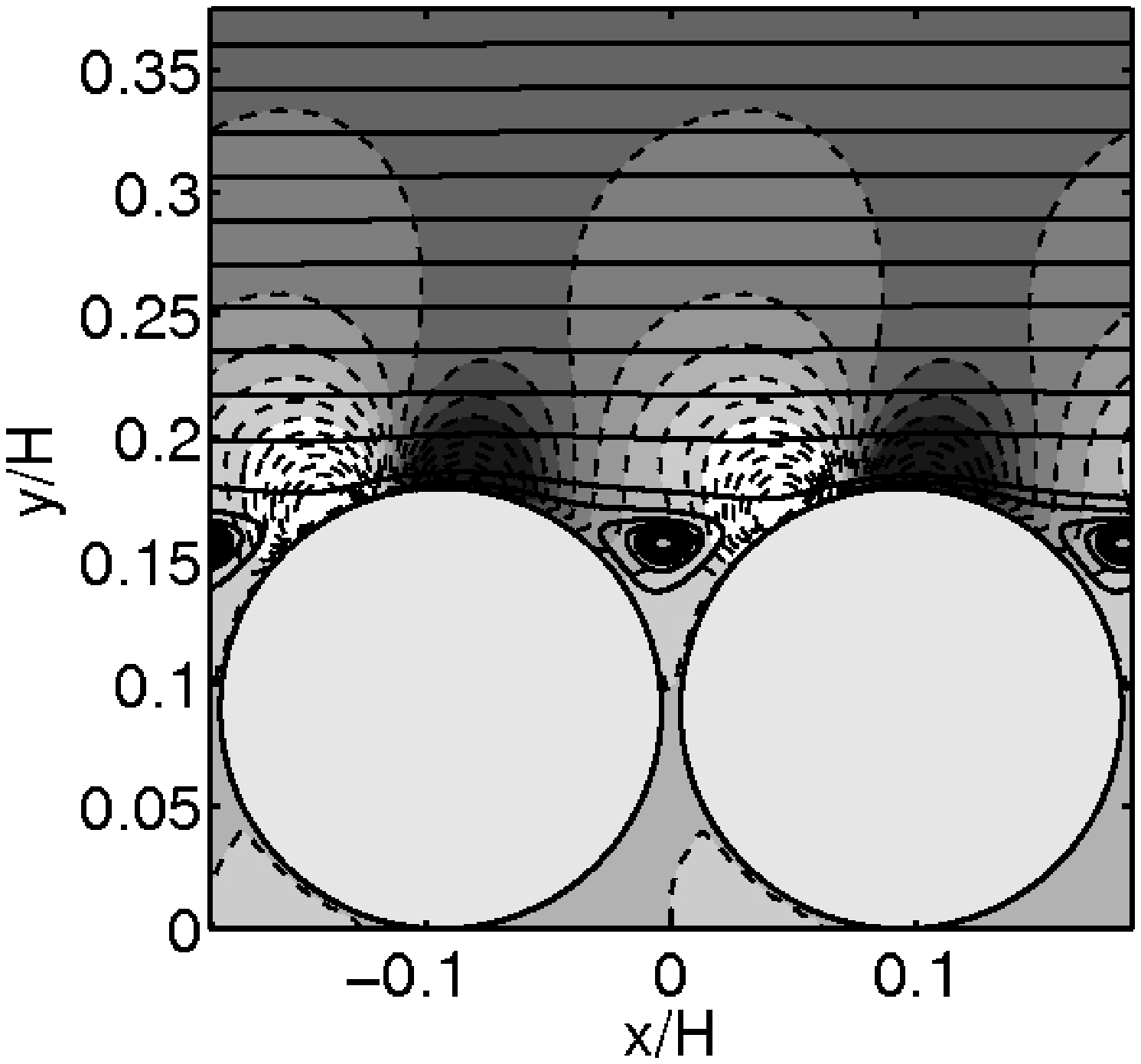}
      \hspace{-0.75\linewidth}\raisebox{0.75\linewidth}
      {\colorbox{white}{$(c)$}}
      \\
      \centerline{$x/H$}
    \end{minipage}
    \begin{minipage}{.45\linewidth}
      \includegraphics[height=0.85\linewidth,
      clip, bb= 0 27 437.76 412]
      {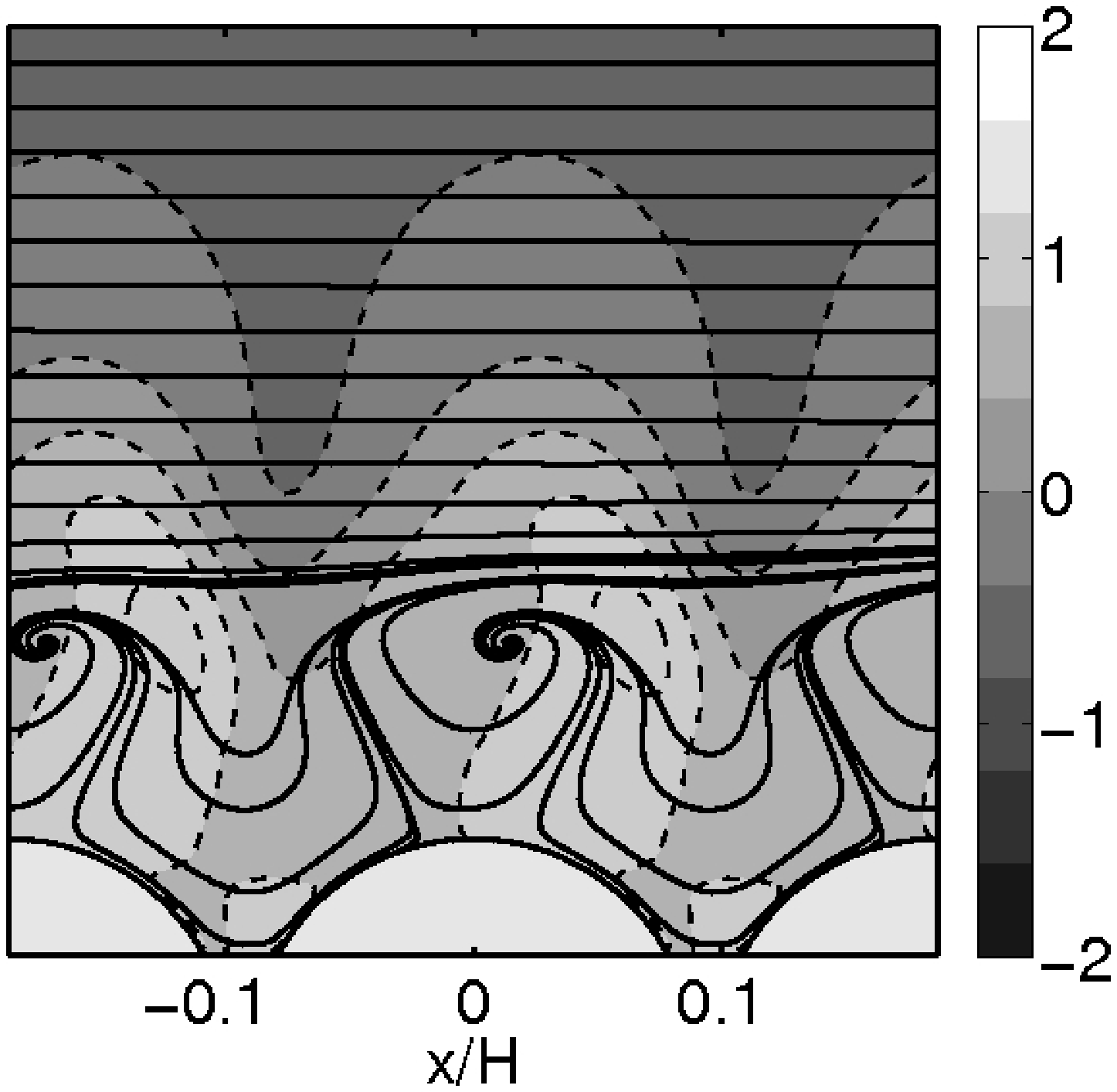}
      \hspace{-0.92\linewidth}\raisebox{0.75\linewidth}
      {\colorbox{white}{$(d)$}}
      \\
      \centerline{$x/H$\hspace{1.5cm}}
    \end{minipage}
  \end{center}
  
  \caption{
    As figure~\ref{fig:3d-mean} but showing the time-averaged pressure
    field and corresponding streamlines; 
    dashed lines: iso-contour lines of pressure,
    $\langle p\rangle_{tp} / \left(\rho u_\tau^2\right)$,
    from values of -4 to 4 in steps of 0.4;
    continuous lines: streamlines in the plane computed from $\langle u
    \rangle_{tp}$ and $\langle v \rangle_{tp}$.
    The direction of the bulk velocity is from left to right in all panels.
  } 
  \label{fig:3d-stream}
\end{figure}

The sheltering effect of the neighbouring particles
causes the flow velocity to decrease rapidly close to the roughness
tops and leads to marginal flow velocities within the roughness
layer. 
The highest velocity
gradients are produced in the vicinity of the roughness tops.
Similar observations were made in the experiments of
\cite{Pokrajac_Manes_TPM_2009} who studied 
a comparable particle arrangement at bulk Reynolds numbers of
order $10^4$ and a ratio of $h / D \approx 3.5$. 
Also similar to their results is the formation of a recirculation 
between two spanwise rows of spheres that extends over the
entire spanwise direction.
The shape of the recirculation is similar in both of our present
cases, however the strength differs.  
In case F50 the backflow velocities reach values 
as low as 
$\langle u \rangle_{tp}\approx -0.4u_\tau$ 
in figure \ref{fig:3d-mean}$(c)$ and $\langle u \rangle_{tp}\approx
-0.2u_\tau$ in figure \ref{fig:3d-mean}$(d)$. In case F10 the
magnitude of the backflow 
velocity is below $0.05u_\tau$.  
The recirculation can also be observed in figure \ref{fig:3d-stream} that 
shows streamlines of the mean flow projected into 
the same planes 
shown in figure \ref{fig:3d-mean}, i.e.~by computing the streamlines
using only $\langle u \rangle_{tp}$ and $\langle v
\rangle_{tp}$, together with contours of the time-averaged pressure field. 
The pressure distributions in both simulations are similar. 
However, in case F10 the magnitude of the pressure, $\langle
p\rangle_{tp} / (\rho u^2_\tau)$,
is a factor of two smaller compared to case F50.
Please note that the three-dimensional time-averaged flow is not fully
converged and at the location of the planes shown in figure
\ref{fig:3d-stream} there is a
weak net flow in the spanwise direction with a maximum amplitude of 
$\langle w \rangle_{tp}\approx 4 \cdot 10^{-4} U_{bh}$ ($ 6 \cdot 10^{-4}
U_{bh}$) in case F10 (F50).  
This net flow is within the range of the statistical
uncertainty. 
  
%

\begin{figure}
\begin{center}
    \begin{minipage}{2ex}
      \rotatebox{90}{\hspace{3ex}$\left( y-y_0 \right) / D$}
    \end{minipage}
    \begin{minipage}{.6\linewidth}
      \includegraphics[width=1.\linewidth]
      {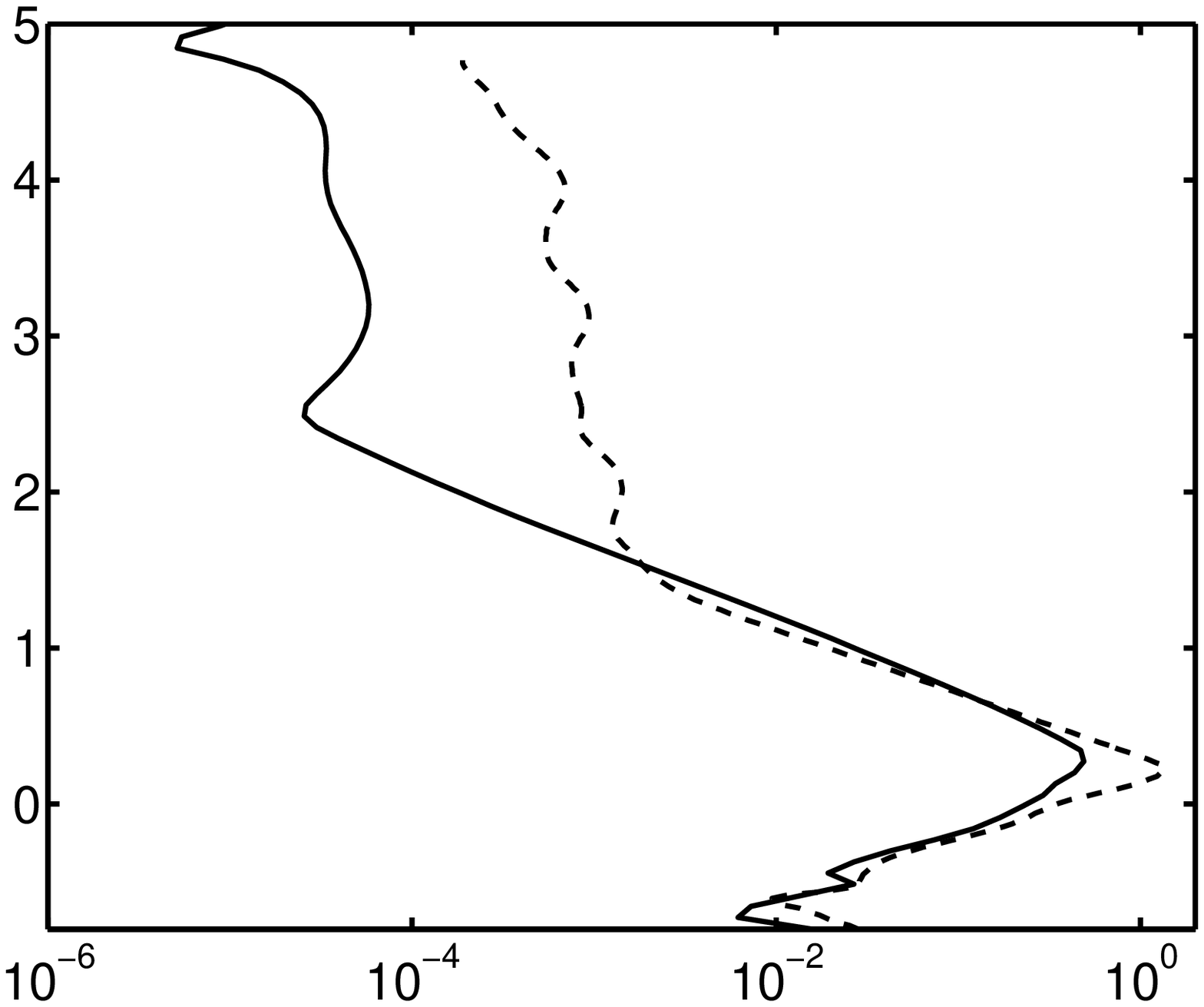}\\ 
      \centerline{$p_{rms}^{\prime\prime} / (\rho u_\tau^2)$}
    \end{minipage}

    \caption{Three-dimensionality of the time-averaged flow field as a
      function of wall distance quantified via
      $p^{\prime\prime}_{rms}$ (defined in \ref{eqn:diff-3d-plane}). 
      Solid lines: case F10 ; dashed lines: case F50.} 
    \label{fig:3d-effect}
  \end{center}
\end{figure}

A question of interest is how far the three-dimensionality of the
bottom wall directly influences the flow. 
Figure \ref{fig:3d-stream} already shows that one particle diameter above the
roughness tops the time-averaged pressure field is still visibly
affected. 
In order to quantify the effect of three-dimensionality, the
difference between the time-average of a field
$\langle\phi\rangle_{tp}$
(where $\phi$ 
can stand for either pressure or one of the velocity components) and
its time and plane-averaged value, 
$\langle\phi\rangle$, can be defined, viz.
\begin{equation}\label{eqn:diff-3d-plane}
  \phi^{\prime\prime}=\langle \phi\rangle_{tp}-\langle \phi\rangle
  \,.
\end{equation}
Note that the quantity defined by (\ref{eqn:diff-3d-plane}) is
sometimes called spatial disturbance in the context of the
double-averaging methodology \citep[][]{Nikora_etal_JHE_2001}.
From equation (\ref{eqn:diff-3d-plane})
the corresponding standard deviation 
$\phi^{\prime\prime}_{rms}=\sqrt{\langle
  \phi^{\prime\prime}\phi^{\prime\prime}\rangle}$ can be computed.  
In both cases, F10 and F50, the standard deviation of pressure
$p^{\prime\prime}_{rms}$ drops by several orders of magnitude in between $y=D$
and $y=2D$, as shown in figure \ref{fig:3d-effect}.
The same is true for the velocity field (not shown). 
Therefore, the time-averaged flow statistics appear to be essentially
one dimensional beyond wall distances of $2D$.
This is somewhat smaller than the values reported in previous
investigations of flow 
over rough walls \citep[cf.][]{Jimenez_ARFM_2004} which might be
related to the low values of $D^+$ and $Re_\tau$ considered.
%

%% file: results_for_mod_v1.tex
\subsection{Statistics of particle forces}
\label{ssec:stat_forces}
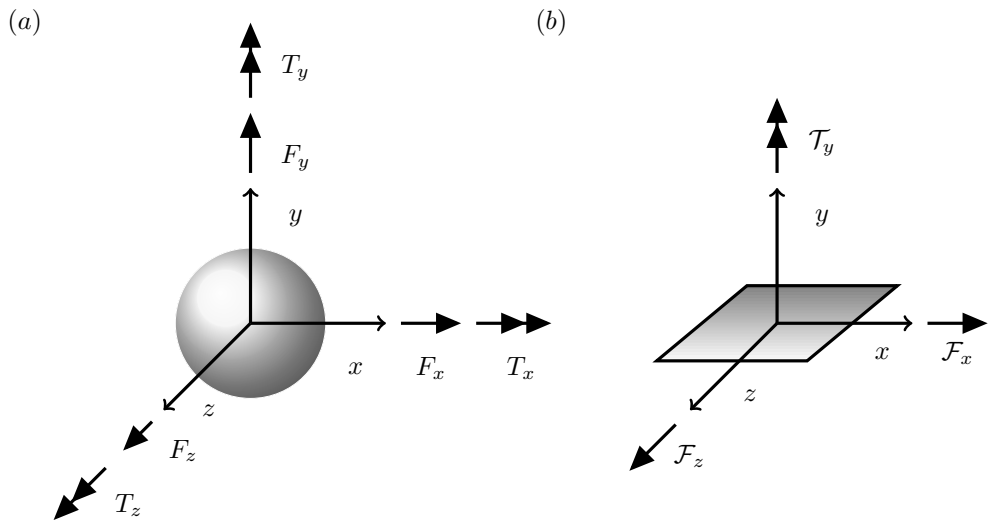
\begin{figure}
  \begin{center}
    \begin{minipage}{1.\linewidth}
      \pgfarrowsdeclarecombine[0pt]
      {tor}{tor}{triangle 45}{triangle 45}{triangle 45}{triangle 45}%
      
      \begin{tikzpicture}
        \coordinate (o) at (-7,0,0);
        
        \definecolor{ccc}{rgb}{0.8, 0.8, 0.8} 
        \shadedraw[shading=ball,ball color=gray!10, white] (o) circle
        (1.);
        
        \draw[] (o)++(-3,4,0) node[] {($a$)} ;
        
        \draw[very thick,->] (o) -- +(1.8,0,0) 
        node[yshift=-.6cm, xshift=-.4cm] {$x$} ;
        \draw[very thick,-triangle 45] (o) ++(2,0,0)--+(0.8,0,0) 
        node[yshift=-.6cm,xshift=-.4cm]{$F_x$} ; 
        \draw[very thick,-tor] (o) ++(3,0,0)--+(1.0,0,0) 
        node[yshift=-.6cm,xshift=-.4cm]{$T_x$} ; 
        
        \draw[very thick,->] (o) -- +(0,1.8,0) 
        node[yshift=-.4cm, xshift=.6cm]{$y$} ; 
        \draw[very thick,-triangle 45] (o) ++(0,2,0)--+(0,0.8,0) 
        node[yshift=-.6cm,xshift=.6cm]{$F_y$} ; 
        \draw[very thick,-tor] (o) ++(0,3,0)--+(0,1.0,0) 
        node[yshift=-.6cm,xshift=.6cm]{$T_y$} ;
        
        \draw[very thick,->] (o) -- +(0,0,3)   
        node[yshift=0cm,xshift=.6cm]{$z$} ;
        \draw[very thick,-triangle 45] (o) ++(0,0,3.4)  --+(0,0,1.0) 
        node[yshift=0cm,xshift=.8cm]{$F_z$} ; y
        \draw[very thick,-tor] (o) ++(0,0,5)  --+(0,0,1.8)
        node[yshift=0.2cm,xshift=1.cm]{$T_z$} ; 
        
        \coordinate (o) at (0,0,0);
        
        \begin{scope}[every node/.append style={xslant=1.2},xslant=1.2,
          ]
          \path[fill=lightgray, shade,draw=black, very thick]
          (-1,-0.5,0) rectangle +(2,1); 
        \end{scope}
        
        \draw[] (o)++(-3,4,0) node[] {($b$)} ;
        
        \draw[very thick,->] (o) -- +(1.8,0,0) 
        node[yshift=-.4cm, xshift=-.4cm] {$x$} ;
        \draw[very thick,-triangle 45] (o) ++(2.,0,0)--+(0.8,0,0)
        node[yshift=-.4cm, xshift=-.4cm]{$\mathcal{F}_x$} ;  
        
        \draw[very thick,->] (o) -- +(0,1.8,0) 
        node[yshift=-.4cm, xshift=.6cm]{$y$} ; 
        \draw[very thick,-tor] (o) ++(0,2,0)--+(0,1.0,0) 
        node[yshift=-.6cm, xshift=.6cm]{$\mathcal{T}_y$} ;
        
        \draw[very thick,->] (o) -- +(0,0,3)   
        node[yshift=.2cm,xshift=.8cm]{$z$} ;
        \draw[very thick,-triangle 45] (o) ++(0,0,3.5)  --+(0,0,1.6) 
        node [yshift=.2cm,xshift=.8cm]{$\mathcal{F}_z$} ; 
      \end{tikzpicture}
    \end{minipage}
    
    \caption{
        Sketch illustrating the definition of force
        (---$\blacktriangleright$) and 
        torque (---$\blacktriangleright\blacktriangleright$) 
        %
        %
        on a particle ($a$) and a square surface element in a smooth-wall
        channel ($b$)
      }
    \label{fig:sketch}
  \end{center}
\end{figure}

%
%
%
The hydrodynamic force, $\mathbf{F}$, acting on a particle is defined as
\begin{eqnarray}
  \label{eqn:tot-force}
  \mathbf{\mathbf{F}} &=& \int_\Gamma {
    \boldsymbol{\tau}
    \cdot \mathbf{n}}~\mathrm{d}\Gamma  
  - \int_\Gamma{ p^{tot} \mathbf{n} }~\mathrm{d}\Gamma,
\end{eqnarray}
where $\Gamma$ is the sphere's surface, $\mathbf{n}$, is 
the surface normal vector, $ \boldsymbol{\tau}
= \rho\nu \left( \partial_j u_i  + \partial_i u_j \right)$ 
is the viscous stress tensor and $p^{tot}$ is the pressure. The latter
can be split into two parts $p^{tot}=p + p_l$, where $p_l$ represents
the linear
variation in streamwise direction which results from the imposed
pressure-gradient that drives the flow, and $p$ corresponds to the 
three-dimensional instantaneous fluctuation.
The first term on the right hand side of equation (\ref{eqn:tot-force}) is
the force due to viscous stresses, the second term is the force due to
pressure. 
A sketch that illustrates the definition of the force on a
particle can be seen in figure \ref{fig:sketch}($a$).

%
%
In order to scale the hydrodynamic forces, reference
quantities need to be defined. 
%
For the present case of particles within a roughness layer the
subject is a matter of discussion and several definitions have been proposed
in the literature \citep[see][]{Hofland_etal_JHE_2005}. 
%
Here, the reference force is defined 
as $F_R = \rho u_\tau^2  A_R$ with the reference area $A_R = L_x L_z /N_p$.
%
%

%
\begin{table}
\begin{center}
\begin{tabular}{ lccccccccccccc}
Case & $C_{F}^x$ & $C_{F}^y$ & $C_{F}^z$ &  $\alpha_F$ &
 $ {\sigma_F^x} / F_R$ & $ \sigma_F^y /F_R$ & $
 {\sigma_F^z}/F_R $ &  $S_F^x$ & $S_F^y$ & $S_F^z $ &
$K_F^x$ & $K_F^y$ & $K_F^z$ \\
 F10 & 1.04 & 0.19 & 0.00 & 11$^\circ$  &  0.57 &  0.20 & 0.66 &  0.18
 &  1.80 & 0.01 &  10.13 &  19.08 & 9.92 \\  
 F50 & 1.15 & 0.37 & 0.00 & 18$^\circ$  &  1.32 &  0.66 & 1.26 &  0.06
 &  0.26 & 0.01 &  4.98 &  5.68 & 4.29 \\  
\end{tabular}
\caption{Statistics of particle forces in case F10 and case F50,
  where $C_F^{x_i}=\langle F^{x_i} / F_R \rangle$ is the normalised
  mean force component in the $x_i$-direction,
  $\alpha=\arctan(C_F^y/C_F^x)$ is the angle of the resulting force
  with respect to
  the $x$-axis, $\sigma_F^{x_i}$ is the normalised standard deviation of the
  force in $x_i$, $S_F^{x_i}$ and $K_F^{x_i}$ are the skewness and
  kurtosis of the respective force component.}
\label{tab:stat_forces}
\end{center}
\end{table}

\begin{figure}
\begin{center}
  \begin{minipage}{2ex}
    \rotatebox{90}{$\tilde{z}/D$}
  \end{minipage}
  \begin{minipage}{.46\linewidth}
    \includegraphics[width=1.\linewidth]
    {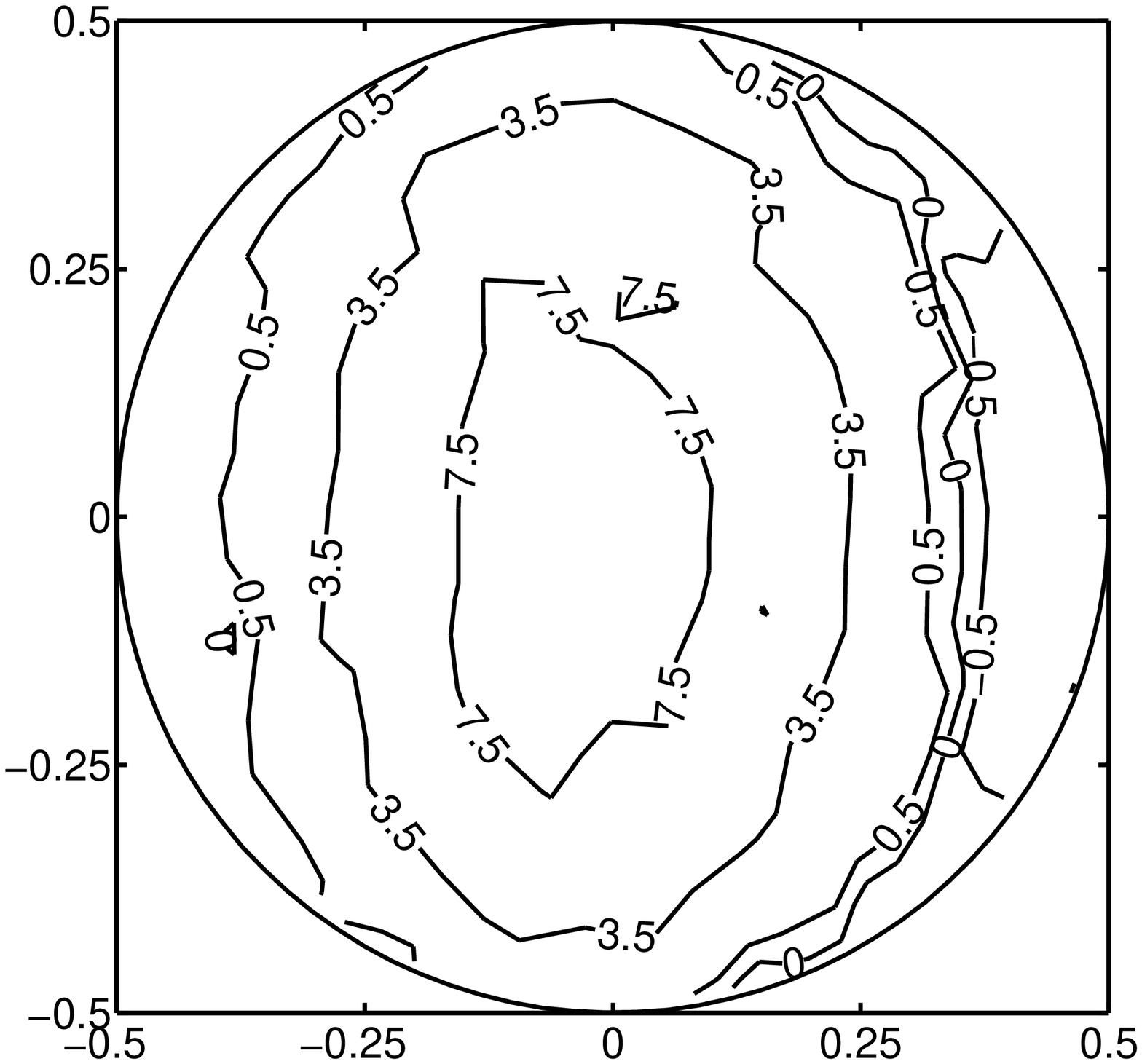}
    \hspace{-0.9\linewidth}\raisebox{0.82\linewidth}{$(a)$}
  \end{minipage}
  \begin{minipage}{2ex}
    \rotatebox{90}{$y/D$}
  \end{minipage}
  \begin{minipage}{.46\linewidth}
    \includegraphics[width=1.\linewidth]
    {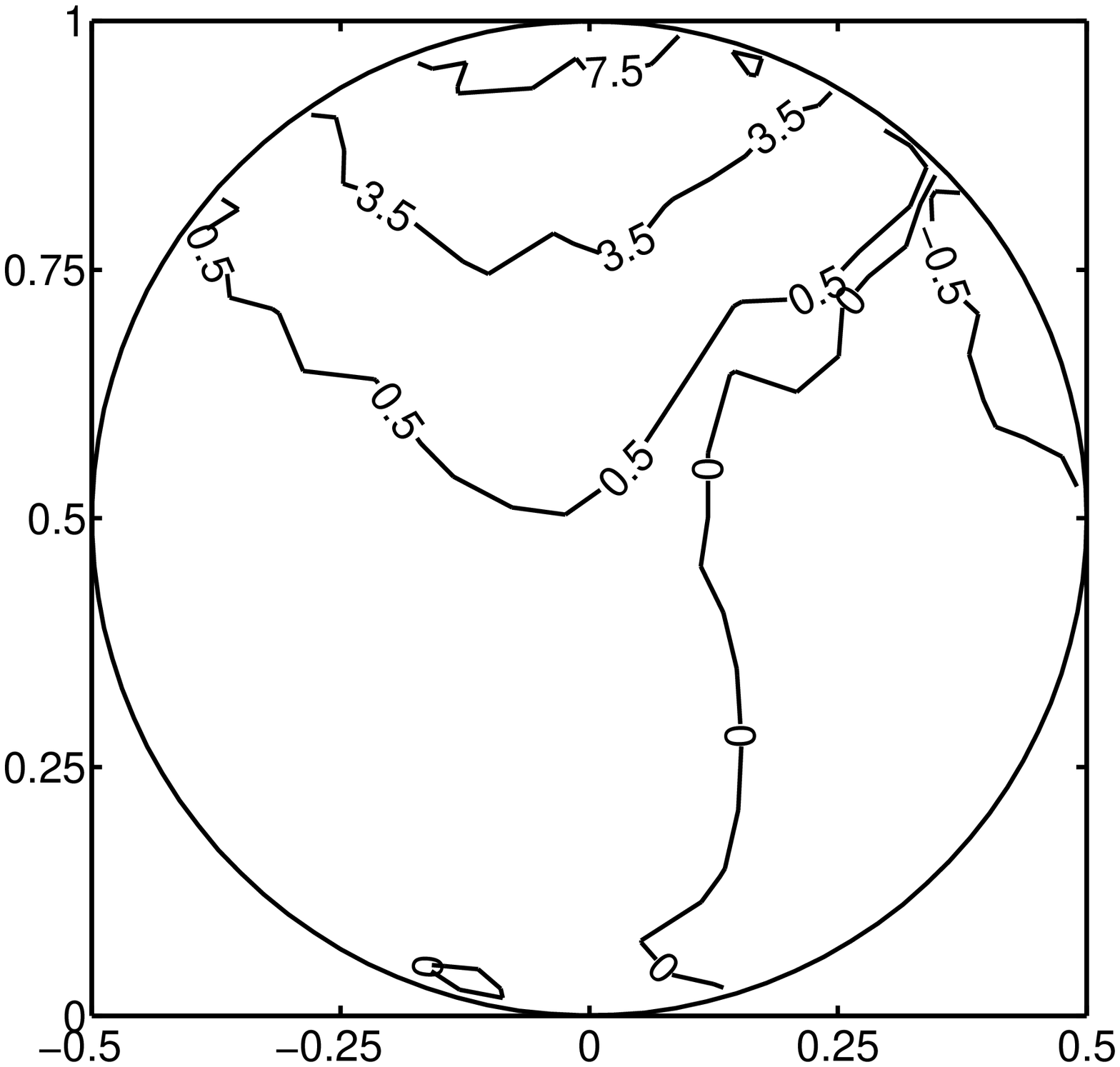}
    \hspace{-0.9\linewidth}\raisebox{0.82\linewidth}{$(b)$}
  \end{minipage}

  \begin{minipage}{2ex}
    \rotatebox{90}{\hspace{3ex}$\tilde{z}/D$}
  \end{minipage}
  \begin{minipage}{.46\linewidth}
    \includegraphics[width=1.\linewidth]
    {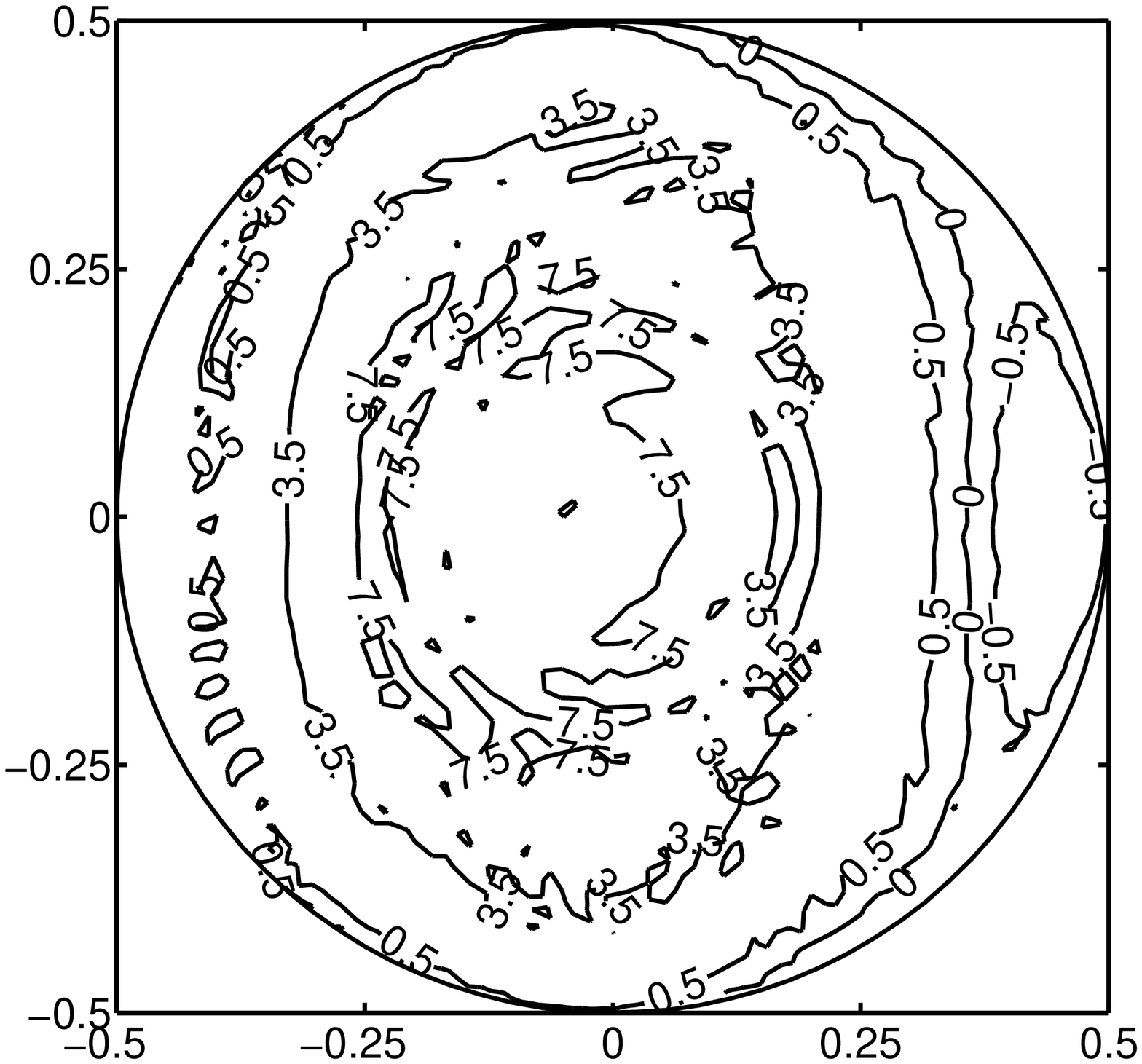}
    \hspace{-0.9\linewidth}\raisebox{0.82\linewidth}{$(c)$}\\
    \centerline{$\tilde{x}/D$}
  \end{minipage}
  \begin{minipage}{2ex}
    \rotatebox{90}{\hspace{3ex}$y/D$}
  \end{minipage}
  \begin{minipage}{.46\linewidth}
    \includegraphics[width=1.\linewidth]
    {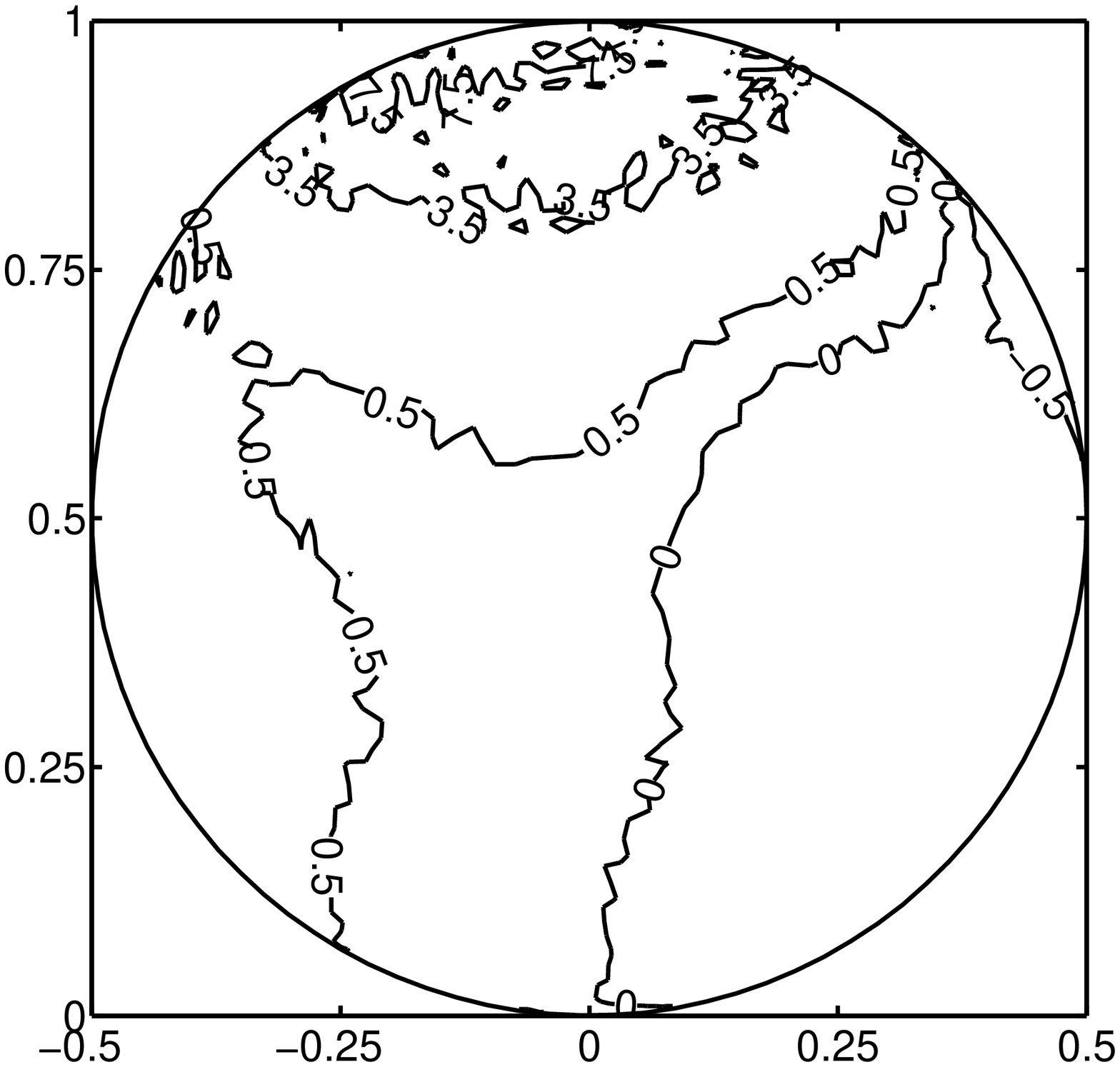}
    \hspace{-0.9\linewidth}\raisebox{0.82\linewidth}{$(d)$}\\ 
    \centerline{$\tilde{x}/D$}
  \end{minipage}
  
  \caption{Spatial distribution of $\tau_D$, normalised by $F_R /
    A_{sph}$. 
    $\tilde{x}$ and $\tilde{z}$ are the coordinates with respect to the
    particle centre.
    The contour lines shown correspond to  
    [-0.5, 0.0, 0.5, 3.5, 7.5];
    Panels $(a)$ and $(b)$ show case F10, 
    panels $(c)$ and $(d)$ show case F50.}
  \label{fig:drag_dist_sph}
\end{center}
\end{figure}

\begin{figure}
\begin{center}
  \begin{minipage}{2ex}
    \rotatebox{90}{$\tilde{z}/D$}
  \end{minipage}
  \begin{minipage}{.46\linewidth}
    \includegraphics[width=1.\linewidth]
    {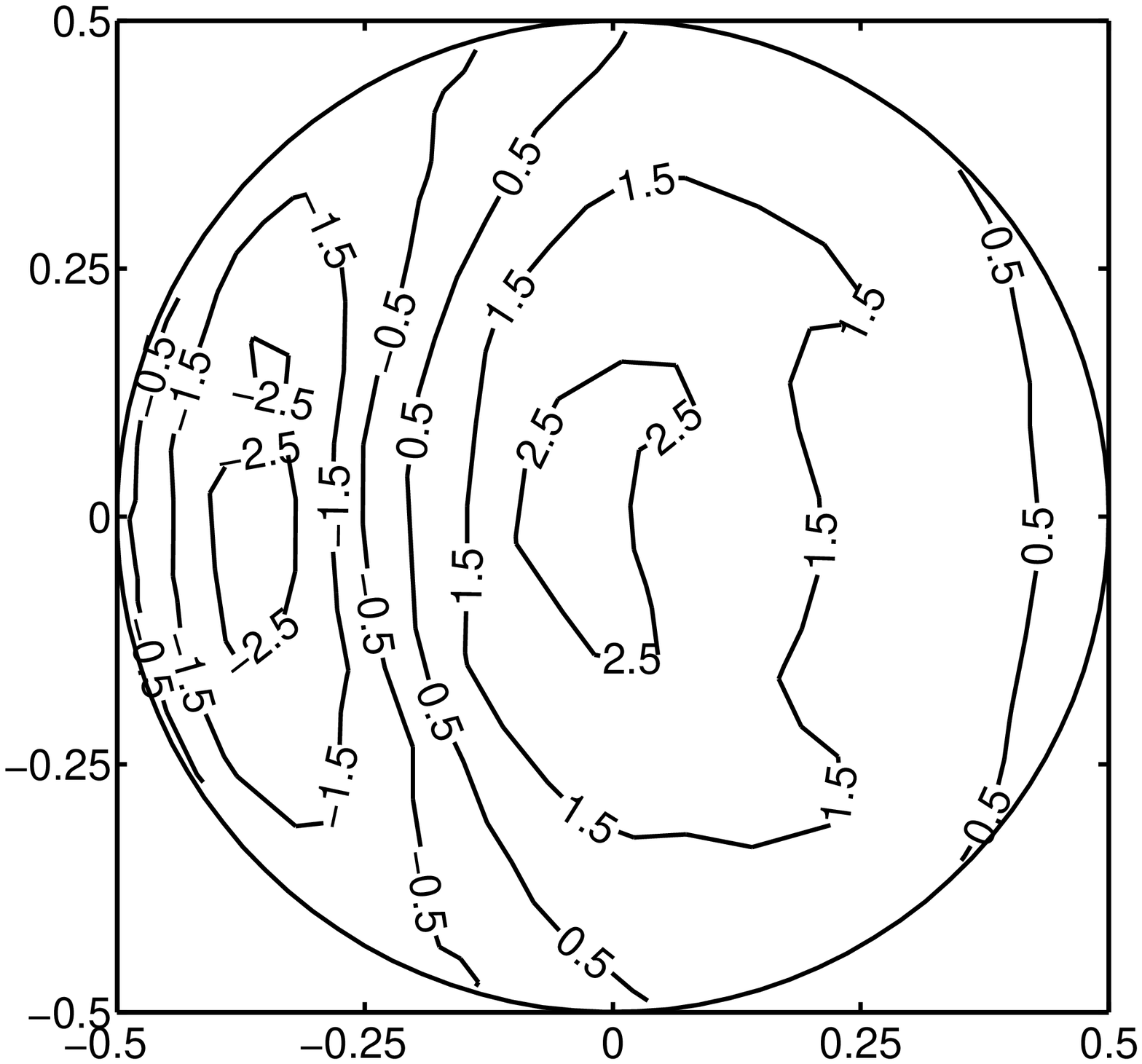}
    \hspace{-0.9\linewidth}\raisebox{0.82\linewidth}{$(a)$}
  \end{minipage}
  \begin{minipage}{2ex}
    \rotatebox{90}{$y/D$}
  \end{minipage}
  \begin{minipage}{.46\linewidth}
    \includegraphics[width=1.\linewidth]
    {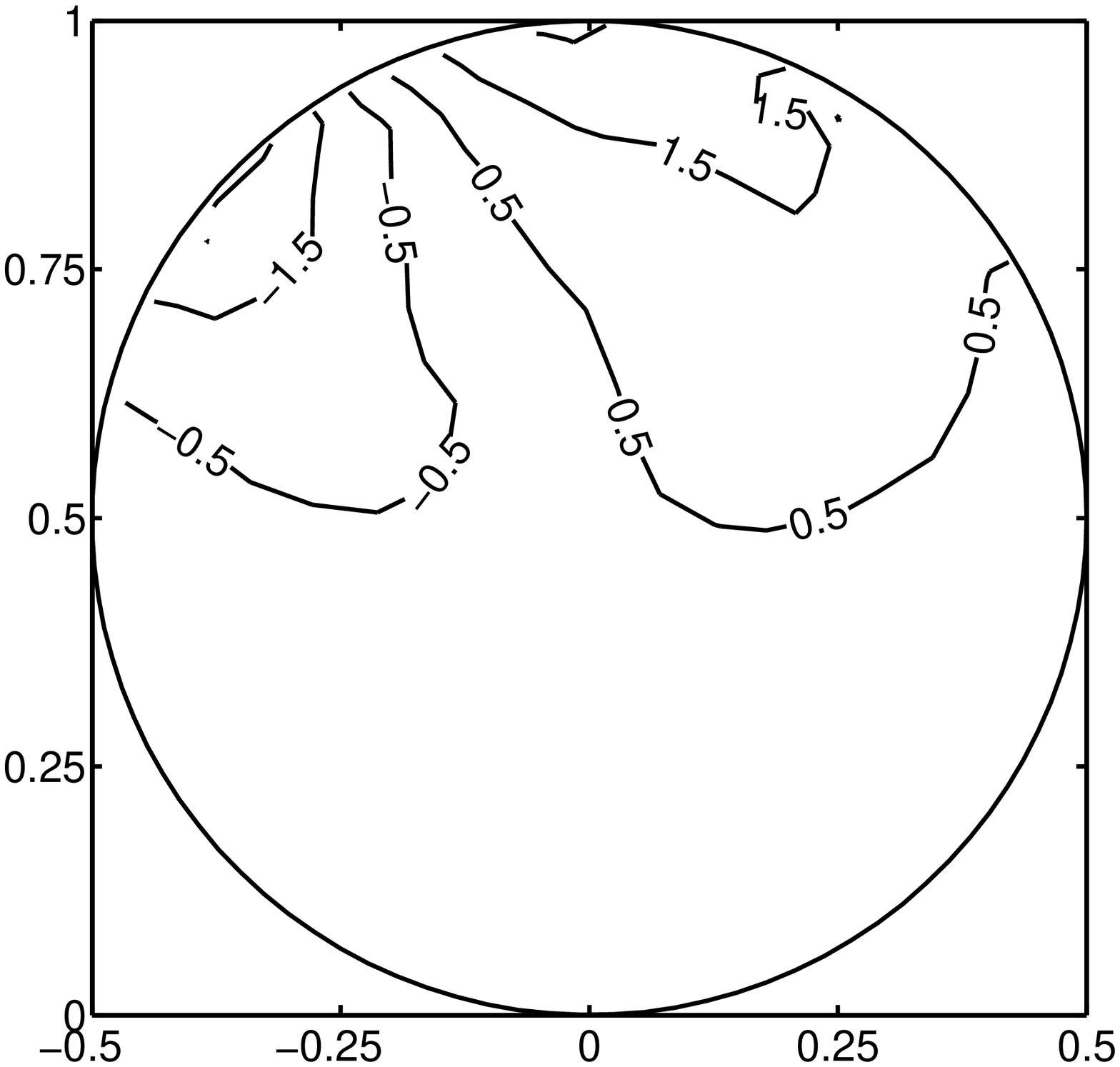}
    \hspace{-0.9\linewidth}\raisebox{0.82\linewidth}{$(b)$}
  \end{minipage}

  \begin{minipage}{2ex}
    \rotatebox{90}{\hspace{3ex}$\tilde{z}/D$}
  \end{minipage}
  \begin{minipage}{.46\linewidth}
    \includegraphics[width=1.\linewidth]
    {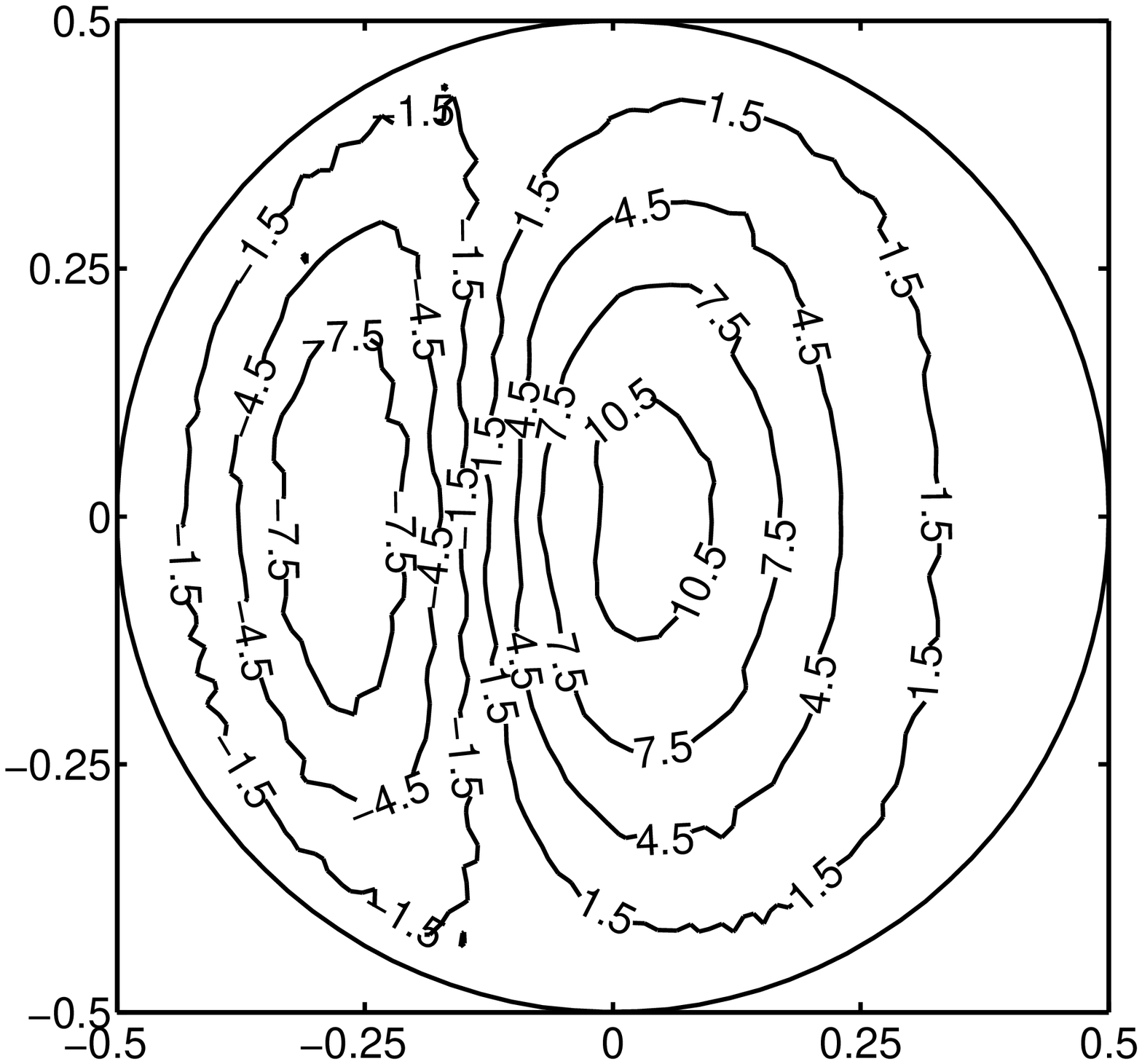}
    \hspace{-0.9\linewidth}\raisebox{0.82\linewidth}{$(c)$}\\
    \centerline{$\tilde{x}/D$}
  \end{minipage}
  \begin{minipage}{2ex}
    \rotatebox{90}{\hspace{3ex}$y/D$}
  \end{minipage}
  \begin{minipage}{.46\linewidth}
    \includegraphics[width=1.\linewidth]
    {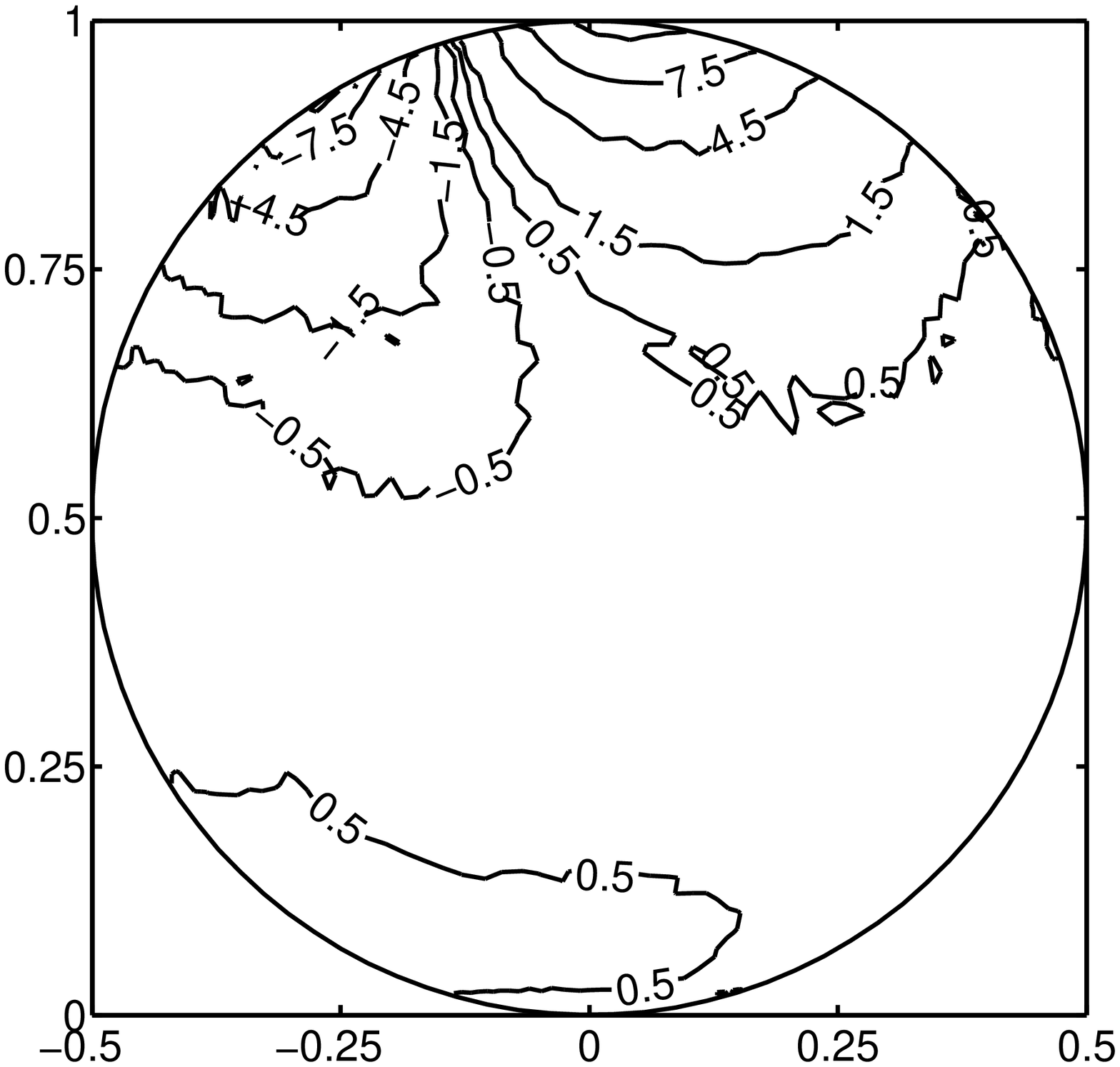}
    \hspace{-0.9\linewidth}\raisebox{0.82\linewidth}{$(d)$}\\ 
    \centerline{$\tilde{x}/D$}
  \end{minipage}

\caption{
  As figure \ref{fig:drag_dist_sph}, but for $\tau_L$. 
  The contours lines shown in $(a)$ and $(b)$ are at values from -2.5
  to 2.5 in steps of 1, in $(c)$ and $(d)$ at values from -7.5 to 10.5
  in steps of 3; in $(d)$ additionally the contours at the values -0.5
  and 0.5 are shown. 
}
\label{fig:lift_dist_sph}
\end{center}
\end{figure}

%
Table \ref{tab:stat_forces} summarises the particle force statistics of
the two cases, where  $C_F^{x_i}$ is the mean force on a particle in
$x_i$-direction normalised by $F_R$.
%
%
As can be seen, the mean values of
the forces acting in the streamwise direction (henceforth also called
``drag'') and the wall-normal direction (``lift'') are positive. 
Since the mean forces are directly related to the
mean flow through the time-averaged version of equation (\ref{eqn:tot-force}),
it is possible to shed some light onto the 
mechanisms that lead to drag and lift by analysing 
figures \ref{fig:3d-mean} and \ref{fig:3d-stream}. 
A more detailed picture can be obtained from
figure \ref{fig:drag_dist_sph} and figure \ref{fig:lift_dist_sph} which
show the distribution 
on the sphere's surface
of the stress leading to drag, $\tau_D$, and lift, $\tau_L$, viz.
\begin{eqnarray}\label{equ:stress-contrib-to-drag}
\tau_D&=&(\langle \boldsymbol{\tau} \rangle_t \cdot \mathbf{n}
- \langle p^{tot} 
\rangle_t \mathbf{n})\cdot  \mathbf{e}_1\,,\\
\label{equ:stress-contrib-to-lift}
\tau_L&=&(\langle \boldsymbol{\tau} \rangle_t \cdot \mathbf{n}
- \langle p^{tot} 
\rangle_t \mathbf{n})\cdot  \mathbf{e}_2\,,
\end{eqnarray}
where $\mathbf{e}_i$ is the unit vector in the $x_i$-direction. 
The stresses in figures \ref{fig:drag_dist_sph} and
\ref{fig:lift_dist_sph} 
are normalised by $F_R / A_{sph}$, where $A_{sph}=
\pi D^2$ is the surface area of the sphere; by virtue of this
normalisation the total integral of the quantities shown in the
figures yields the force coefficients $C_F^{x_i}$ given in
table \ref{tab:stat_forces}. 
Please note that the results of case F50 appear less smooth due to the
smaller number of particles, and therefore a smaller number of
samples.

%
Figure \ref{fig:drag_dist_sph} shows that the local stress contributing
to drag is similarly
distributed over the particle surface in both of our present flow
cases F10 and F50. One can observe a region of strong positive values with
the largest magnitude centred around a position slightly upstream of the
particle tops. From 
figure \ref{fig:drag_dist_sph}($a,c$) we can see that this region of
high positive local contributions to drag is 
slightly elongated in the spanwise direction. 
It results from the wall-normal gradients of the average streamwise
velocity component which are particularly important in the upper part
of the sphere as well as from the high pressure values found near the
upstream side of each sphere (cf.\ figures \ref{fig:3d-mean} and
\ref{fig:3d-stream}). 
On the downstream side of the particles, still in the upper
hemisphere, a smaller region with weak negative contributions to drag is
found, as a result of the recirculation region. 
In most of the lower (near-wall) half of the spheres, the contour lines of the
local drag contribution are roughly oriented in the wall-normal
direction, changing sign slightly downstream of the cross-stream plane
passing through the particle centre. 
In this context it should be noted, that the driving pressure gradient 
$\mbox{d}p_l/\mbox{d}x<0$ makes a weak but non-negligible
contribution to the drag which can be quantified as
approximately 2\% (9\%) of $C_F^x$ in case F10 (F50). 
Therefore, non-negligible values of local contributions to drag are
expected even in relatively quiescent regions, as is the case inside
the roughness layer. 
The qualitative and quantitative similarity of the distribution of
$\tau_D$ in both cases F10 and F50 results in
similar values for the drag coefficient 
in both cases (cf.\ table \ref{tab:stat_forces}). 
In particular, the overall drag coefficient $C_F^x$  in case F10
is close to unity, increasing to $1.15$ in case F50. 
These values are a result of the weak contribution of the drag on
the rigid wall below the layer of spheres to the total drag on
the wall and the choice of the reference force.
The drag coefficient as defined in the present study can be
approximated as 
\begin{equation}\label{eqn:aprox_cd}
  C_F^x 
  \approx\frac{V_f^{(tot)}+V_{sph}}{h A_R}
  \,,
\end{equation}
where $V_f^{(tot)}$ is the total volume occupied by fluid in a
periodic cell around a particle, $V_{sph}$ is
the volume occupied by a particle.
The approximation (\ref{eqn:aprox_cd}) neglects the
streamwise component of the shear force acting on the bottom wall
in addition to the drag due to the
periodic part of the pressure acting on the spherical caps. 
Evaluating this geometrical relation (\ref{eqn:aprox_cd}) yields $1.04$
($1.15$) for case F10 (F50).

%
%
Positive values for the lift coefficient, as observed in the present
simulations 
(cf.~table \ref{tab:stat_forces}), can be explained by two mechanisms.  
The approaching flow accelerates in the frontal part until the 
top of the sphere and from then on it decelerates. This
fact is reflected in the curvature of the streamlines 
(figures \ref{fig:3d-mean}$a,c$),
yielding a pressure distribution which exhibits lower values of
pressure near the particles tops (figures \ref{fig:3d-stream}$a,c$),
and therefore a positive lift.
%
In addition to pressure, shear might lead to a positive lift. 
As can be seen in figure \ref{fig:3d-mean}, the flow field above the
spheres is asymmetric  
(with respect to a cross-sectional plane through the particle centres)
as a result of the 
recirculation behind the particles. 
Therefore, the friction on the upstream side of the particle (in the upper
hemisphere) is expected to be higher
compared to the corresponding friction on the downstream side,
contributing positively to the lift on the particle. 
The pressure differences as well as the asymmetry of the flow
seem to be more pronounced in case F50 than in case F10 and might 
explain the observed increase in the lift coefficient. 

Figure \ref{fig:lift_dist_sph} shows
the distribution  on the sphere's surface of the stress leading to
lift, $\tau_L$. 
The shape of the contours is again similar
in both cases, however, the magnitude of the stress $\tau_L$ seems to
be significantly larger 
in case F50, leading after integration to the factor of two presented
in table \ref{tab:stat_forces}.
The spatial distribution is characterised by one dominant patch of
each positive and negative values of $\tau_L$, the maximum of both
being located on the $(x,y)$ symmetry plane, the former (positive)
near the particle top, the latter (negative) shifted upstream by
approximately $45^\circ$ ($30^\circ$) in case F10 (F50).
From the contours, it appears that the flow below the virtual wall
contributes little to the lift. 

\begin{figure}
  \begin{center}
  \begin{minipage}{2ex}
    \rotatebox{90}{\hspace{3ex}$y/D$}
  \end{minipage}
  \begin{minipage}{.46\linewidth}
    \includegraphics[width=1.\linewidth]
    {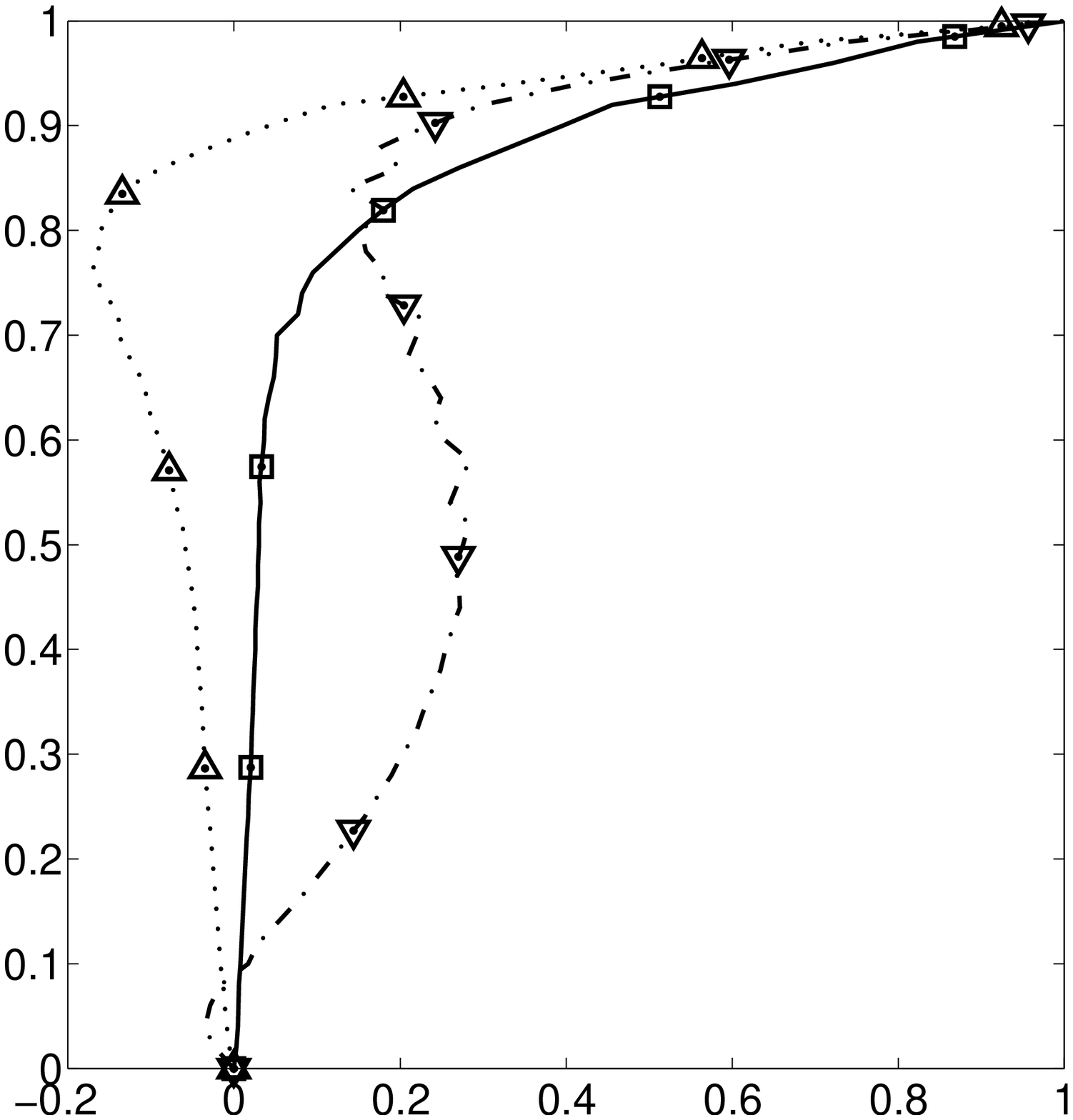}
    \hspace{-.9\linewidth}\raisebox{.92\linewidth}{$(a)$}\\ 
    \centerline{${\cal S}_\phi/{\cal S}_\phi(D)$}
  \end{minipage}
  \begin{minipage}{2ex}
    \rotatebox{90}{\hspace{3ex}$y/D$}
  \end{minipage}
  \begin{minipage}{.46\linewidth}
    \includegraphics[width=1.\linewidth]
    {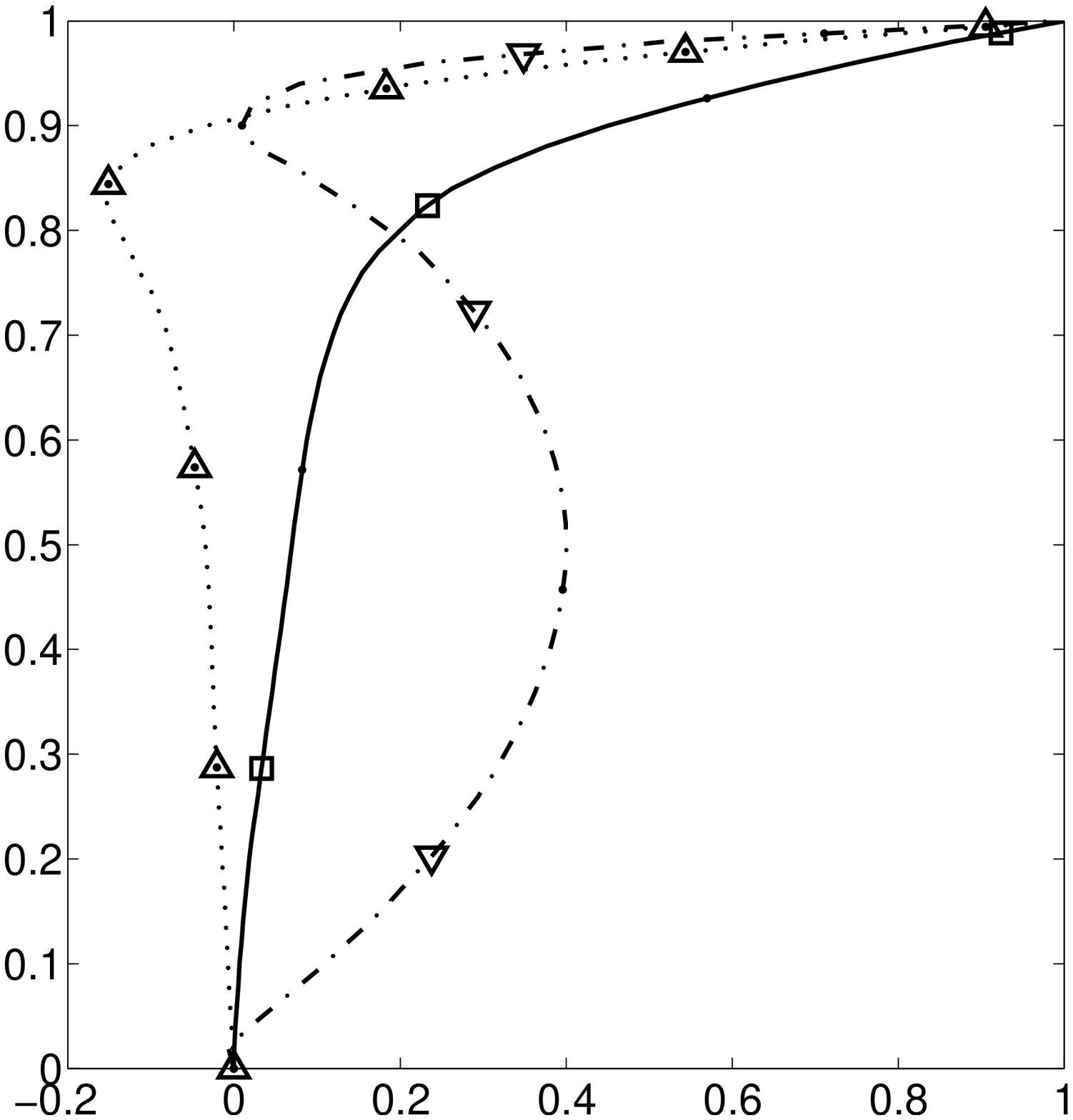}
    \hspace{-.9\linewidth}\raisebox{.92\linewidth}{$(b)$}\\ 
    \centerline{${\cal S}_\phi/{\cal S}_\phi(D)$}
  \end{minipage}
  
  \caption{Cumulative function 
    $\mathcal{S}_\phi$ of the stress 
    contribution to the mean 
    value of drag, lift and spanwise torque on a particle
    as a function of $y$ and normalised by
    its maximum value. 
    $- \square -$, drag;
    $- \cdot - \bigtriangledown - \cdot -$, lift;
    $\cdots \bigtriangleup \cdots$, spanwise torque. 
    Panel $(a)$ shows case F10, panel $(b)$ shows case F50.
}
\label{fig:cum_contr_F10-F50}
\end{center}
\end{figure}

In order to quantify the contribution
integrated from the bottom of a particle up to a certain fraction of
its diameter,
we can define a  cumulative function  
\begin{equation}
  \label{eqn:cumulative-contrib-to-stress}
  \mathcal{S}_\phi(y)=\frac{D}{2}\,\int_0^y
  \int_{0}^{2\pi} \tau_\phi(y,\theta) ~\mathrm{d}\theta \mathrm{d}y\,,
\end{equation}
where $\tau_\phi(y,\theta)$ stands for either $\tau_L$ or $\tau_D$ 
evaluated at a position on the sphere's surface given by the
wall-distance $y$ and an azimuthal angle $\theta$ in the wall-parallel
plane. 
Figure \ref{fig:cum_contr_F10-F50} shows $\mathcal{S}_L$ and $\mathcal{S}_D$ 
normalised by the net values of lift and drag, respectively. 
The contribution to the net drag by the flow in  
the lower half of the sphere is small in both cases, 
the cumulative drag value increasing monotonically and with increasing
slope from the wall to the top of the sphere. 
Conversely, the cumulative contribution to the lift first
increases with increasing wall-distance up to values of approximately
25\% (40\%) of the total in case F10 (F50) at $y\approx0.5D$, before decreasing
again to a small value at $y\approx0.9D$. Beyond that, in a small area
surrounding the top of the sphere, is where most of the net lift
is generated. 
%
%
In case F50, the lift increases with respect to case F10 more than the
drag, which leads to a higher angle, $\alpha=\arctan(C_F^y/C_F^x)$,
of the resulting force (cf.\ table \ref{tab:stat_forces}). 
%

%
The spanwise force should be zero for symmetry reasons.
In both cases the calculated mean spanwise force coefficient is
more than two orders of magnitude lower than the drag coefficient. This fact
provides confidence in the convergence of the statistics.  

\begin{figure}
  \begin{center}
    \begin{minipage}{3ex}
      \rotatebox{90}{\hspace{3ex}
        $\langle F_y \rangle / (\rho \nu^2)$}
    \end{minipage}
    \begin{minipage}{.55\linewidth}
      \includegraphics[width=1.\linewidth,
      clip=true,
      viewport=56 30 526 424]
    {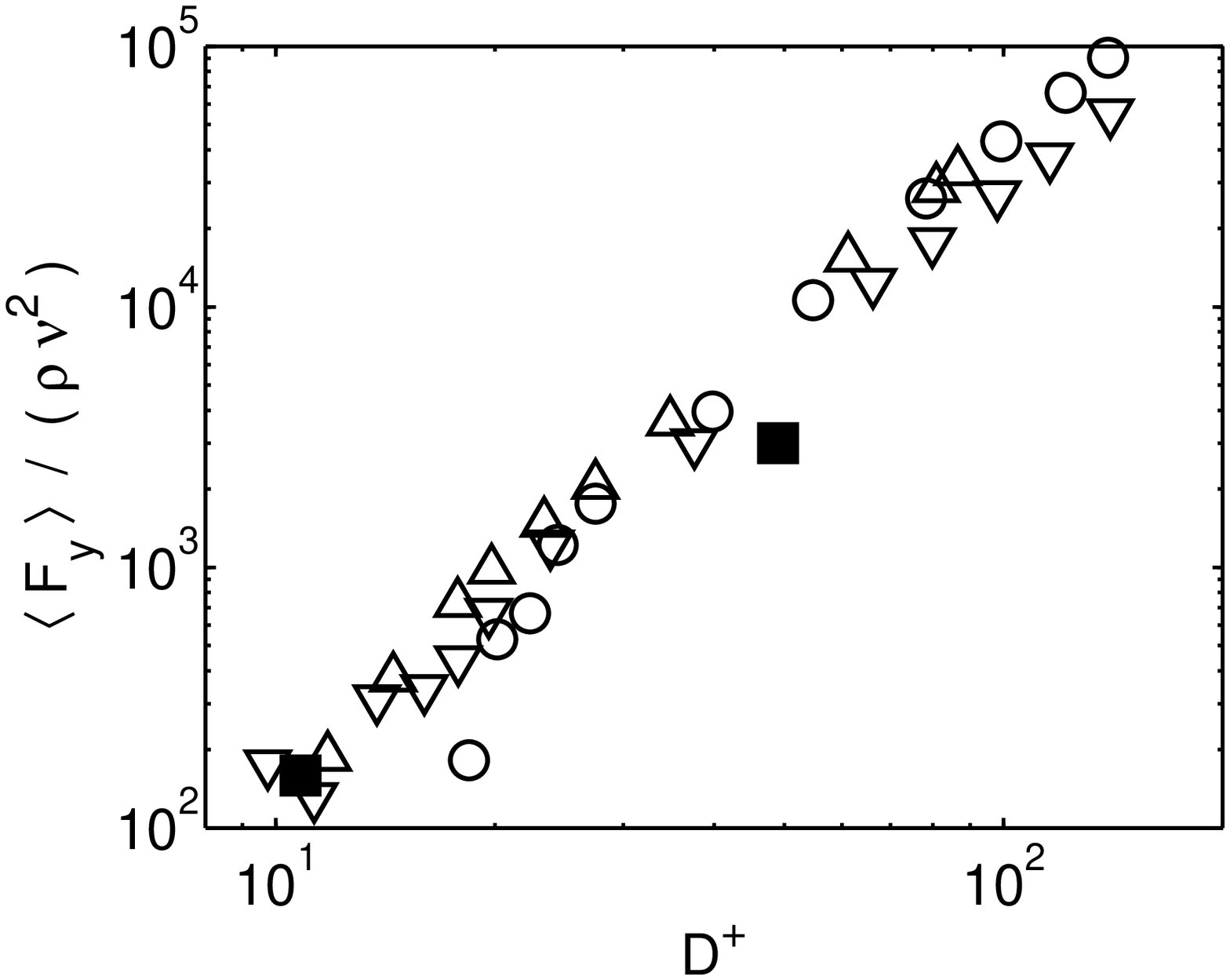}
    \\ 
    \centerline{$D^+$}
  \end{minipage}

\caption{Comparison of $\langle F_y \rangle / \left( \rho \nu^2
  \right)$ as a function of $D^+ $ of case F10 and case F50 (solid
  symbols) with mean lift on a sphere 
  placed in between roughness elements in a boundary layer by
  \cite{Hall_JFM_1988} (open symbols);  present
  simulations: $\blacksquare$; experiments \cite{Hall_JFM_1988}:
  $\bigtriangledown$: $D_r=5/3 D$, $\bigcirc$: $D_r = D$,
  $\triangleleft$: $D_r = 2/3 D$, where $D_r$ is the radius of the
  rods spaced with $D_r$ upstream and downstream of the sphere.}
\label{fig:comp_hall}
\end{center}
\end{figure}

%
Additional support to the mean forces just discussed is provided by
comparison 
to experimental measurements performed in a somewhat similar
configuration by \cite{Hall_JFM_1988}. In that study, the mean lift
on a particle near a boundary 
was measured in a wind tunnel with smooth as well as rough walls.
The interesting case for the present discussion consisted of a sphere
of diameter $D$ placed in-between spanwise rods of diameter
$D_r$ evenly spaced out with a distance $D_r$.  
Figure \ref{fig:comp_hall} presents the comparison of the mean lift
normalised by $\rho \nu^2$ as a function of $D^+$,
between the values obtained in the experiments of
\cite{Hall_JFM_1988} and the present simulations. In spite of the
different setups, the lift 
obtained in case F10 is perfectly consistent with the measurements while
the lift obtained in case F50 is somewhat lower. The reason for this
might be  
that in the setup of the simulations the neighbouring spheres are
closer producing 
an increased sheltering effect. This is also supported by the
experimental observation that
lower lift values are obtained when 
the value of $D_r/D$ is increased \citep{Hall_JFM_1988}. 

%

%
%
%
In contrast to the direct relation between the mean flow field and the
mean forces on a particle, a similar straightforward relation between
the statistics of the fluid velocity fluctuations and of the
particle force fluctuations cannot be derived 
from equation (\ref{eqn:tot-force}). This is due to two factors. First,
the definition of the standard deviation (and higher order moments) of
the force fluctuations 
is non-linear. Second, the
integrals in equation (\ref{eqn:tot-force})
act like a filter in the sense that not all scales participate in creating
force fluctuations on a particle. For example, flow scales much smaller than
$D$ might cancel out in the integral sense
as will be discussed in detail in \S\ref{ssec:stat_torq}. 
In spite of this observation, a direct relation between flow
velocity statistics above the bed or behind an obstacle is often assumed in the
literature in order to estimate the intensity of force fluctuations on
a particle \citep[cf.][]{Papanicolaou_etal_JHE_2002,
  Garcia_ASCE_manual_08}.

In the present simulations we observe that the standard deviations for
the streamwise and spanwise components of the particle forces are of
similar magnitude in both cases F10 and F50 (cf.\ 
table \ref{tab:stat_forces}). 
It is also found that the standard deviation of lift in both cases is
roughly half the value of the other two components. 
Overall the intensity of the fluctuations in case F50 is more 
than a factor of two larger than in case F10. 
Thus, the particle force fluctuations in the present case do not seem to scale
directly with the intensity of the plane- and time-averaged fluid velocity
fluctuations (cf.\ figure \ref{fig:velrms_norm}),  
since $u_{rms}$ is larger than $w_{rms}$ over most of the 
flow depth, and especially close to the wall. Furthermore, the
difference in the fluid velocity fluctuation intensities between case
F10 and F50 is very small compared to the above stated difference in
the particle force fluctuation intensities. 
It can therefore be concluded that a direct link between fluid and
particle force fluctuation intensities cannot be inferred in the
present cases. 

%
%
%

\begin{figure}
  \begin{center}
    \begin{minipage}{2ex}
      \rotatebox{90}{\hspace{3ex}$\sigma_F \cdot pdf$}
    \end{minipage}
    \begin{minipage}{.45\linewidth}
      \includegraphics[width=1.\linewidth]
      {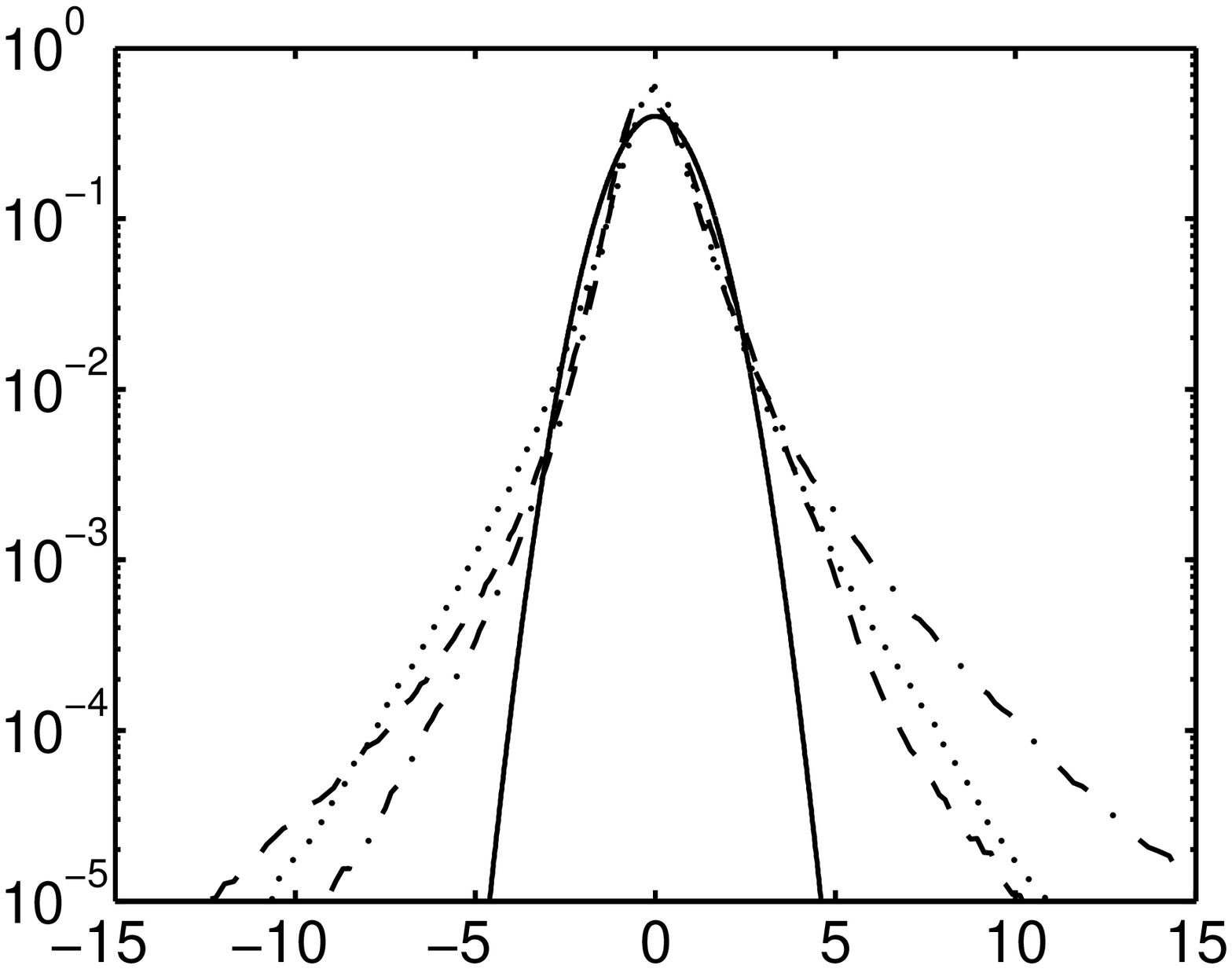}
      \hspace{-.85\linewidth}\raisebox{.65\linewidth}{$(a)$}
      \\ 
      \centerline{$F^\prime / \sigma_F$}
    \end{minipage}
    \begin{minipage}{.45\linewidth}
      \includegraphics[width=1.\linewidth]
       {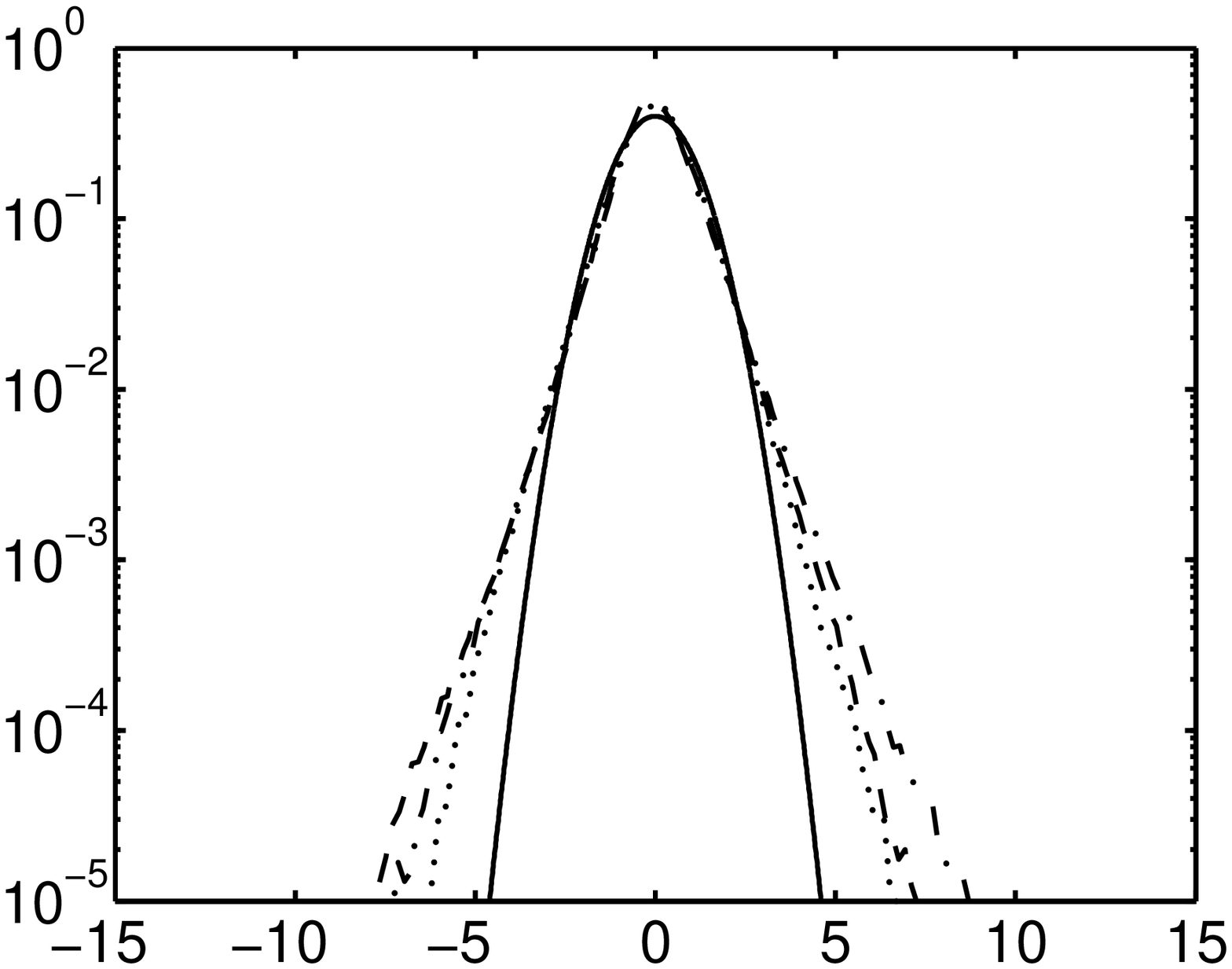}
      \hspace{-.85\linewidth}\raisebox{.65\linewidth}{$(b)$}
      \\ 
      \centerline{$F^\prime / \sigma_F$}
    \end{minipage}
  \end{center}
 
  \caption{Normalised probability density functions of force fluctuations. 
      $(a)$: case F10; $(b)$: case F50; 
      continuous line: Gaussian distribution, 
      dashed line: $F^\prime_x /\sigma_F^x$; 
      dash-dotted line: $F_y^\prime/\sigma_F^y$; 
      dotted line: $F_z^\prime/\sigma_F^z$.
    }
    \label{fig:pdf_force}
\end{figure}

%
%
The results of skewness and kurtosis of the force distributions 
(cf.~table \ref{tab:stat_forces})
are now
discussed jointly with the probability density function (pdf) of the
particle force fluctuations shown in figure \ref{fig:pdf_force}. 
For both cases F10 and F50 the highest skewness is obtained 
for the lift, i.e.\ $S_F^y$.
In other words, large positive lift fluctuations are significantly
more likely to 
occur than large negative lift fluctuations. 
This is clearly visible in figure \ref{fig:pdf_force}($a$),
where lift events of several standard deviations
higher than the mean have a non-negligible probability of occurrence.
In case F50, this effect 
is not as strong as in case F10 (cf.\ figure \ref{fig:pdf_force}$b$),
and accordingly the value of 
the skewness  $S_F^y$ is lower in the former case. 
The small positive skewness of the drag
indicates similarly that instantaneous high drag events are more
likely compared to low drag events. For this component, however, the
effect appears to be much weaker as compared to lift. 
Finally, symmetry arguments again lead to the conclusion that $S_F^z$
should be zero, and this is indeed the case.

The kurtosis of all profiles is rather large indicating a strong
intermittency of the forces, 
i.e.\ the pdfs in figure \ref{fig:pdf_force} exhibit much longer tails
than a Gaussian distribution.
However, as the spheres become larger the values of skewness and
kurtosis approach the Gaussian values of zero and three. 
This trend might be due to the fact that the force
on the particle is an integral quantity, and as mentioned before,
small intermittent events might be averaged out. 
This argument is further elaborated in
\S\ref{ssec:stat_torq} below. 

The present results might be compared to the experimental data provided
by \cite{Mollinger_Nieuwstadt_JFM_1996} for lift fluctuations on a
single sphere with $D^+=2.9$ positioned on top of a smooth wall. 
Although their flow configuration is somewhat different (no sheltering
effect, turbulent boundary layer) they also
report positive values for the skewness ($S_F^y =1.2 $) and high
values of flatness ($K_F^y = 7.0$).  
Furthermore, the pdf of the lift fluctuations in their study 
is of similar shape to the one 
obtained in the present case F10.  
This qualitative agreement suggests that the present results might be 
relevant to a broader range of flow configurations, e.g.\ different
sphere arrangements or packing densities. 

%% file: results_torq_mod_v1.tex
\subsection{Statistics of particle torque}
\label{ssec:stat_torq}
%
The hydrodynamic torque $\mathbf{T}$ acting on a spherical particle
with respect to its centre is defined as follows:
\begin{eqnarray}
  \label{eqn:torque}
  \mathbf{\mathbf{T}} = \int_\Gamma \mathbf{r}_c 
  \times \left(\boldsymbol{\tau} \cdot \mathbf{n} 
  \right) ~\mathrm{d}\Gamma\,,
\end{eqnarray}
where $\mathbf{r}_c=(x_c,y_c,z_c)$ is the distance vector from the
particle centre 
to an element of the surface $\Gamma$. 
It should be noted that -- contrary to the definition of the total particle
force (\ref{eqn:tot-force}) -- the pressure does not enter the
integral (\ref{eqn:torque}), since in the present case the differential
pressure force $- p^{tot}\mathbf{n} ds$ acting on a surface element
$ds$, is always directed towards the particle centre.
Based on the reference force $F_R$ given in
\S\ref{ssec:stat_forces} the reference torque is defined as 
$T_R = F_R r_R$, where $r_R$ is the distance from the particle centre to the 
virtual wall, $r_R =y_0 - D / 2$. The quantity $T_R$ will be used in
the following for the normalisation of the various torque-related
statistical values.  
A sketch that illustrates the definition of the torque on a
particle can be seen in figure \ref{fig:sketch}($a$).

%
%
%
%
\begin{table}
\begin{center}
\begin{tabular}{ lcccccccccccc}
Case & $C_T^x$ & $C_T^y$ & $C_T^z$ & $\ {\sigma_T^x} /T_R$ &
 ${ \sigma_T^y }/T_R$ & $\ {\sigma_T^z} /T_R$  &  $S_{T}^x$ & $S_T^y$ & $S_T^z
 $ & $K_{T}^x$ & $K_{T}^y$ & $K_{T}^z$ \\ 
 F10 & 0.00 & 0.00 & -0.98 &  0.21 &  0.04 & 0.36 &  0.01 & -0.01 &
 -1.04 &  6.46 &  6.17 & 4.72 \\  
 F50 & 0.00 & 0.00 & -0.73 &  0.17 &  0.11 & 0.27 & -0.01 & -0.01 &
 -0.76 &  3.75 &  4.91 & 3.37
\end{tabular}
\caption{
  Statistical moments of torque on particles in case
  F10 and F50. 
  $C_T^{x_i}=\langle T^{x_i}\rangle/T_R$ is the normalised
  mean torque component in the $x_i$-direction, $\sigma_T^{x_i}$ is the
  standard deviation of the 
  torque in $x_i$ direction, $S_T^{x_i}$ and $K_T^{x_i}$ are the
  skewness and kurtosis of the respective torque component.
} \label{tab:stat_torque}
\end{center}
\end{table}

\begin{figure}
\begin{center}
  \begin{minipage}{2ex}
    \rotatebox{90}{$\tilde{z}/D$}
  \end{minipage}
  \begin{minipage}{.46\linewidth}
    \includegraphics[width=1.\linewidth]
    {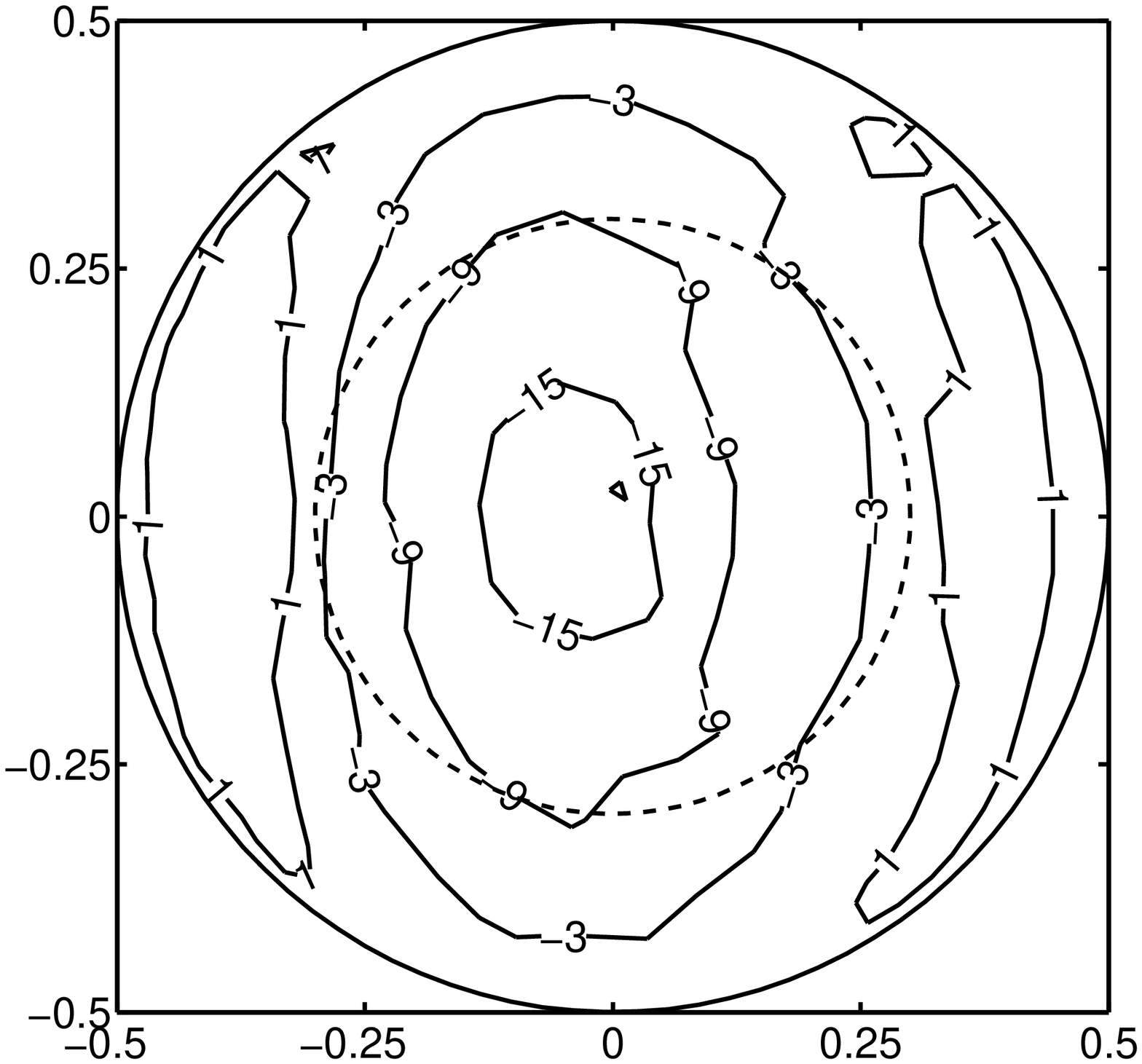}
    \hspace{-0.9\linewidth}\raisebox{0.82\linewidth}{$(a)$}\\
  \end{minipage}
  \begin{minipage}{2ex}
    \rotatebox{90}{$y/D$}
  \end{minipage}
  \begin{minipage}{.45\linewidth}
    \includegraphics[width=1.\linewidth]
    {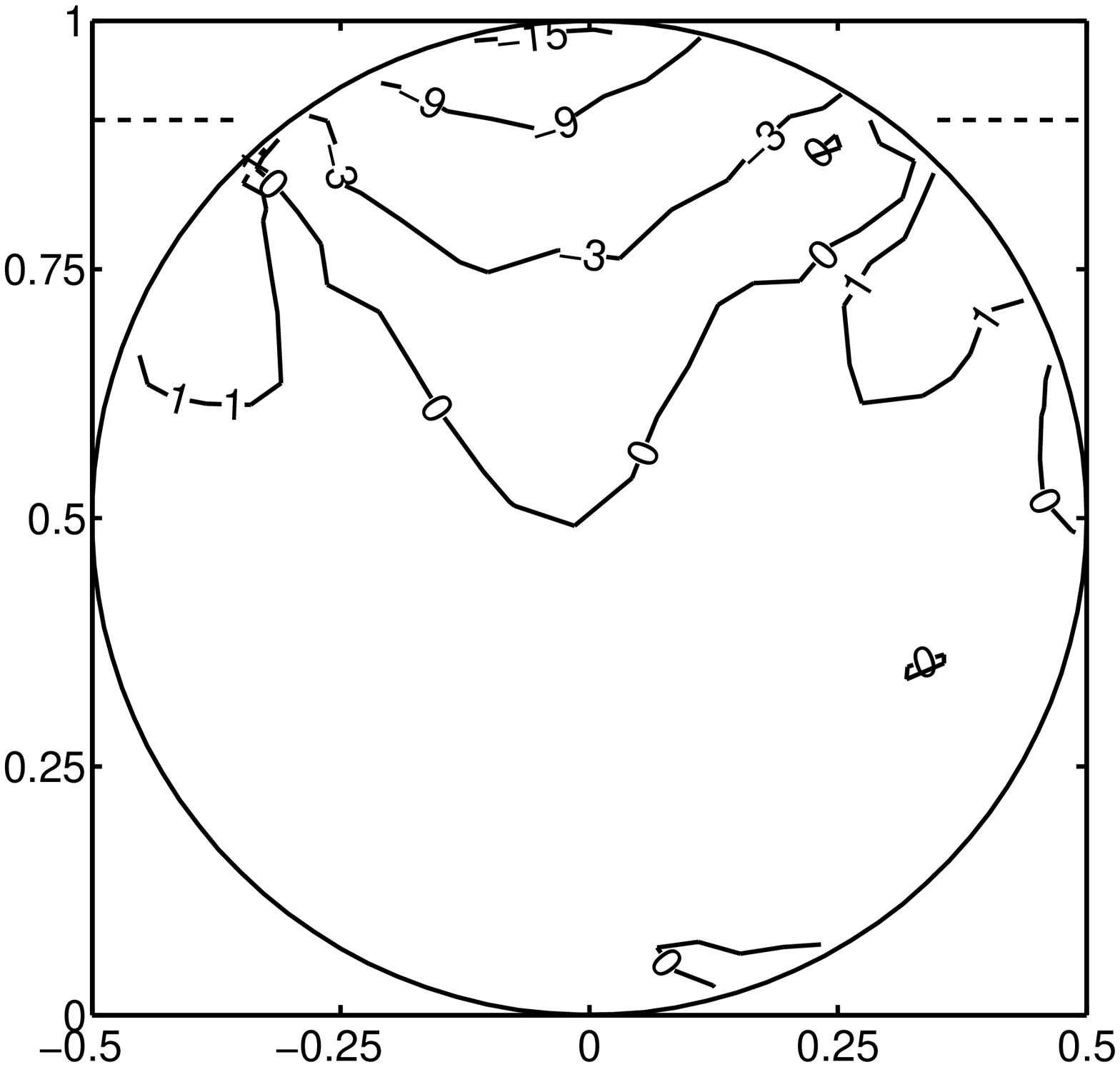}
    \hspace{-0.9\linewidth}\raisebox{0.85\linewidth}{$(b)$}\\
  \end{minipage}

  \begin{minipage}{2ex}
    \rotatebox{90}{\hspace{3ex}$\tilde{z}/D$}
  \end{minipage}
  \begin{minipage}{.46\linewidth}
    \includegraphics[width=1.\linewidth]
    {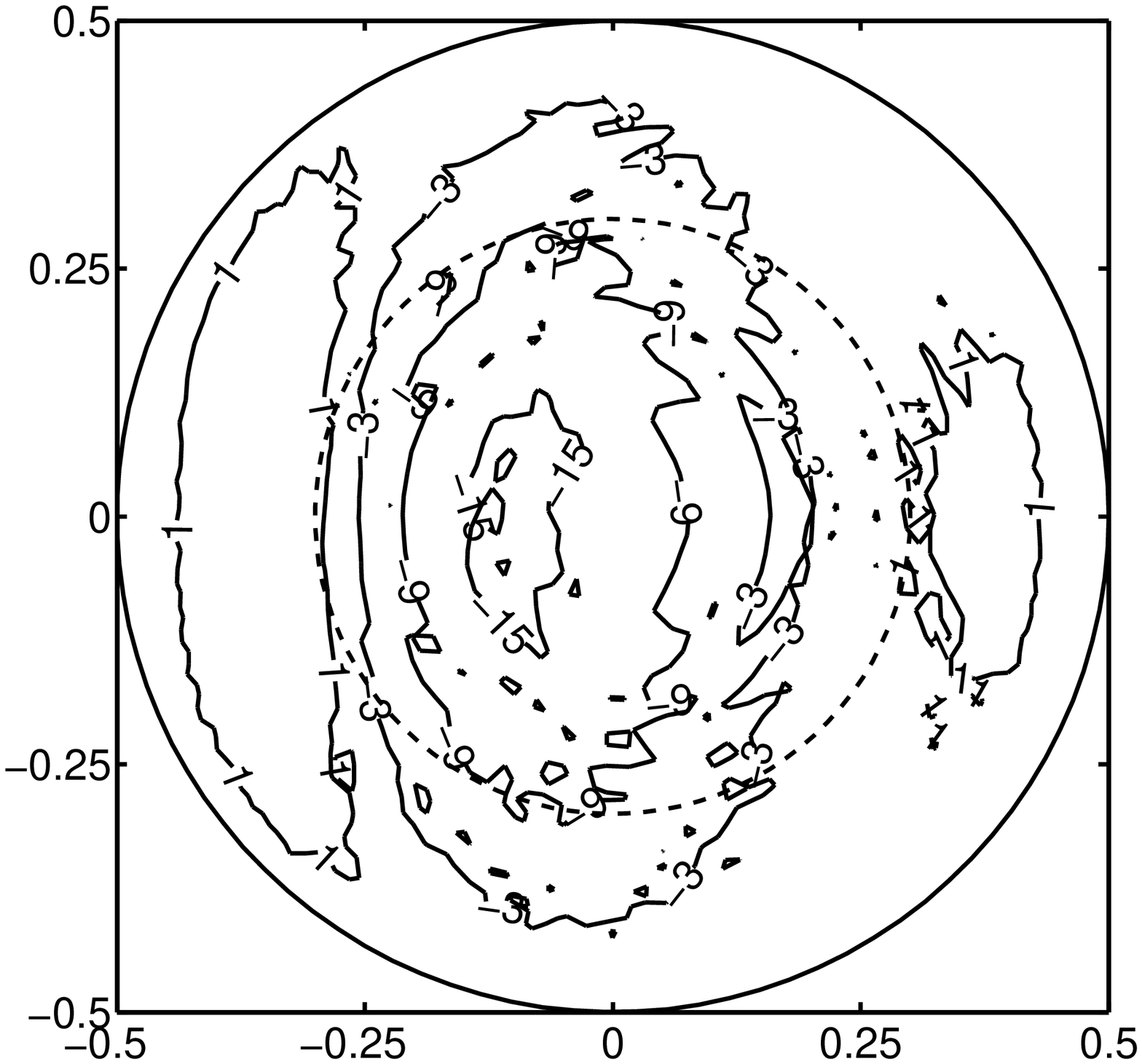}
    \hspace{-0.9\linewidth}\raisebox{0.82\linewidth}{$(c)$}\\
    \centerline{$\tilde{x}/D$}
  \end{minipage}
  \begin{minipage}{2ex}
    \rotatebox{90}{\hspace{3ex}$y/D$}
  \end{minipage}
  \begin{minipage}{.46\linewidth}
    \includegraphics[width=1.\linewidth]
    {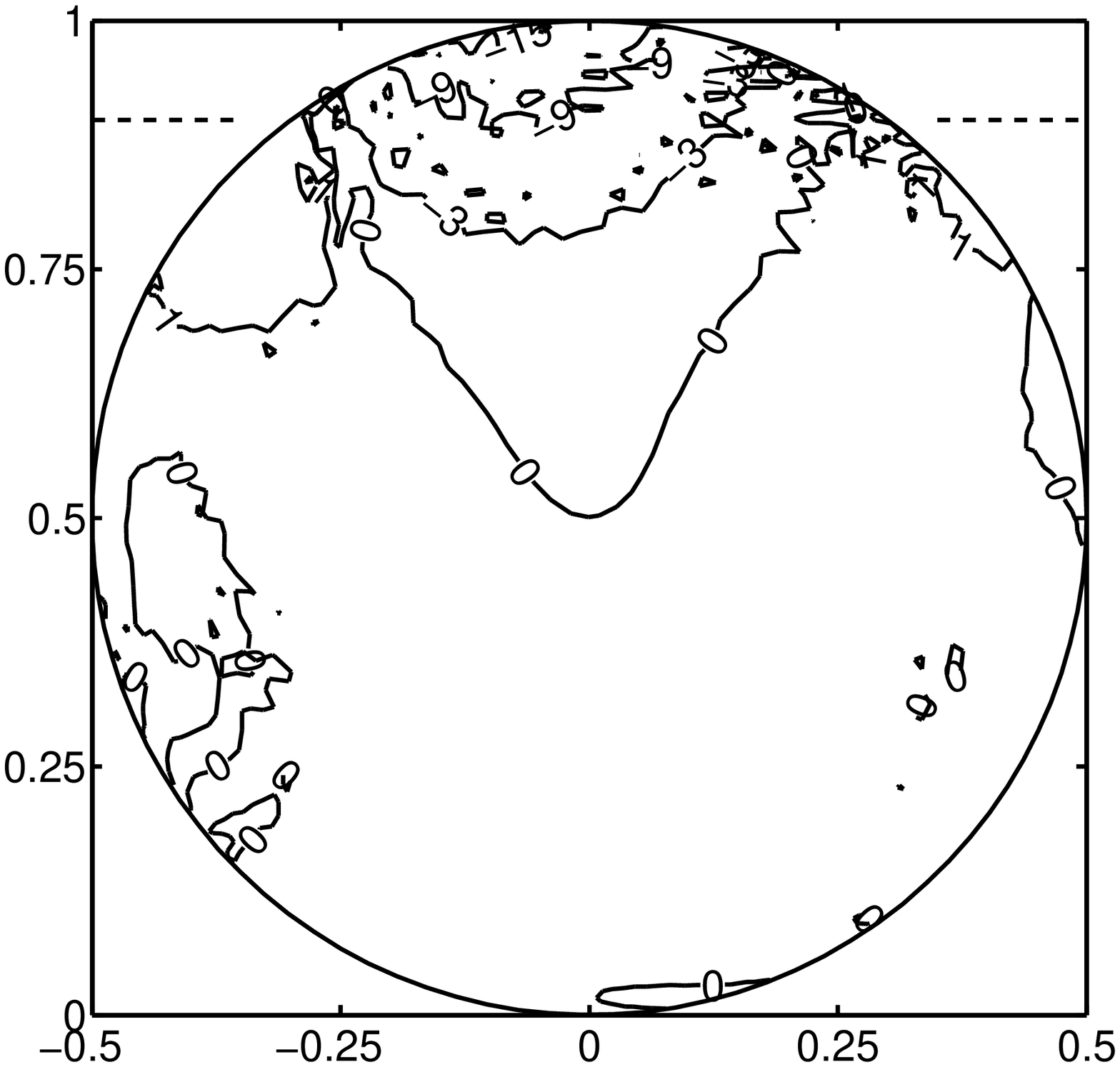}
    \hspace{-0.9\linewidth}\raisebox{0.85\linewidth}{$(d)$}\\ 
    \centerline{$\tilde{x}/D$}
  \end{minipage}
  
\caption{Spatial distribution of $\tau_T$, normalised by $T_R /
  A_{sph}$;
 $\tilde{x}$ and $\tilde{z}$ are the coordinates with respect to the
  particle centre.
  Contour lines are shown at values of [-15, -9, -3, 1] in all
  plots; in $(b)$ and $(d)$ additionally the contour line at zero
  value is shown. 
  The dashed line indicates the location of 
  $y=0.9D$. 
  Panels $(a)$ and $(b)$ show case F10, 
  panels $(c)$ and $(d)$ show case F50.
}
\label{fig:torq_dist_F10-F50}
\end{center}
\end{figure}

%
%
Table \ref{tab:stat_torque} shows the statistical moments of the torque
acting on the particles. Here $C_T^{x_i}$ is the mean torque in the
$x_i$-direction normalised by $T_R$.  
Once more, due to symmetry the only
non-zero component of the mean torque is expected to be $C_T^z$.
The table shows that negative mean values for the
spanwise component are obtained. 
These negative values of $C_T^z$ are expected for the torque on a
particle in positive shear
(cf.~figure \ref{fig:umean_h} and figure \ref{fig:3d-mean}).
The torque coefficient as it is defined above takes values close
to $-1$ for case F10, while it is approximately 25\% lower in
magnitude in case F50.  

In order to analyse these integral results in more detail
figure \ref{fig:torq_dist_F10-F50} shows
the distribution on the sphere's surface
of the stresses leading to spanwise torque, 
\begin{equation}\label{equ:stress-contrib-to-spanwise-torque}
  \tau_T=\tau_L x_c - \tau_D y_c\,.
\end{equation}
The distribution of $\tau_T$ in figure \ref{fig:torq_dist_F10-F50} 
is in both cases similar in shape and values of the contours. 
As for the distribution of $\tau_D$ (cf.~figure \ref{fig:drag_dist_sph}),
a shift towards the particle front can 
be observed for the minimum values of $\tau_T$ near the particle
top. This shift is more pronounced in case F50. 
Negative values of $\tau_T$ occur almost exclusively in the upper part
of the particle.
Thus over most part of the sphere $\tau_T$ is positive, but low in magnitude. 
The cumulative contribution function of $\tau_T$, denoted by
$\mathcal{S}_T$ (cf.\ equation \ref{eqn:cumulative-contrib-to-stress}), 
which is also shown in figure \ref{fig:cum_contr_F10-F50},
reveals that when integrating the contribution of $\tau_T$ in the
lower part of the sphere it 
adds up to approximately $-0.15 C_T^z$ in the vicinity of the
virtual wall.  
In both cases the values of $\tau_T$ are predominantly negative for
wall-distances above $y\approx 0.8D$, such that $S_T$ vanishes around
$y=0.9 D$.
It can therefore be argued that the net spanwise torque $C_T^z$ is
generated in the surface area between a wall distance of $y=0.9D$ and
the particle top (i.e.\ the region highlighted by a dashed line in
figure \ref{fig:torq_dist_F10-F50}). 
%

\begin{figure}
\begin{center}
  \begin{minipage}{2ex}
    \rotatebox{90}{\hspace{3ex}$\sigma_i / \textnormal{norm}$}
  \end{minipage}
  \begin{minipage}{.46\linewidth}
    \includegraphics[width=1.\linewidth,
    clip=true,
    viewport=43 28  546 419]
    {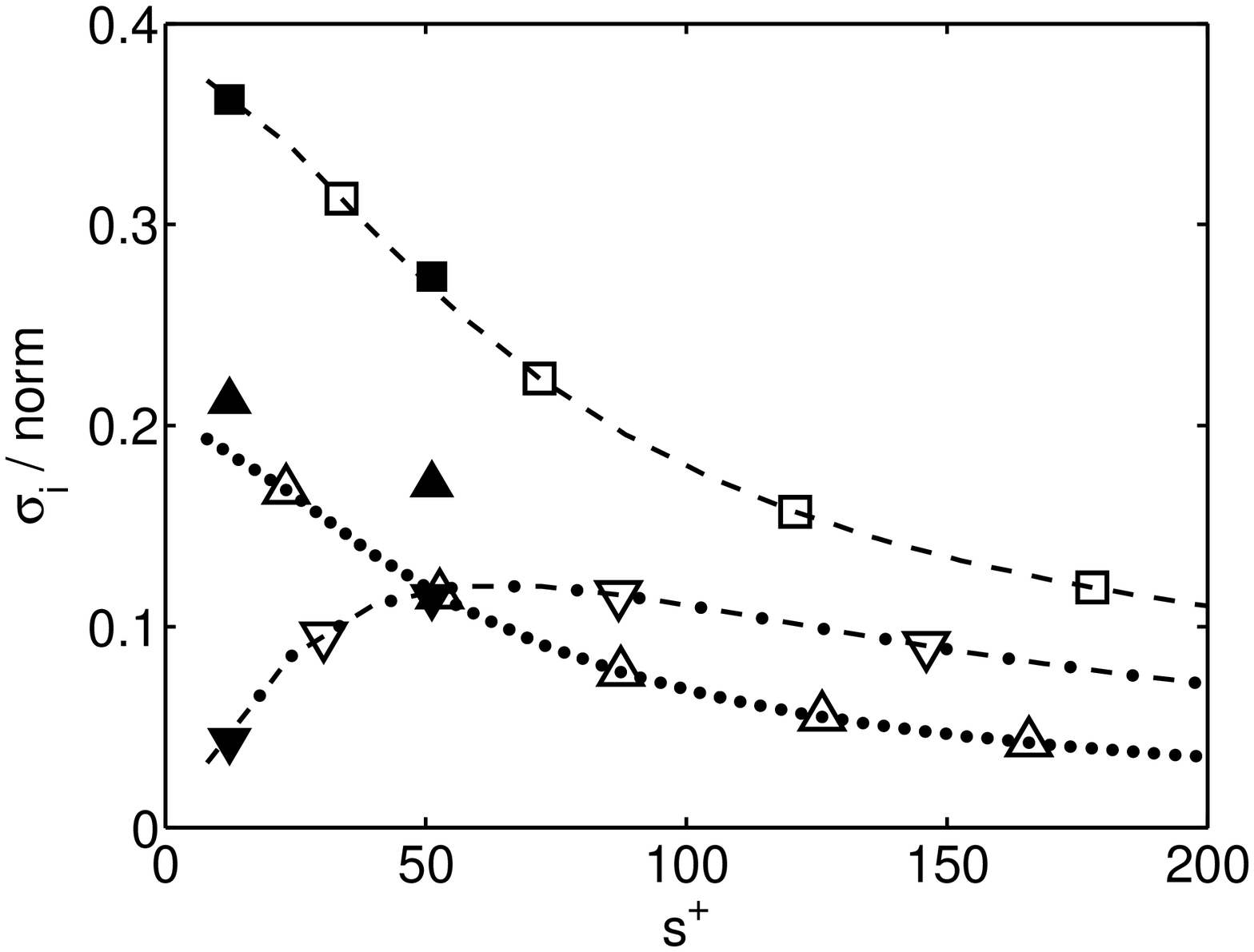}
    \hspace{-0.22\linewidth}\raisebox{0.63\linewidth}{$(a)$}\\
    \centerline{$s^+$}
  \end{minipage}
  \begin{minipage}{2ex}
    \rotatebox{90}{\hspace{3ex}$K$}
  \end{minipage}
  \begin{minipage}{.45\linewidth}
    \includegraphics[width=1.\linewidth,
    clip=true,
    viewport=27 28  519 419]
    {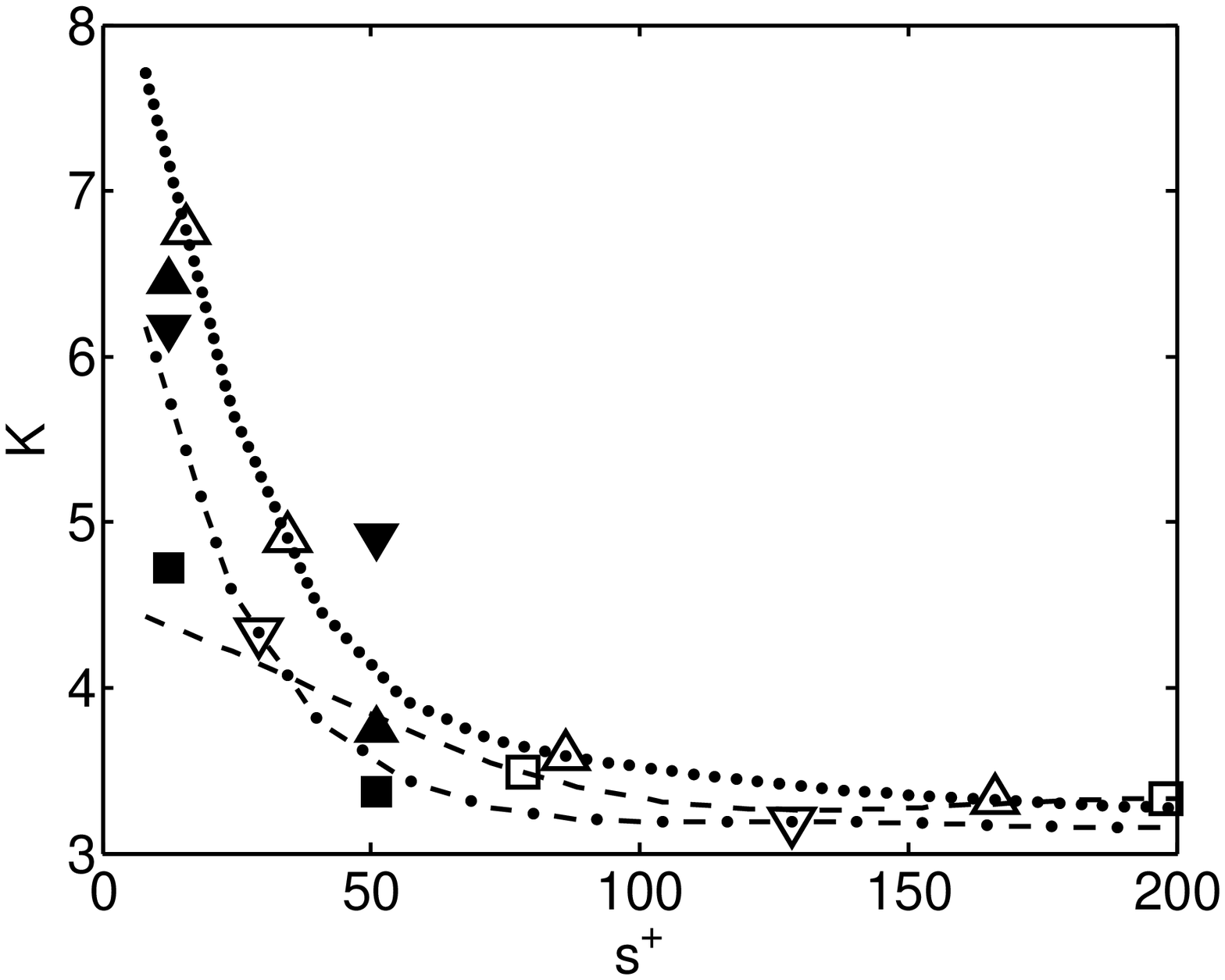}
    \hspace{-0.22\linewidth}\raisebox{0.63\linewidth}{$(b)$}\\
    \centerline{$s^+$}
  \end{minipage}

\caption{
  Statistical moments ($a$ showing the root-mean-square
  value, $b$ the kurtosis) of force and torque obtained from
  the simple model described in the text and of actual values of the
  corresponding torque obtained on the particles in case F10 and
  F50. 
  In the former case (smooth wall) the integration is performed over
  a square wall element of side length $s$, the open symbols and lines
  corresponding to: 
  $ - - \square - -$, $\mathcal{F}_x$;
  $- \cdot - \bigtriangledown - \cdot -$, $\mathcal{T}_y$; 
  $ \cdots \bigtriangleup \cdots$, $\mathcal{F}_z$. 
  The forces are normalised by $\rho u_\tau^2 s^2$, torque is
  normalised by $1/2 \rho u_\tau^2 s^3 $. 
  In the latter case (rough wall) the integration is taken over the
  particle surface (as given in equation \ref{eqn:torque}) with the
  filled 
  symbols corresponding to: 
  $\blacktriangle$, $T_x$; 
  $\blacktriangledown$, $T_y$;
  $\blacksquare$, $T_z$. 
  Note that in cases F10 and F50
  the reference length $s$ is taken
  as $\sqrt{A_R}$. 
  }
\label{fig:smooth_ave}
\end{center}
\end{figure}

%
Before turning to the discussion of the torque fluctuations, first a
simple model is introduced, which allows us 
to elucidate some of the characteristics of
the torque fluctuations by considering the scales of flow motion
that lead to the generation of torque on a particle. 
It should be noted that other authors have previously investigated
the relation between flow structures at different scales and the
forces/torque exerted upon sediment particles \citep[e.g.\
][]{Hofland_diss_05}. 
Here we employ a somewhat different approach which allows us to use
data from a smooth-wall flow.  
In particular, we first analyse drag and torque fluctuations
experienced by a square wall-element in channel flow with a
geometrically smooth wall, systematically varying the linear
dimension of the wall-element. Subsequently, the obtained
statistical results
are related to the corresponding statistics of the components of the
torque acting on a spherical particle in our main simulations. 

Analogously to the definition of force and torque on a particle
the force in the $x$ and $z$ direction on a square surface element
in a smooth-wall channel with
area $A_s = s^2$ can be defined as
\begin{eqnarray} 
\mathcal {F}_x = \int_{-s/2}^{+s/2}\int_{-s/2}^{+s/2}
{\left.\tau_{xy}\right|_{y=0} ~\mathrm{d}x \mathrm{d}z} \,, \quad
\mathcal {F}_z = \int_{-s/2}^{+s/2}\int_{-s/2}^{+s/2}
{\left.\tau_{zy}\right|_{y=0} ~\mathrm{d}x \mathrm{d}z} \,,
\label{eqn:def_area_forc}
\end{eqnarray}
where $s$ is the side length of the element and $\left.\tau_{ij}\right|_{y=0}$
are the components of the stress tensor at the wall.
The torque on the element with respect to its centre can be
defined as
\begin{equation} 
\mathcal {T}_y = 
\int_{-s/2}^{s/2}\int_{-s/2}^{s/2} \left( r^s_x
  \left.\tau_{zy}\right|_{y=0}- r_z^s \left.\tau_{xy}\right|_{y=0} \right)
~\mathrm{d}x \mathrm{d}z \,, 
\label{eqn:def_area_torq}
\end{equation}
where ${\mathbf r}^s$ is the direction vector with respect to the
centre of the area element.
A sketch that illustrates the definition of the force and torque on a
square element in a smooth-wall channel can be seen in figure
\ref{fig:sketch}($b$).

Drag and spanwise force on the smooth-wall element are expected to be
mostly affected by velocity 
scales in streamwise and spanwise direction, respectively, that are of
sizes similar or larger than $s$. 
The effect of velocity fluctuations at length scales much smaller than
$s$ will tend to cancel out due to the integral character of the force
(\ref{eqn:def_area_forc}).  
Thus the highest value of force fluctuation 
should be expected for smallest values of $s$, 
as the contribution of the smaller scales is lost for larger values of
$s$. 
Conversely, due to the cross-product in (\ref{eqn:def_area_torq})
the torque on a smooth-wall surface element
is mostly affected by wall normal vortical motions
of sizes comparable to $s$. 
The effect of much smaller and much larger scales will cancel out or
lead to only small values of torque. 
Thus for small as well as high values of 
$s$ small values of $\mathcal{T}_y$ 
are expected. 
At some intermediate value of $s$, the characteristics of wall normal
vortical motions should be most efficient in generating torque, leading
to maximum values of $\mathcal{T}_y$.
%
Figure \ref{fig:smooth_ave}$(a)$ supports that
hypothesis. It shows the normalised standard deviation of the forces
and torque on the surface element, 
$\sigma_{\mathcal{F}}^x$, $\sigma_{\mathcal{F}}^z$ and
$\sigma_{\mathcal{T}}^y$  normalised by $\rho u_\tau^2 s^2$ and $1/2
\rho u_\tau^2 s^3$, respectively. 
On a square element with $s^+\approx70$ torque appears to be most efficiently
produced. This value is somewhat larger than the average
distance between the low speed and high speed streak close to the wall
which is commonly found to be of order $50 \nu / u_\tau$.
As can be seen in figure \ref{fig:smooth_ave}$(b)$ the kurtosis of the
above quantities monotonically decreases with the size of the surface
element indicating that the intermittency of the small scales is
larger than that of the large scales. 
%

A direct analogy 
between this smooth-wall model and the force and torque on
a particle is not fully justified,
as the flow and the geometry are more
complex in the present case. 
However, some of the characteristics of the particle torque statistics
obtained for the present cases can be explained with the aid of such a
simple model as will be discussed in the following.  
Please note that only one torque component can be defined for a plane
wall element (here ${\cal T}_y$), in addition to the two in-plane forces
considered (${\cal F}_x$ and ${\cal F}_z$). 
A correspondence with the three torque components acting on a
spherical particle is established when considering the plane wall
element as being located at the top (i.e.\ the pole located at $y=D$)
of the particles in 
 case F10 and F50.
The componentwise correspondence is then: 
${\cal F}_x\rightarrow -T_z$, 
${\cal T}_y\rightarrow T_y$,
${\cal F}_z\rightarrow -T_x$
(cf.\ figure~\ref{fig:sketch}).

%
%
%
The normalised standard deviations of the particle torque components shown in
table \ref{tab:stat_torque} are all non-zero as can be expected. 
The amplitudes of the fluctuations of the streamwise and spanwise
torque components are found to be the largest, while the wall-normal
component is significantly weaker.
Compared to the small-sphere case (F10), the streamwise and spanwise
components are smaller in the large sphere case (F50), by 20\%
($\sigma_T^x/T_R$) and 25\% ($\sigma_T^z/T_R$), respectively. 
Contrarily, the wall-normal value $\sigma_T^y/T_R$
is significantly larger in case F50 than in case F10 (nearly by a
factor of three). 

Figure \ref{fig:smooth_ave}, which has already been partially
discussed above, also shows the second and fourth statistical moments
of particle torque fluctuations as a function of particle size. 
As can be seen the standard deviation (figure \ref{fig:smooth_ave}$a$)
of the wall normal torque acting on the particles, $T_y$, matches
rather well the standard deviation of the wall normal torque exerted
on a comparable-size square element in the reference smooth-wall flow,
${\cal T}_y$.  
In addition, the figure shows that the standard deviation of the
spanwise torque, $T_z$, compares very well to the standard deviation
of the drag exerted on a square element in the smooth wall case,
${\cal F}_x$. 
Concerning the streamwise component of particle torque, $T_x$, it is
found that its standard deviation is somewhat larger than the 
standard deviation of the spanwise force fluctuations in the smooth
wall model, ${\cal F}_z$; however, both exhibit a similar decreasing
trend with increasing values of the length scale. 
The overall good agreement between fluctuation intensity of
forces/torque acting on an element of a smooth wall and the
corresponding torque components 
of the particle in case F10 and case F50
is interesting for several reasons. 
First, it suggests that the significant torque fluctuations are
generated in a rather limited region around the particle tops where
apparently to some extent the analogy with the hydrodynamic action
on a wall-parallel square element holds. 
In particular, the present simulations F10 and F50
provide two data points in the hydraulically smooth and transitionally
rough flow regime, which are
fully consistent with the existence of a length scale/particle size of
maximum wall-normal torque generation, as suggested by the
simplified model. 
Secondly, if the above analogy is accepted, then it implies that the
response of the particle torque fluctuations to the near-wall
turbulent flow can indeed be described as a selective filtering
effect, mainly characterised by a single length scale (the particle
diameter). 
%

\begin{figure}
  \begin{center}
    \begin{minipage}{2ex}
      \rotatebox{90}{\hspace{3ex}$\sigma_\phi \cdot pdf$}
    \end{minipage}
    \begin{minipage}{.45\linewidth}
      \includegraphics[width=1.\linewidth]
      {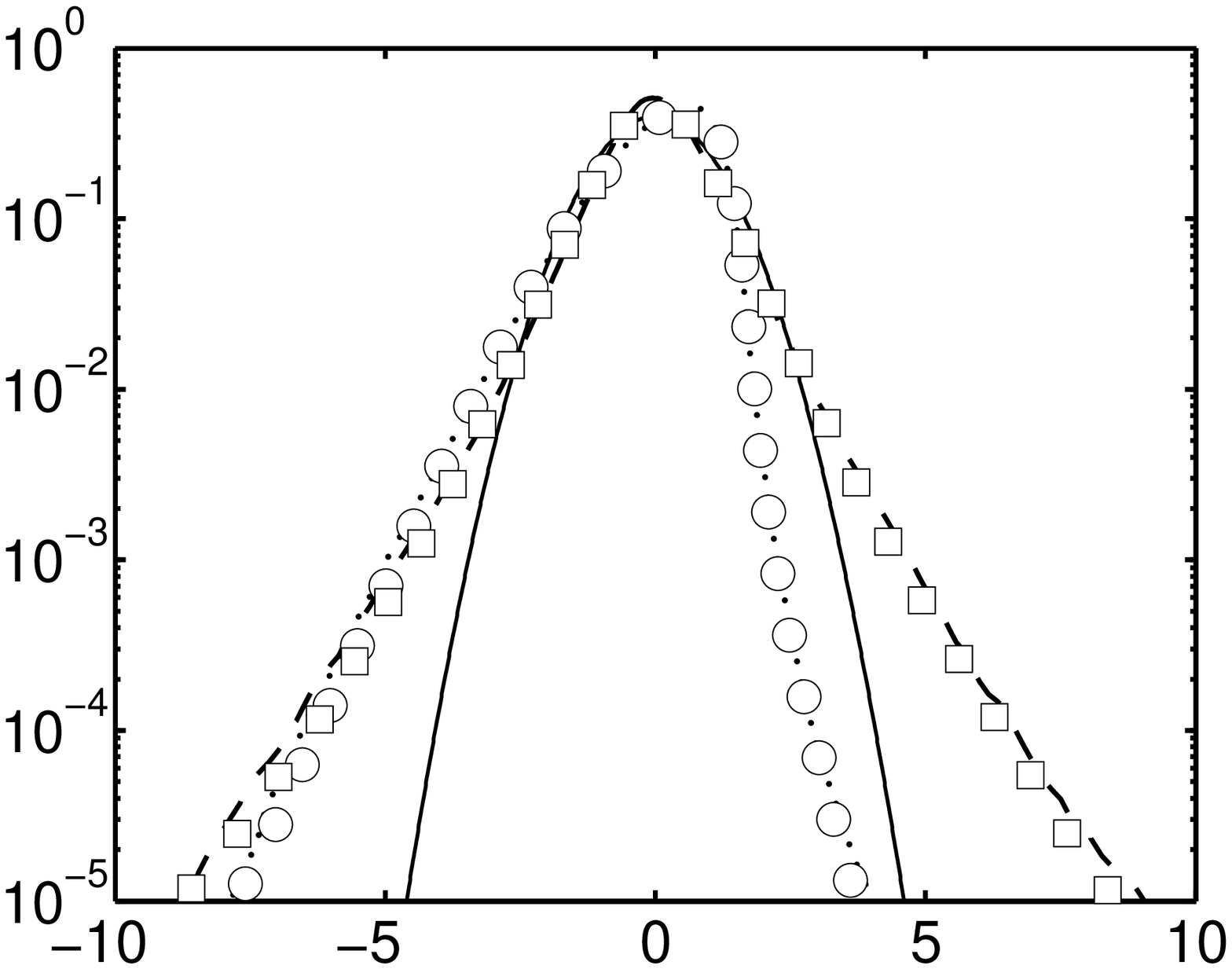}
      \hspace{-.85\linewidth}\raisebox{.65\linewidth}{$(a)$}
      \\ 
    \end{minipage}
    \begin{minipage}{.45\linewidth}
      \includegraphics[width=1.\linewidth]
      {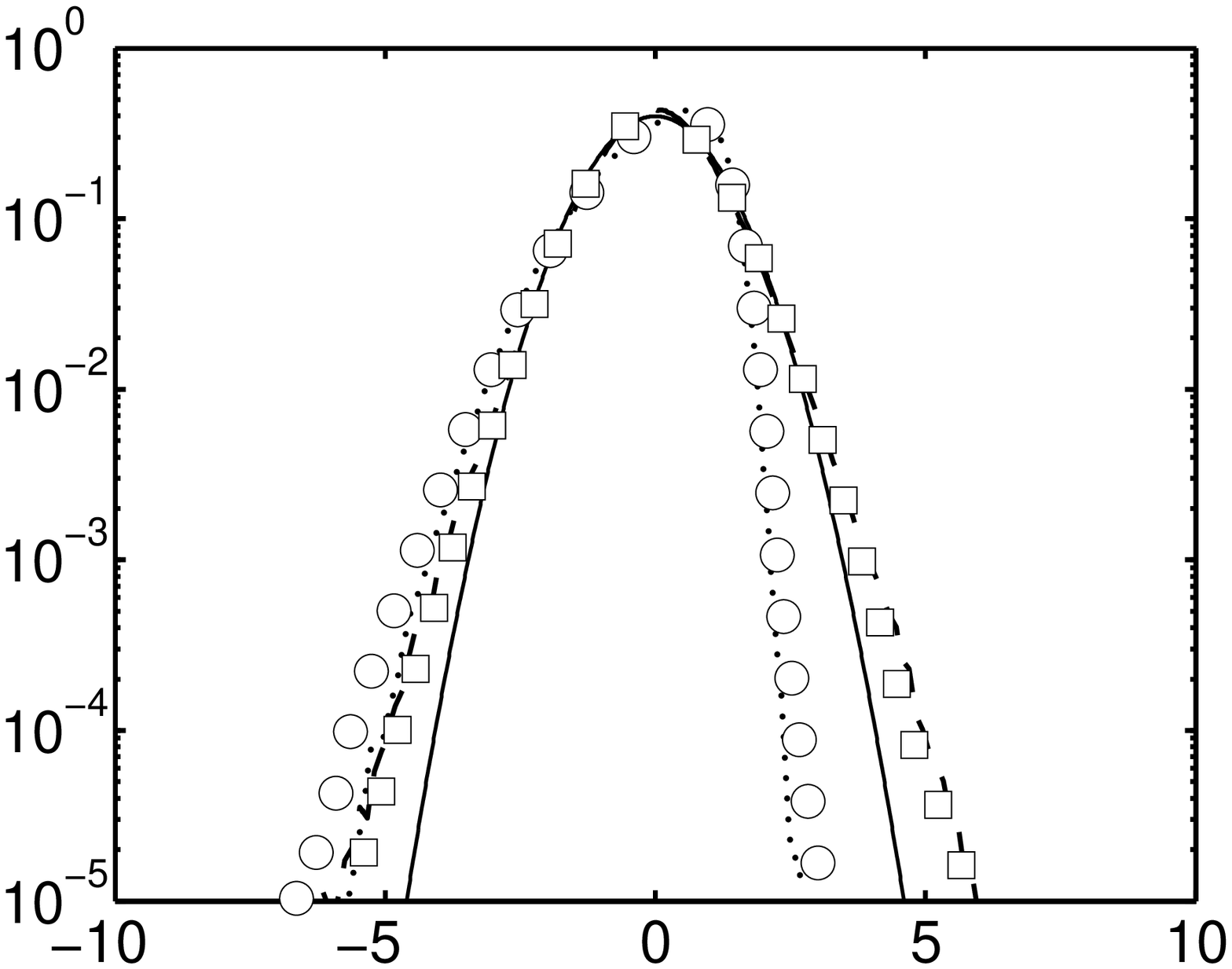}
      \hspace{-.85\linewidth}\raisebox{.65\linewidth}{$(b)$}
      \\ 
    \end{minipage}
    
    \begin{minipage}{2ex}
      \rotatebox{90}{\hspace{3ex}$\sigma_\phi \cdot pdf$}
    \end{minipage}
    \begin{minipage}{.45\linewidth}
      \includegraphics[width=1.\linewidth]
      {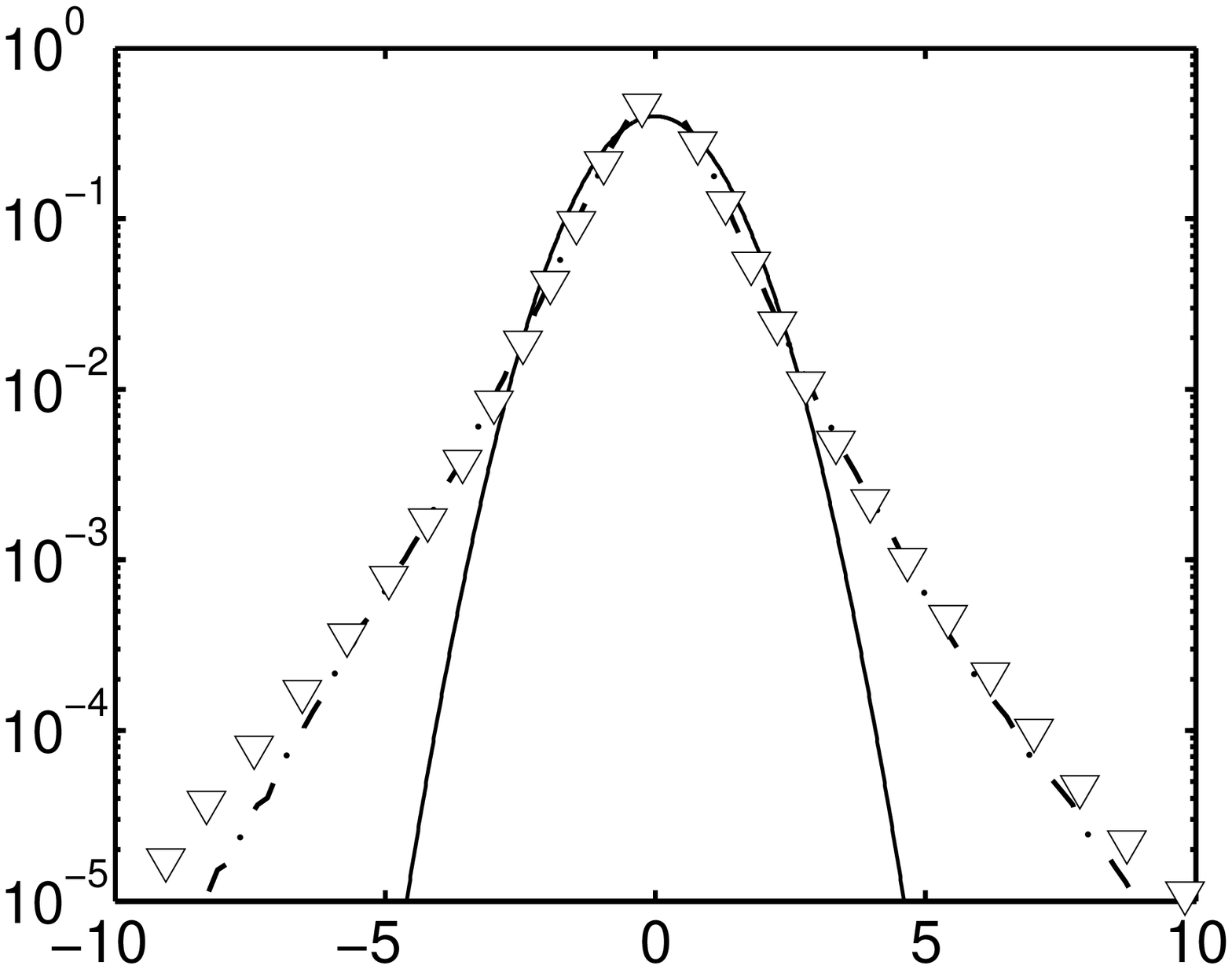}
      \hspace{-.85\linewidth}\raisebox{.65\linewidth}{$(c)$}
      \\ 
      \centerline{$\phi^\prime / \sigma_\phi$}
    \end{minipage}
    \begin{minipage}{.45\linewidth}
      \includegraphics[width=1.\linewidth]
      {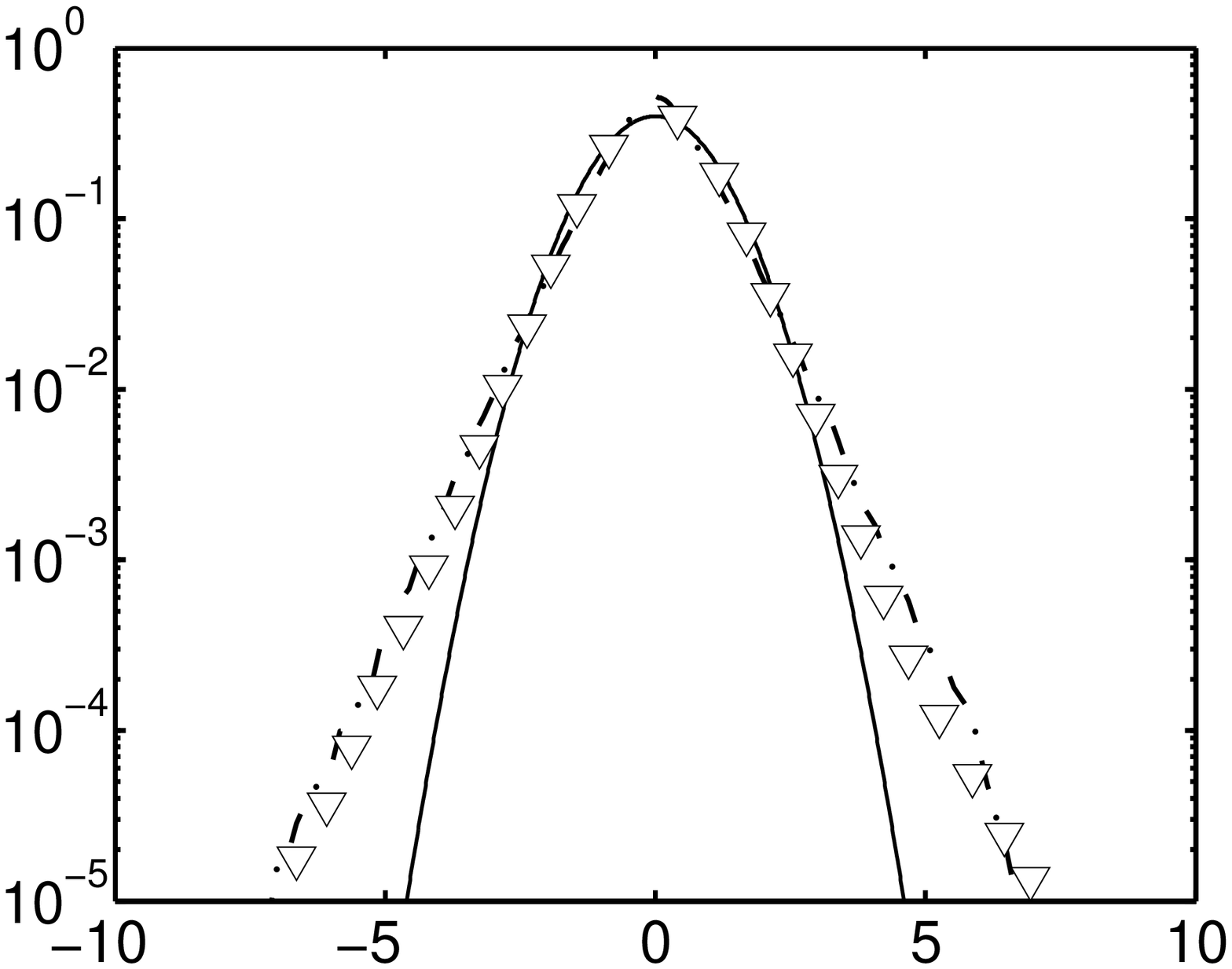}
      \hspace{-.85\linewidth}\raisebox{.65\linewidth}{$(d)$}
      \\ 
      \centerline{$\phi^\prime / \sigma_\phi$}
    \end{minipage}
    
    \caption{Normalised pdfs of the quantities for which statistical
      moments have 
      been shown in figure~\ref{fig:smooth_ave}. 
      The panels $(a)$ and $(c)$ show data from case F10 compared to data
      from the simple model with $s^+=12$; 
      panels $(b)$ and $(d)$ show data from case F50 compared to data
      from the simple model with $s^+=52$. 
      In Panels $(a)$ and $(b)$ the lines and symbols correspond to: 
      dashed line, $T^\prime_x / \sigma_T^x$; 
      dotted line, $T^\prime_z / \sigma_T^z$;
      $\bigcirc$, $-\mathcal{F}^\prime_x / \sigma_{\mathcal{F}}^x$;
      $\square$, $-\mathcal{F}^\prime_z / \sigma_{\mathcal{F}}^z$ (please
      note the negative signs).  
      Panels $(c)$ and $(d)$ show: 
      dash-dotted line, $T^\prime_y / \sigma_T^y$; 
      $\triangledown$, $\mathcal{T}^\prime_y / \sigma_{\mathcal{T}}^y$. 
      The solid line corresponds to a Gaussian distribution. 
    }
    \label{fig:pdf_torque}
  \end{center}
\end{figure}

Normalised pdfs of the particle torque fluctuations are shown in
figure \ref{fig:pdf_torque}. 
It can be seen that the curves for all
three torque components in both cases F10 and F50 approximately match
the curves of the corresponding force/torque components of the
smooth-wall model (evaluated with a side-length $s$ matching the
respective length $\sqrt{A_R}$), thereby further corroborating
the analogy. 
Concerning the shape of the particle torque pdfs themselves, it is
observed that the two symmetric components (streamwise $T_x$ and
wall-normal $T_y$) have significantly longer tails than a Gaussian
function, and consequently exhibit higher than Gaussian values of kurtosis (cf.\
table \ref{tab:stat_torque}). The kurtosis is found to decrease with
increasing particle size, consistent with the above filtering argument
(also cf.~figure \ref{fig:cum_contr_F10-F50}$b$).  

The pdf of the fluctuations of the spanwise component of particle
torque, $T_z$, is clearly asymmetric with a pronounced negative
skewness. Now, it is well established that the pdf of streamwise
velocity fluctuations $u^\prime$ in smooth-wall channel flow is positively
skewed close to the wall \citep{Kim_Moin_Moser_JFM_1987,
  Jimenez_Hoyas_JFM_2008}.  
In the limit of a wall-element with vanishing size, the pdf of
$\mathcal{F}_x$ is directly related to the pdf of the streamwise
velocity fluctuations just above the wall. Since, as found
above, the particle torque component $T_z$ behaves similarly as the
smooth-wall force  $-\mathcal{F}_x$ (note the changed sign), the
observed negative skewness of the former is consistent with the
positively skewed streamwise velocity pdf previously found in
smooth-wall channel flow.  

 It should be noted that the analogy drawn between shear forces
 acting upon a square element of a smooth-wall channel flow and
 the corresponding hydrodynamic torque components of spherical
 particles in the roughness layer can be expected to lose its appeal
 in the fully rough regime. In that case, which is outside the scope
 of the present study, pressure-induced forces will by far outweigh
 viscous forces. Although the torque around the particle center will
 still by definition be devoid of a pressure contribution
 (\ref{eqn:torque}), other quantities of interest to the onset of
 particle motion, such as the torque around the line connecting the
 contact points with the downstream neighbor particles, 
 might be 
 dominated by the contribution from pressure forces.
 Therefore, the main utility of the proposed simple model is 
 presumably limited to the regime of transitional roughness. 
%

%% file: discussion_mod_v1.tex
\subsection{Implications for the onset of sediment erosion}
Sediment erosion is often parametrised in terms of the non-dimensional
Shields number $\theta$ which is defined as follows:
\begin{equation}\label{equ:definition-shields}
  \theta=\frac{\tau_w}{(\rho_p-\rho)\,g\,D}\,,
\end{equation}
where $\tau_w=\rho u_\tau^2$ is the wall shear-stress, $\rho_p$ the
density of the sediment particles and $g$ the value of the
gravitational acceleration \citep{Shields_1936, vanRjin_93,
  Garcia_ASCE_manual_08}. 
If we suppose that erosion is initiated by lift forces alone, then we
can characterise the onset of erosion by a balance between
hydrodynamic lift force, $F_y$, and buoyant
weight of the particle, $F_B=(\rho_p-\rho)gD^3\pi/6$, yielding
the following expression for the critical Shields parameter:
\begin{equation}\label{equ:critical-shields}
  \theta_c=\frac{2}{3c_L}\,,
\end{equation}
where the lift coefficient is defined as $c_L= F_y / (\rho u_\tau^2 D^2\pi/4)$.
Please note, that in the above definition of $c_L$ a
slightly different normalisation than in \S\ref{ssec:stat_forces} is chosen. 
It can be seen that for this erosion scenario the critical value of
the Shields number is inversely proportional to the lift coefficient
at the onset of erosion. 

\begin{figure}
  \begin{center}
    \begin{minipage}{2ex}
      \rotatebox{90}{$pdf$}
    \end{minipage}
    \begin{minipage}{.5\linewidth}
      \includegraphics*[width=\linewidth]
      {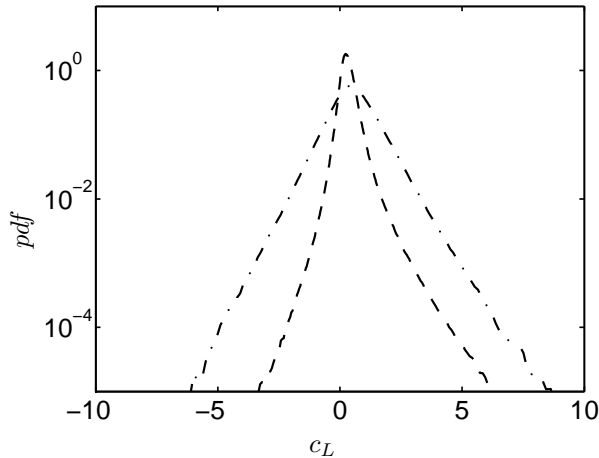}
      \\
      \centerline{$c_L$}
    \end{minipage}
  \end{center}
  \caption{Non-normalized pdf of the lift coefficient defined as
    $c_L=F_y/(\rho u_\tau^2 D^2\pi/4)$, dashed line: case F10,
    dash-dotted line: case F50. 
  } 
  \label{fig:pdf-lift-coeff}
\end{figure}

Figure~\ref{fig:pdf-lift-coeff} shows the pdf of the lift coefficient
$c_L$ for the two present cases. In order to determine the smallest
value of the Shields parameter for which sediment erosion can be
initiated, the largest occurring value of $c_L$ needs to be considered
in each case. Since the pdfs exhibit exponential tails, it is
difficult to determine a precise upper bound of $c_L$. However, when a
given (small) minimum probability of observation is fixed, it is clear
that the larger spheres (case F50) will yield a larger maximum value
of $c_L$ than the smaller spheres (F10). 
Consequently, a smaller critical Shields number is obtained for 
the spheres in case F50. 
Other modes of erosion (sliding parallel to the contact point with the
downstream neighbour particles; rotation around the contact point) have
also been analysed with similar conclusions 
(graphs of pdfs omitted).
%

Although exhibiting considerable scatter, experiments and field
observations seem to indicate an increase with $D^+$ of the critical
Shields number $\theta_c$ over the current range of particle diameters 
\citep{vanRjin_93, Garcia_ASCE_manual_08}.
One has to be cautious, however, since results obtained in our
idealised flow configuration are compared to experimental
observations in flows involving a wide range of different irregular
particle arrangements as well as varying particle shapes and size
distributions. 
With this caveat in mind, we can take 
the different trend found in the present simulations ($\theta_c$
decreases from $D^+\!=\!10$ to $D^+\!=\!50$) as compared to experiments
($\theta_c$ increases from $D^+\!\approx\!10$ to $D^+\!\approx\!100$)
as an indication that extreme force- and torque-generating events recorded in
fixed-particle configurations might not immediately yield a criterion
which is sufficient to judge whether erosion will indeed occur as
predicted when particles are freely mobile (under otherwise identical
conditions).  
In particular, the question of the influence of the interaction
between the incipient particle motion and the surrounding flow field
as well as the related question of the necessary duration of
force-/torque-generating flow events cannot be answered with certainty
based upon fixed-particle data alone. 
In order to answer these queries, additional studies involving an
analysis of high-fidelity data of the actual process of sediment
erosion need to be carried out. 

%% file: conclusions_mod_v1.tex
\section{Summary and conclusions}
\label{sec:conclusion}
Direct numerical simulation of open channel flow over a 
geometrically
rough wall has been performed at a bulk Reynolds number of $Re_b
\approx 2900$. 
The wall consisted of a layer of spheres in a square 
arrangement 
touching a solid wall. Two particle diameters
were considered: case F10 with $D^+= 10.7$ ($Re_\tau=188$), and case
F50 with $D^+=49.3$ ($Re_\tau=235$). 
In case F10 the effect of roughness on the 
flow field statistics was small, and the limit of the
hydraulically smooth flow
regime is approached; in case F50 the roughness
effect was stronger, and the flow is 
in the transitionally rough flow regime.

The complexity of the time-averaged three-dimensional flow field
within the roughness layer was discussed in detail.
In both cases a recirculation forms downstream of the spheres that
is connected over the entire spanwise direction, 
being more pronounced in the large sphere case. 
Three-dimensionality above the roughness layer
is lost rapidly with wall-distance, yielding a 
time averaged flow field which is essentially
one-dimensional beyond a distance of two particle diameters.
 The main result of this paper is
the characterisation of the
force and torque acting on a particle due to the turbulent flow.
It was found that in the present cases the mean drag 
on a sphere is 4\% (case F10) and 15\% (case F50) higher than the
reference force $F_R=\rho u_\tau^2 A_R$,  
where $A_R$ is the wall-normal projected area of the 
wall per particle. 
Given our definition of the friction velocity, these numbers reflect
the fact that the drag force on the roughened bottom wall (below the
fixed spheres)
is small.  
In both cases a strong positive lift was obtained in agreement with
previous experiments, exceeding values of 18\% and 32\%, respectively,
of the corresponding drag.
The values of the mean spanwise torque on a particle are comparable to
$-F_R r_R$, where $r_R$ is the distance from the particle centre to
the position of the virtual wall, located at 
a distance of $y_0=0.8D$ from the plane part of the solid wall.
It was shown that in both cases the mean drag, lift and spanwise
torque are to a large extent produced 
in a region of the particle surface which is located above the virtual
wall ($y>y_0$).  
The spatial distribution over the particle surface of the stresses that lead
to time-averaged drag, lift and spanwise torque 
are found to be similar in shape in the two cases.
%

We have observed that the intensity of particle force fluctuations
(when normalised by $F_R$) is significantly larger in the large-sphere 
case. 
Conversely, when analysing the torque it is found that only the
fluctuation intensity of the spanwise component is larger in the
large-sphere case, whereas the two remaining components exhibit
smaller fluctuation intensities when the sphere is larger. 
By means of a simplified model we were able to show that 
the torque fluctuations might be explained by 
the spheres acting as a filter
with respect to the size of the flow scales which can effectively
generate torque fluctuations. 
As a model we have considered the shear-forces and torque exerted by the
flow on a square wall-element in a smooth-wall
configuration. By systematically varying the linear dimension of the
wall-element we are able to analyse the influence of the length
scale. 
Here we find that the normalised fluctuation intensity of the
streamwise and 
spanwise shear-forces monotonously decreases with the filter size,
while the wall-normal torque experiences a maximum of 
normalised fluctuation
intensity for intermediate filter sizes of approximately 70 wall
units.  
By assuming that the largest part of the particle torque fluctuations
is generated in a small area around the particle tops, the results
from the simplified model carry over to the corresponding components
of the particle torque. 
We obtain indeed a reasonable agreement between standard deviation and
kurtosis of shear-forces and torque acting on a square wall-element on the one
hand and respective particle torque components on the other hand. 
However, since we have only considered two particle sizes, we cannot
state with certainty that wall-normal particle torque fluctuations are
indeed most intense at the above mentioned scale of 70 wall
units. 
Similarly, based on the current data it is not possible to judge
whether the model is capable of providing insight in the fully rough
flow regime. These points should be clarified in future studies.

Fluctuations of both force and torque were found to exhibit strongly
non-Gaussian pdfs with particularly long tails. The deviation from
a Gaussian distribution
was significantly smaller in the large-sphere case, which
was attributed to the smaller effect that highly intermittent small
scales have on the larger particle surface area. 
Moreover, it was observed that the spanwise torque component has a
marked negative skewness. In the light of the analogy with a
wall-element in a smooth-wall configuration, this finding is
consistent with the positive skewness of the streamwise velocity
fluctuations near the wall in a smooth wall channel flow. 
%

Concerning the potential for sediment erosion, it was concluded from
the present data that the largest recorded values of hydrodynamic lift
lead to critical Shields numbers which are smaller in the large-sphere
case as compared to the small-sphere case. The same trend was found
when considering the forces projected onto the direction tangential to
the downstream contact point between spheres in neighbouring positions
as well as when evaluating the balance of angular moments around the
contact point. 
Measurements in experiments with truly mobile particles seem to
indicate the opposite trend (increasing critical Shields number with
increasing particle size in the range of $D^+\approx10$ to $100$).
However, these opposite trends do not necessarily imply a
contradiction, since additional effects which might play a role in the
dynamical process of erosion have not been addressed in the present
study. 
Two important idealisations with respect to real-world sediment
erosion have been made in the present work: 
the regularity of the geometrical arrangement
and the immobility of the particles.
Concerning the geometry, it is expected that different
particle arrangements, size distributions and shapes will lead to a
modification of the forces acting upon sediment particles. In
particular, varying the protrusion of individual particles
has been shown to have a significant effect
on the onset of erosion
\citep{Fenton_Abbott_PRSCA_1977, Cameron_phd_2006}.
In this respect, the configuration studied in the present work can
be considered as a
case where mutual sheltering of
particles is high
due to their uniform diameter, spherical
shape and regular arrangement.

Concerning the immobility of the particles, we see several
consequences arising from this idealisation which could potentially
affect the implications for sediment erosion:  
(i) the modification of the flow field by the particles during the
incipient motion; 
(ii) the determination of the temporal duration of force- and
torque-generating flow events which is necessary in order to achieve
irreversible onset of particle motion;
(iii) the influence of collective mobility. 
In order to evaluate the importance of these mobility effects, 
additional data from configurations with truly eroding particles needs
to be analysed.  
It might then still be possible to devise a refined erosion
criterion which allows for the determination of erosion
probabilities based upon data from fixed-particle configurations.  

Both of these additional aspects (the bed geometry and the particle
mobility) should be addressed in future studies. 

\vspace*{1ex}
This work was supported by the German Research Foundation (DFG) under
project JI 18/19-1.
The computations have been carried out at the Steinbuch Centre for
Computing (SCC) of Karlsruhe Institute of Technology and at the
Leibniz Supercomputing Centre (LRZ) of the Bavarian Academy of
Sciences and Humanities. The support from these institutions is
gratefully acknowledged. 
We also thank the anonymous referees for useful suggestions
leading to the improvement of the manuscript.

%% file: appendix_paper_mod_v1.tex
\begin{appendix}
\label{sec:appendix_pap}
\section{The position of the virtual wall and the friction velocity} 
\label{ssec:appendix_utau_y0}
Several common methods exist for the definition of an origin $y_0$ of
the wall-normal coordinate when analysing turbulent flow statistics
over rough walls. 

A priori definitions can be based on geometrical considerations.
Examples 
are the volume of the roughness
elements divided by the area of the virtual 
wall \citep[cf.][]{Schlichting_IngArch_1936}, which for the present
geometry leads to $y_0/D=0.44$ (0.56) in 
case F10 (F50), 
or the average of the maximum surface elevation, which leads to 
$y_0/D=$ 0.54 (0.65) in case F10 (F50). 
A posteriori methods employ the data from measurements or simulations
to define $y_0$. 
\citet{thom_JRMS_1971} and \citet{jackson_JFM_1981} 
propose to define $y_0$ by the wall-normal position of the centroid of the drag
profile on the roughness elements. 
In the present study such a definition would lead to 
values of $y_0/D=0.88$ (0.84) in case F10 (F50).
It should be noted that in case of a porous sediment layer,
this definition is biased by the inter-porous flow. 
Most researchers, however, use methods which 
involve the adjustment of a logarithmic law to the mean velocity profile
\citep{raupach_antonia_rajagopalan_AMR_1991}, 
especially for high Reynolds number flows.
Based on these methods, several studies on turbulent flow over spherical
roughness (for various Reynolds numbers, particle arrangements and
flow geometries) can be found that provide the value of $y_0$ for a given
particle diameter (cf.\ reviews in
\citealp{Bayazit_EUROMECH_1983}; 
\citealp{Nezu_Nakagawa_93}, p.\ 26;
\citealp{Dittrich_habil_1998}, p.\ 29;
\citealp{Detert_Nikora_Jirka_JFM_2010}), including also
studies that match well with the present flow conditions
\citep{Nakagawa_Nezu_JFM_1977,
  Grass_Stuart_Mansour-Tehrani_PRSA_1991, Cameron_phd_2006,
  Singh_sandman_williams_JHR_07}.
In these studies the virtual wall is positioned 
at $y_0/D$ in the interval 0.61 to 0.82. 
In some studies the virtual wall is defined at the location of the
roughness crests (i.e.\ $y_0/D=1$) such as in
\citet{manes_pokrajac_mcewan_JHE_2007}. 

In the present work we have selected to fix the position of the
virtual wall at a given level $y_0/D=0.8$ inside the range of values
determined in relevant experiments. 

Turning now to the definition of the velocity scale
$u_\tau$, we will discuss three common approaches in the following. 
Again, a widely used method is to obtain $u_\tau$ by adjusting
a logarithmic law to the mean velocity profile. 
Assuming the values $\kappa=0.40$ and $y_0/D=0.80$ for the K\'arm\'an
constant and the offset of the virtual wall, respectively, a fit
over the range $50\delta_\nu\leq (y-y_0)\leq 0.5h$ yields
$u_\tau/U_{bh}=0.062$ (0.081) in case F10 (F50). 
However, it should be recalled that in the present low-Reynolds
number flow the limited extent of the logarithmic region makes this
approach relatively error-prone. 

Alternatively, the global momentum balance can be used in order to
relate the driving force (either due to a pressure gradient or
gravity) to the different contributions to the drag force generated
at the fluid-solid interfaces.
While the mean momentum balance is uniquely defined, it does not
immediately provide a velocity scale. 
In some studies the velocity scale is defined from the
volumetric force integrated from the virtual wall-distance to the
free surface \citep[for
example][]{Nakagawa_Nezu_JFM_1977,Detert_Nikora_Jirka_JFM_2010}, 
i.e.\ in our notation $u_\tau^2=-\langle
\mathrm{d}p_l/\mathrm{d}x\rangle h/\rho$.
This definition leads to 
$u_\tau/U_{bh}=0.066$ (0.081) in
case F10 (F50). 
Other authors choose to integrate the driving force exclusively over
the volume occupied by the fluid 
\citep{Manes_Pokrajac_Nikora_Ridolfi_Poggi_GRL_2011}, leading to 
$u_\tau^2\,A_w=-\langle
\mathrm{d} p_l /\mathrm{d}x\rangle V_f/\rho$ 
(where $A_w$ is the area of the
wall-parallel cross-section of the considered control volume and
$V_f$ the corresponding volume occupied by the fluid).  
This latter definition yields $u_\tau/U_{bh}=0.066$ (0.081) in case
F10 (F50).  

Finally, let us consider definitions based on the total shear stress
profile. In smooth-wall flow, the total shear stress $\tau_{tot}$ is
linear with wall-distance and the appropriate velocity scale is
given by $u_\tau^2=\tau_{tot}(0)/\rho$. 
In rough-wall flow, $\tau_{tot}$ in general deviates from a linear
relation below the roughness crests which prevents the use of a
similar definition, e.g.\ based upon $\tau_{tot}(y=y_0)$. 
Instead, some researchers propose to determine the velocity scale
independently of the position of the virtual wall 
by using the total shear stress at the roughness crests, i.e.\
$u_\tau^2=\tau_{tot}(y=D)/\rho$ \citep{Pokragac_etal_RF_2006}. 
This definition leads to values of 
$u_\tau/U_{bh}=0.066$ (0.080) in case F10 (F50). 
Note that this latter definition makes a direct comparison of
different data sets difficult, since the total shear stress
profiles $\tau_{tot}/(\rho u_\tau^2)$ represented as a function of
$(y-y_0)/h$ will in general not collapse. 
Alternatively, $u_\tau$ can be computed from the total shear 
stress extrapolated from the region where it varies linearly (i.e.\ 
above the roughness crests) down to the position of the virtual 
wall, yielding the defining relation
\begin{equation}\label{equ-app-a1-def-utau}
  \tau_{tot}=\rho u_\tau^2\left(1-\frac{y-y_0}{h}\right)
  \,,
\end{equation}
valid for $y>D$. 
This definition leads to $u_\tau/U_{bh}=0.066$ (0.082) in case F10 (F50). 
Incidentally it can be deduced from the global momentum balance that
our definition implies $u_\tau^2=-\langle \mathrm{d}p_l /
\mathrm{d}x\rangle h/\rho$, 
i.e.\ it turns out that the definition of $u_\tau$ through
(\ref{equ-app-a1-def-utau}) is equivalent to the above mentioned
definition used by \cite{Nakagawa_Nezu_JFM_1977} and
\cite{Detert_Nikora_Jirka_JFM_2010} based 
upon an integral of the driving force.   
%

\section{Details on averaging procedures}\label{ssec:appendix_ave1}
In the present paper two definitions
have been used for averaging a discrete flow field, $\phi(i,j,k)$, in
$x$-$z$ planes:  
\begin{eqnarray}
\langle \phi \rangle^A_{xz} (j) &=& 
\frac{1}{N_x ~ N_z}\sum_{i=1}^{N_x} \sum_{k=1}^{N_z} 
  \phi(i,j,k)
  \,, \label{eqn:op_ave1}\\
\langle \phi \rangle^B_{xz} (j) &=& 
\frac{1}{N_m(j)}\sum_{i=1}^{N_x} \sum_{k=1}^{N_z} 
\phi(i,j,k)\,
m(i,j,k)
\,,
\label{eqn:op_ave2}
\end{eqnarray}
where $N_x$ and $N_z$ are the number of grid points in $x$ and $z$ directions,
respectively, and $m$ is a field that works as a mask for computing
the averages. 
If a given point lies within the fluid domain, 
at that point $m=1$, otherwise $m=0$.
$N_m(j)$ is the sum of $m$ over a wall parallel plane at wall-distance
$y_j$, i.e.~$N_m(j) =
\sum_{i=1}^{N_x} 
\sum_{k=1}^{N_z} 
m(i,j,k)
$.  
$N_m(j)$ equals $N_x N_z$ above the roughness layer such that both
expressions (\ref{eqn:op_ave2}) and (\ref{eqn:op_ave1}) are
equal to each other away from the roughness elements. 
Within the roughness layer $N_m(j)$ and thus the number of samples for
each wall parallel plane decreases. 
In the context of the
double-averaging methodology these two quantities are generally
referred to as
superficial and intrinsic spatial average, respectively
\citep[cf.][]{Nikora_etal_JHE_2007}.

Note that the zero velocity condition is forced only at
the surface of the 
particles due to reasons of efficiency \citep{Uhlmann_JCP_2005}.
This leads to fictitious non-zero velocities at the grid points that
lie within the 
particles. \cite{Fadlun_etal_JCP_2000} demonstrated that the external
flow is essentially unchanged by this procedure which has been 
confirmed later by \cite{Uhlmann_JCP_2005}.
Since the internal fictitious flow
affects the value of $\langle \phi \rangle^A_{xz} (j)$ (according to
\ref{eqn:op_ave1}) in the roughness layer, we present averages
computed according to (\ref{eqn:op_ave2}) where the flow within
the roughness layer is discussed (i.e.\
figure~\ref{fig:velrms_norm_D} and figure~\ref{fig:3d-effect}).  
When focusing upon the flow above the roughness layer (i.e.\ in
figure~\ref{fig:umean_h} and figure~\ref{fig:velrms_norm}), we
choose to present data computed according to (\ref{eqn:op_ave1}),
because the number of available samples is larger, as explained in
\S~\ref{ssec:appendix_ave2}. 
%

\section{Consistency of runtime and a posteriori statistics}
\label{ssec:appendix_ave2} 
For the flow field statistics presented in this paper, two different
sets of data have been used. 
The first set of flow field statistics was collected
during the runtime of the simulation employing
equation (\ref{eqn:op_ave1}). 
This leads to a number of the order of $10^{11}$ samples per
wall-normal grid point in case F10 and
F50, collected over the entire observation interval. 
The second set of data was obtained from
analysing stored snapshots of the flow field of which 90 were used
in each case. 
The latter set has been used to compute some additional statistical
quantities not stored during runtime. 
Since it provides a smaller number of
samples (roughly a factor of $10^3$ less), we will in the following check its
consistency with the more complete set accumulated at runtime. 

\begin{figure}
\begin{center}
    \begin{minipage}{3ex}
      \rotatebox{90}{\hspace{3ex}$y/H$}
    \end{minipage}
    \begin{minipage}{.45\linewidth}
      \includegraphics[width=1.\linewidth]
      {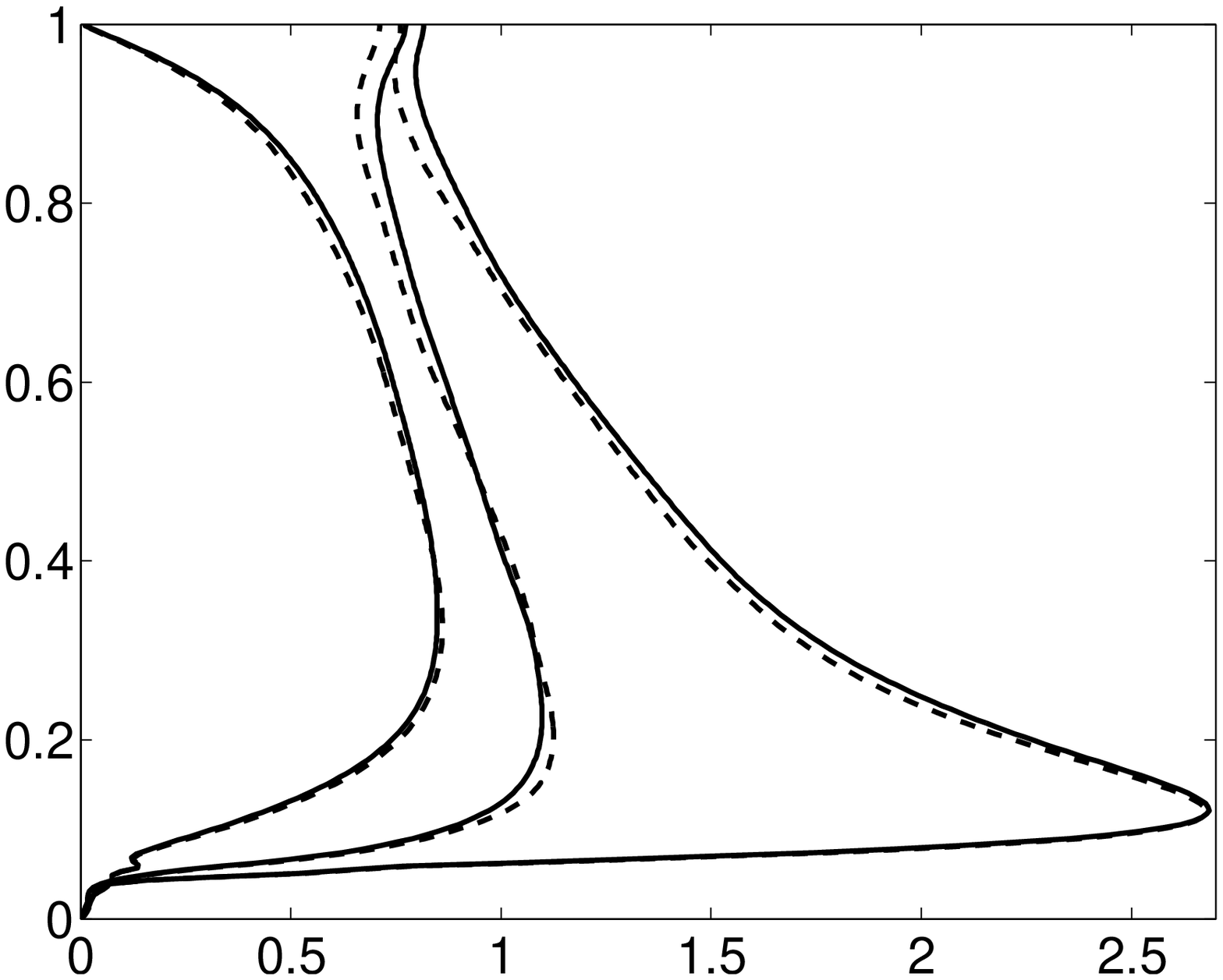}
      \hspace{-.2\linewidth}\raisebox{.65\linewidth}{$(a)$}
      \\ 
      \centerline{$u_{rms}^i / u_\tau$}
    \end{minipage}
    \begin{minipage}{.45\linewidth}
      \includegraphics[width=1.\linewidth]
      {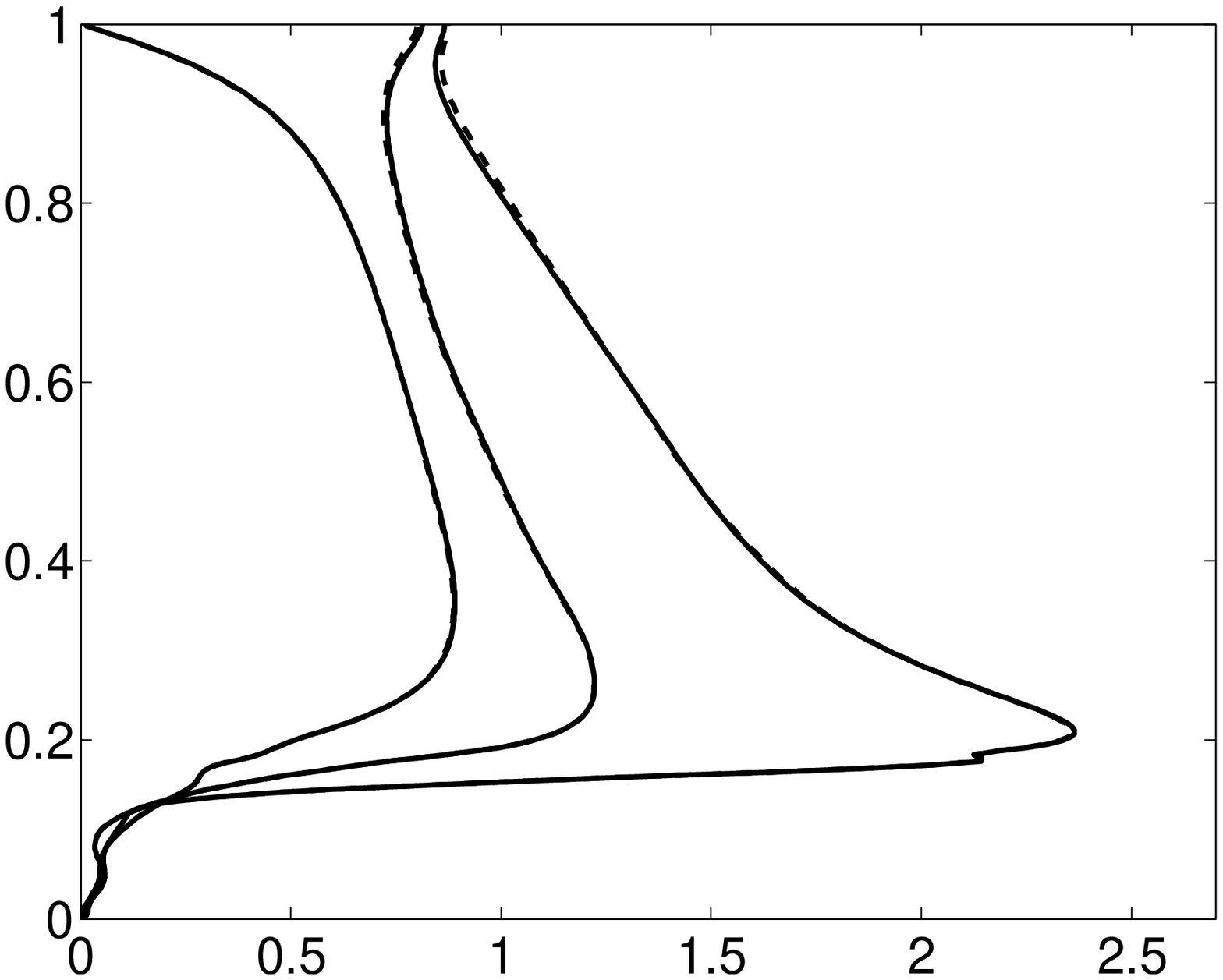}
      \hspace{-.2\linewidth}\raisebox{.65\linewidth}{$(b)$}
      \\ 
      \centerline{$u_{rms}^i / u_\tau$}
    \end{minipage}
    
    \caption{Comparison of velocity fluctuations normalised by $u_\tau$
      obtained from run-time (solid line) and from snapshots (dashed line)
      as a function of $y / H$. Curves from left to right are the
      components in 
      wall-normal ($v_{rms}/u_\tau$), spanwise
      ($w_{rms}/u_\tau$) and streamwise ($u_{rms}/u_\tau$) direction.
      Panel $(a)$ shows case F10, panel $(b)$ shows case F50.}
    \label{fig:comp_stat_runtime-snap-shots}
  \end{center}
\end{figure}

Figure \ref{fig:comp_stat_runtime-snap-shots} shows for each case the
second order moments of the velocity fluctuations 
obtained at run-time in comparison to the same quantities 
obtained from the snapshots of the simulations applying the
averaging operator as defined in (\ref{eqn:op_ave1}). 
The differences between the two data sets are small, measuring less
than 0.06$u_\tau$ (0.02$u_\tau$) in case F10 (F50). Incidentally, it
can be seen from the figures that the discrepancy is largest near the
open surface. We can therefore conclude, that the data set provided
from the 90 stored snapshots is sufficient for the purpose of
computing the quantities shown in figure \ref{fig:velrms_norm_D},
figure \ref{fig:3d-mean}, 
figure \ref{fig:3d-stream} and figure \ref{fig:3d-effect}. 

\end{appendix}